\documentstyle[12pt,ssi,epsf]{article}

\def\mevc {\ifmmode {\rm MeV}/c \else MeV$/c$\fi}
\def\mevcc {\ifmmode {\rm MeV}/c^2 \else MeV$/c^2$\fi}
\def\gevc {\ifmmode {\rm GeV}/c \else GeV$/c$\fi}
\def\gevcc {\ifmmode {\rm GeV}/c^2 \else GeV$/c^2$\fi}
\def\ra   {\rightarrow}
\newcommand{\Pt} {\ifmmode p_{\rm t} \else $p_{\rm t}$\fi}
\newcommand{\Et} {\ifmmode E_{\rm t} \else $E_{\rm t}$\fi}
\newcommand{\met} {\ifmmode \not\!\!E_{\rm t} \else 
		\mbox{$\not\!\!E_{\rm t}$}\fi}
\newcommand{\sigtt} {\ifmmode \sigma_{t\bar t} \else $\sigma_{t\bar t}$\fi}
\newcommand{\mtop} {\ifmmode m_{\rm top} \else $m_{\rm top}$\fi}
\newcommand{\lxy} {\ifmmode L_{\rm xy} \else $L_{\rm xy}$\fi}
\newcommand{\Ds} {\ifmmode D_{\mbox{\sl s}}^{-}
                       \else $D_{\mbox{\sl s}}^{-}$\fi}
\newcommand{\Bs} {\ifmmode B_{\mbox{\sl s}}^{0}
                       \else $B_{\mbox{\sl s}}^{0}$\fi}
\newcommand{\ds} {\ifmmode D_{\mbox{\sl s}}
                       \else $D_{\mbox{\sl s}}$\fi}
\newcommand{\bs} {\ifmmode B_{\mbox{\sl s}}
                       \else $B_{\mbox{\sl s}}$\fi}
\newcommand{\Dsl} {\ifmmode D_{\mbox{\sl s}}^{-} \ell^+
                       \else $D_{\mbox{\sl s}}^{-} \ell^+$\fi}
\newcommand{\ptrel} {\ifmmode p_{\rm t}^{\rm rel} 
                       \else $p_{\rm t}^{\rm rel}$\fi}
\newcommand{\ed} {\ifmmode \varepsilon D^2 \else $\varepsilon D^2$\fi}
\newcommand{\bjpsi} {\ifmmode B^0 \ra J/\psi K^0_S 
                       \else $B^0 \ra J/\psi K^0_S$\fi}
\newcommand{\bpipi} {\ifmmode B^0 \ra \pi^+ \pi^- 
                       \else $B^0 \ra \pi^+ \pi^-$\fi}
\newcommand{\sta} {\ifmmode \sin 2 \alpha \else $\sin 2 \alpha$\fi}
\newcommand{\stb} {\ifmmode \sin 2 \beta  \else $\sin 2 \beta$\fi}
\newcommand{\stg} {\ifmmode \sin 2 \gamma \else $\sin 2 \gamma$\fi}

\begin{document}

\title{Heavy Flavour Physics \\ From Top to Bottom}

\author{Manfred Paulini \\
Lawrence Berkeley National Laboratory \\
MS 50B-5239, Berkeley, California 94720 \\[0.4cm]
(Representing the CDF and D\O\ Collaboration)
}

\maketitle

\begin{abstract}
We review the status of heavy flavour physics at the Fermilab Tevatron
collider by summarizing recent top quark and $B$~physics
results from CDF and D\O. In particular we discuss the measurement of
the top quark mass and top production cross section as well as 
$B$ meson lifetimes 
and time dependent $B\bar B$ mixing results.
An outlook of perspectives for top and $B$
physics in Run~II starting in 1999 is also given.  
\end{abstract}

\section{Introduction}

In this article we review recent heavy flavour physics results from
the Tevatron $p\bar p$~collider at Fermilab, where heavy flavour
refers to the top quark as well as the bottom quark. After a brief
historical overview we summarize the status of top quark physics at
CDF and D\O\ in Sec.~\ref{top_sec}. In particular we discuss the
measurement of the top production cross section and
the top quark mass.
Section~\ref{b_sec} is devoted to
recent $B$ physics results at a hadron collider, where we concentrate
on $B$ hadron lifetimes and latest time dependent $B\bar B$ mixing
results. A brief look 
to the future, summarizing the prospects of top quark physics as well
as $B$~physics and $CP$~violation in Run~II starting in 1999, is given
in Sec.~\ref{outlook_sec}. We conclude with Section~\ref{conclude_sec}.       

\subsection{Historical Overview}

In 1977 the bottom quark was discovered as a resonance in the dimuon
mass spectrum in 400 GeV proton-nucleus collisions at Fermilab
\cite{yps}. Soon after the discovery the so-called
$\Upsilon$~resonances at a mass of about 9.5~\gevcc\  
were  confirmed in $e^+e^-$ collisions at the
DORIS storage ring at DESY \cite{PlutoDasp}. 
From the narrow resonance observed 
at $(9.46\pm0.01)$~\gevcc\ the PLUTO and DASP detectors were able to
determine the electronic width $\Gamma_{ee}$, which implied a charge
assignment of $-1/3$ for the $b$ quark.    
It took a few years until decays of $B$ mesons, bound states of a $b$~quark 
and a light quark, were observed in 1983 by the CLEO
collaboration \cite{cleob}. The weak isospin of the $b$ quark was
first extracted from the forward-backward asymmetry $A_{FB}$ in
$e^+e^- \ra b\bar b$ at PETRA \cite{afb_petra}, where the
measurement of $A_{FB} = (-22.8\pm6.0\pm2.5)$\% at $\sqrt{s} = 34.6$~GeV was 
found to be consistent with the Standard Model prediction of $A_{FB} = -25$\% 
assuming $I_3 = -1/2$ for the weak isospin of the $b$ quark. 
This indicated that the bottom quark has a partner and it can be
counted as the first evidence for the existence of the top quark. Finally,
concluding this brief historical overview with a link between top and
$B$ physics, $B^0 \bar B^0$ mixing was first 
observed by the ARGUS collaboration in 1987 \cite{bmix}. The
measurement of a large mixing parameter $x_d$ was the first indication
of a large top quark mass.   

Why are we discussing the history of $b$ quark physics? 
A certain pattern can be observed after the discovery of a new
quark. This pattern appears to repeat itself after the discovery of
the top quark. It looks
like right after the discovery of the $b$ quark an attempt was made to
confirm the discovery in a different environment like $e^+ e^-$
collisions. Then the fundamental quantities of the newly discovered
quark, like its charge and isospin, were determined. After sufficient
statistics is accumulated rare phenomena like $B \bar B$ oscillations
were searched for. Nowadays, $B$ physics is fully explored at $e^+e^-$
colliders and rare phenomena like $b \ra u$ transitions or $b \ra s
\gamma$ penguin decays are also studied \cite{PDG}. This raises the
question of why $B$ physics is studied at a $p\bar p$ hadron collider
which is a much more difficult environment to study low \Pt\
physics. We shall discuss this issue further on, in
Section~\ref{b_sec}.  

\subsection{Early Searches for the Top Quark}

The experimental search for the top quark begun soon after the
discovery of the $b$~quark. Between 1979 and 1984 measurements of $R$,
the ratio of the cross sections $\sigma(e^+e^- \ra $ hadrons) to
$\sigma(e^+e^- \ra \mu^+\mu^-)$, were performed at the PETRA $e^+ e^-$
storage ring 
up to a centre-of-mass energy of 46.8 GeV. The value of $R$ was found to
be consistent with Standard Model predictions without a top quark
contribution setting a lower bound on the top quark mass \cite{petra} of 
$> 23.3$ \gevcc. Later searches at $e^+e^-$ colliders were
also negative and limits on \mtop\ of half of the
centre-of-mass energy were set. These limits from TRISTAN \cite{tristan} as
well as SLC \cite{slc_top} and LEP \cite{lep_top} are listed in
Table~\ref{top_limits}. 

With the advantage of higher mass regions being accessible, the search
for the top quark was soon dominated by $p\bar p$ colliders, first
at the SPS collider ($\sqrt{s} = 630$~GeV) at CERN and 
then at the Tevatron ($\sqrt{s} = 1.8$ TeV)
at Fermilab. Initial results reported in 1984 by the UA1
collaboration \cite{UA1_top1} at the SPS 
seemed to be consistent with the production of a top quark of
mass $(40\pm10)$~\gevcc\ in $p\bar p \ra W \ra t \bar b$. The results
were based on the observation of 12 isolated lepton plus 2-jet events
with an expected background of approximately 3.5 events in 200~$n$b$^{-1}$. 
However, these first results were not supported by a
subsequent UA1~analysis \cite{UA1_top2} with a higher statistics data
sample setting a lower limit on \mtop\ of $> 52$~\gevcc. More sensitive
searches were performed later on by UA2 \cite{UA2_top} and with the
start of the Tevatron. CDF
\cite{CDF_top1,CDF_top2} and D\O\ \cite{D0_top} increased the limit on
the mass of the top quark to finally $> 131$ \gevcc\ in 1994 (see
Table~\ref{top_limits}). 

\begin{table}[tb]
\begin{center}
\vspace*{-0.6cm}
\begin{tabular}{|c|rl|c|c|}
\hline
Year & \multicolumn{2}{|c|}{Location} & Mass limit (95\% CL) & Ref.  \\
\hline
\hline
1979 & $e^+e^-$: & PETRA & $> 23.3$ \gevcc\ & \cite{petra} \\
1987 & $e^+e^-$: & TRISTAN & $> 30.2$ \gevcc\ & \cite{tristan} \\
1989 & $e^+e^-$: & SLC \& LEP & $> 45.8$ \gevcc\ & \cite{slc_top,lep_top} \\
\hline
1984 & $p \bar p$: & UA1 & $(40\pm10)$ \gevcc\ & \cite{UA1_top1} \\
1988 & $p \bar p$: & UA1 & $> 52$ \gevcc\  & \cite{UA1_top2} \\
1990 & $p \bar p$: & UA2 & $> 69$ \gevcc\  & \cite{UA2_top} \\
1990 & $p \bar p$: & CDF & $> 77$ \gevcc\  & \cite{CDF_top1} \\
1992 & $p \bar p$: & CDF & $> 91$ \gevcc\  & \cite{CDF_top2} \\
1994 & $p \bar p$: & D\O & $> 131$ \gevcc\  & \cite{D0_top} \\
\hline
\end{tabular}
\caption{Historical overview of searches for the top quark and limits
on the top quark mass.}
\label{top_limits} 
\end{center}
\vspace*{-0.5cm}
\end{table}

However, in April 1994 the CDF collaboration presented evidence for
top quark production \cite{CDF_top_evid,CDF_top_prd} with the
observation of 12 events consistent
with either two $W$~bosons, or a $W$ boson and at least one $b$
jet. The probability that the measured yield was consistent with the
expected background was 0.26\% corresponding to a $2.8\,\sigma$ effect.
Finally, in February 1995 the top quark was discovered by the CDF experiment
\cite{CDF_top_disc} and the D\O\ experiment \cite{D0_top_disc} at the
same time. 
Although top quark physics is still a 
relatively young field at the time of this conference, a lot of
progress has been made in 
understanding the top quark and its characteristics.

\section{Status of Top Quark Physics at CDF and D\O}
\label{top_sec}

\subsection{The Tevatron with the CDF and D\O\ Detectors}

At the Fermilab Tevatron, proton-antiproton collisions take place at a
centre-of-mass energy of $\sqrt{s} = 1.8$~TeV.
The Tevatron Run~I started delivering data in Dec.~1992 and finished
in Feb.~1996. During this period a total of about 110~$p$b$^{-1}$ and
100~$p$b$^{-1}$ 
of data were accumulated by the CDF and D\O\ experiment,
respectively. 
All results presented in this paper refer to the full Run~I statistics
unless otherwise noted.
The running period was devided up in a so-called Run~Ia
from Dec.~1992 through Aug.~1993 and Run~Ib from Dec.~1993 to Feb.~1996.
The collected integrated luminosities by CDF and D\O\ were 
$\approx 19.3$~$p$b$^{-1}$ and 
$\approx 15$~$p$b$^{-1}$ for Run~Ia as well as 
$\approx 90$~$p$b$^{-1}$ and 
$\approx 85$~$p$b$^{-1}$ for Run~Ib, respectively. During that time the
Tevatron operated with six bunches of protons and six bunches of
antiprotons crossing every 3.5 $\mu$s at the experiments interaction
regions. During Run~Ib the highest instantaneous luminosities, which
were reached, 
were around $2.5\cdot10^{31}$~cm$^{-2}$s$^{-1}$. At this luminosity on
average two interactions accured per beam crossing.

\begin{figure}[tb]\centering
\begin{picture}(145,75)(0,0)
\centerline{
\epsfysize=7.0cm
\epsffile[0 0 430 330]{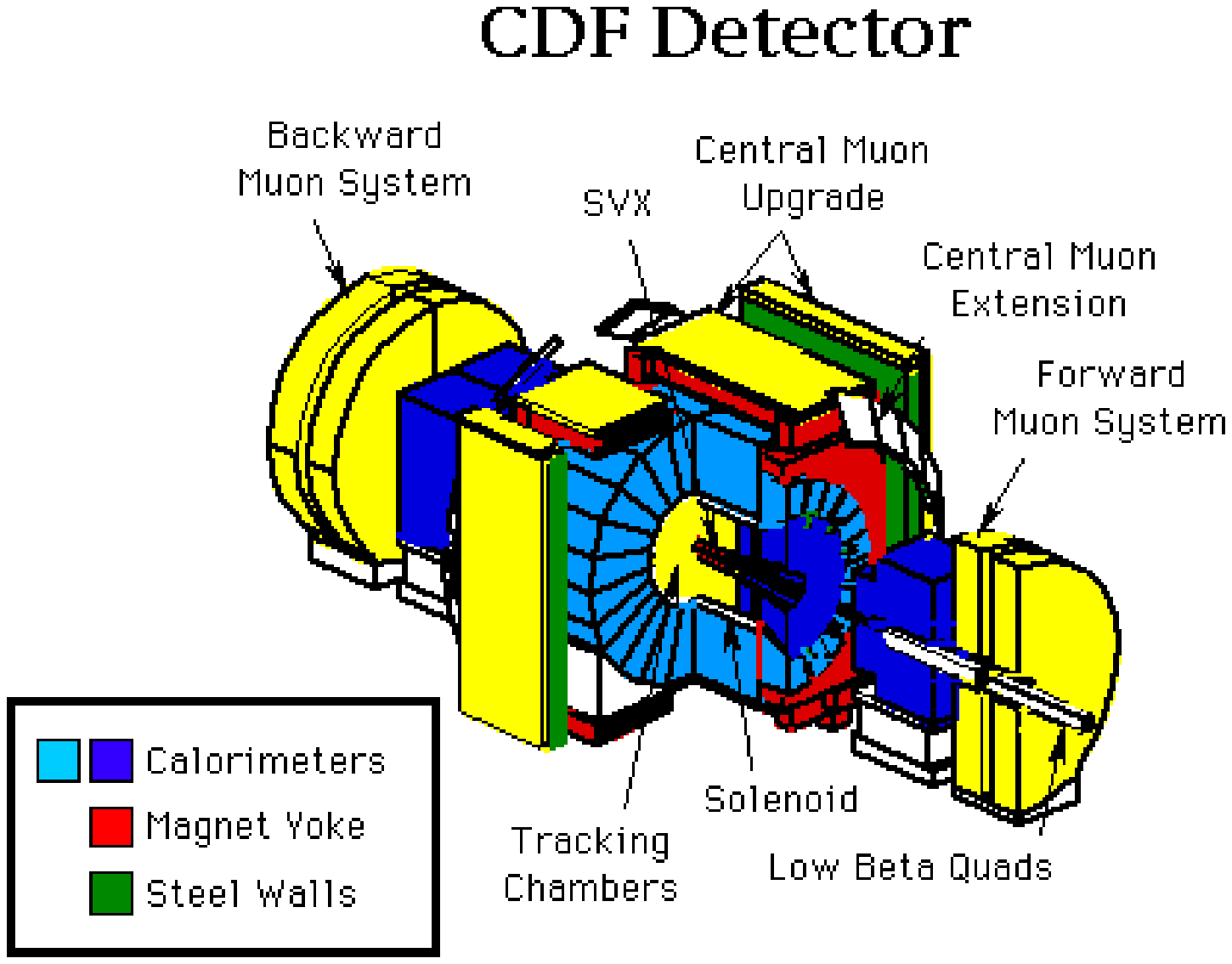}
\epsfysize=7.0cm
\epsffile[80 175 540 680]{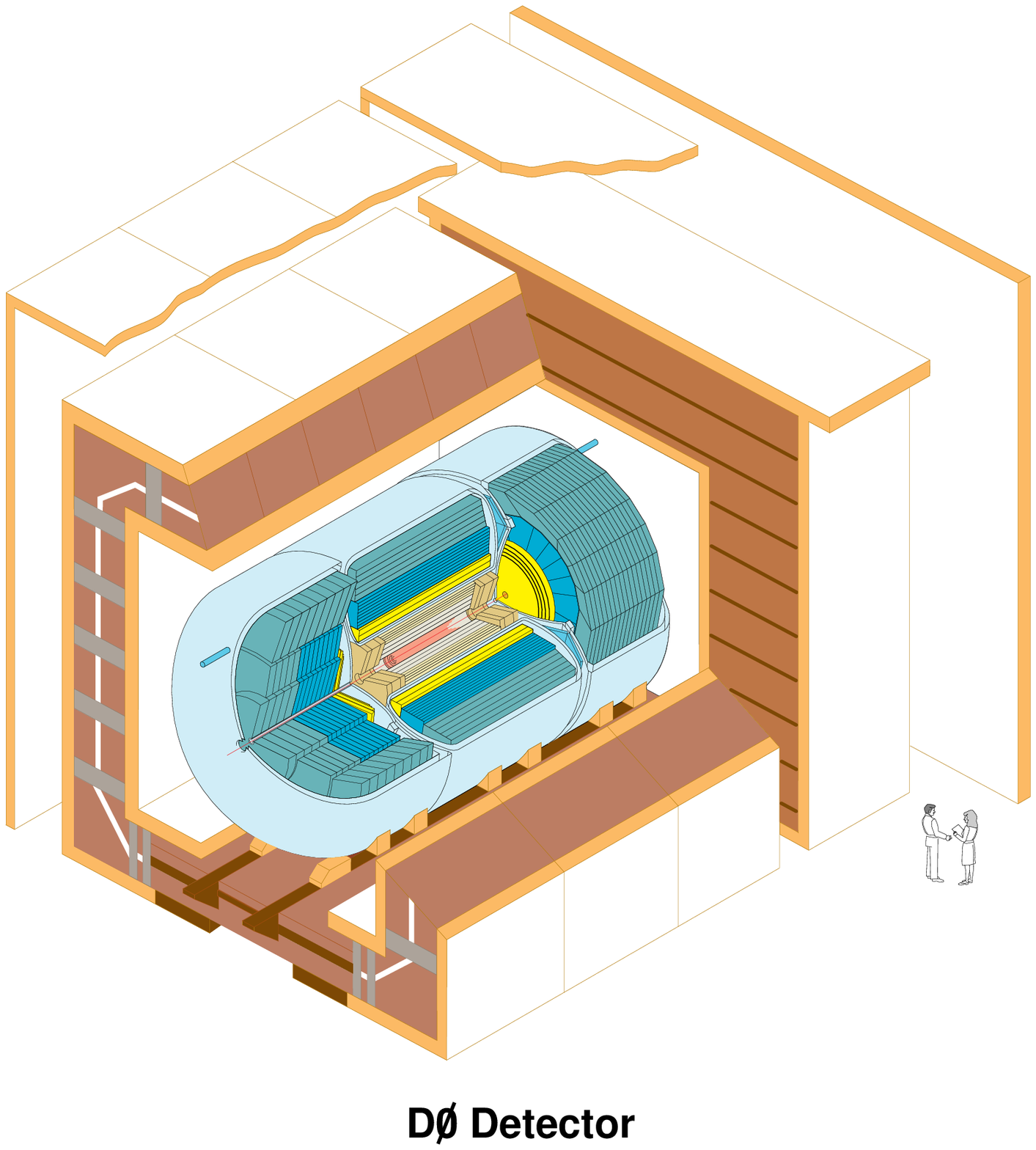}
}
\put(-150,70){\Large\bf (a)}
\put(-60,70){\Large\bf (b)}
\end{picture}
\caption{Schematic view of (a) the CDF detector and (b) the D\O\ experiment.}
\label{cdf_d0}
\end{figure}

\subsubsection{The CDF Detector}

The Collider Detector at Fermilab (CDF) is a general purpose detector
to efficiently identify leptons and hadronic jets, as well as charged
particles in $p\bar p$
collisions \cite{cdf_det}. A schematic view of the CDF detector is
shown in Figure~\ref{cdf_d0}a). 
Three devices inside the 1.4 T solenoid 
are used for the tracking of charged particles: 
the silicon vertex detector (SVX), a set of vertex time projection
chambers (VTX), and the central tracking chamber (CTC).

The SVX 
consists of four layers of silicon microstrip detectors 
located at radii between 3.0~cm and 7.9~cm from 
the interaction point 
and provides spatial measurements in the $r$-$\varphi$ plane
with a resolution of 13 $\mu$m, 
giving a track impact parameter resolution of about 
$(13 + 40/\Pt)~\mu$m~\cite{cdf_svx}, 
where \Pt\ is the transverse momentum of the track in GeV/$c$.
Throughout this article 
$\varphi$ is the azimuthal angle, 
$\theta$ is the polar angle measured from the proton direction, 
and $r$ is the radius from the beam axis ($z$-axis). 
The geometric acceptance of the SVX is $\sim 60\%$ as it 
extends to only $\pm~25$ cm from the nominal
interaction point whereas the Tevatron beam has an 
RMS width of $\sim 30$ cm along the beam direction.
The transverse profile of the beam is circular and has an RMS  
of $\sim 25$-35 $\mu$m.

The VTX reconstructs intermediate tracks in the $r$-$z$ plane and is
used to determine the primary interaction vertex. Surrounding the SVX
and VTX is the 
CTC, a cylindrical drift chamber containing 84 layers grouped into nine
alternating superlayers of axial and stereo wires. 
It covers the pseudorapidity interval $|\eta| < 1.1$, 
where $\eta=-\ln[\tan(\theta/2)]$.
The \Pt\ resolution of the CTC combined with the SVX is 
$\sigma(\Pt)/\Pt = ((0.0066)^2 + (0.0009\,\Pt)^2)^{1/2}$, with \Pt\
measured in \gevc.  

Outside the solenoid are electromagnetic (CEM) and hadronic (CHA)
calo\-rimeters 
($|\eta|<1.1$) that employ a projective tower geometry
with a 
segmentation of $\Delta\eta \times \Delta\varphi \sim 0.1 \times 15^{\circ}$.
The sampling medium is composed of scintillators layered with lead and
steel absorbers.
A layer of proportional wire chambers (CES) 
is located near shower maximum in the CEM and 
provides a measurement of electromagnetic shower profiles 
in both the $\varphi$ and $z$ directions.
Plug and forward calorimeters instrument the region of $1.1 < |\eta| <
4.2$ and consist of gas proportional chambers as active media and lead
and iron as absorber materials. 
The overall resolution for the CDF
central calorimeter is 
$\sigma_E / E = (13.5\% / \sqrt{E_{\rm t}}) + 2\%$ 
for electromagetic showers and
$\sigma_E / E = (75\% / \sqrt{E_{\rm t}}) + 3\%$ for hadrons. 

Several muon subsystems in the central region are used. 
The central muon chambers (CMU) and the central muon upgrade chambers
(CMP) cover 80\% for $|\eta| \leq 0.6$, while the central muon
extention chambers (CMX) extent the coverage up to $|\eta| < 1.1$.
The CMP chambers are  
located behind eight interaction lengths of material.

\subsubsection{The D\O\ Detector}

The D\O\ detector~\cite{d0_det} consists of three primary systems: a
nonmagnetic tracking device, a uranium-liquid argon calorimeter, 
and a muon spectrometer.  
A perspective view of the D\O\ detector can be seen in
Fig.~\ref{cdf_d0}b). 
The tracking system consists of four detector
subsystems: a 3-layer vertex drift chamber, a transition radiation detector
for additional electron identification, a 4-layer
central drift chamber, and two forward drift chambers.  The tracking system
provides charged particle tracking over the region $|\eta|<3.2$.

The hermetic, compensating, uranium-liquid argon sampling calorimeter
is divided into three parts: a central calorimeter and two end calorimeters.
They each consist of an electromagnetic section, a fine hadronic
section, and a coarse hadronic section, housed in a steel cryostat.  The
calorimeter covers the pseudorapidity range $|\eta|<4.2$ with fine
longitudinal segmentation (8 depth segments) and fine transverse segmentation
($\Delta\eta \times \Delta\phi = 0.1 \times 6^{\circ}$,
and $\Delta\eta \times \Delta\phi = 0.05 \times 6^{\circ}$ in the third
depth segment of the electromagnetic calorimeter). 
The overall resolution for the D\O\
calorimeter is 
$\sigma_E / E = (15\% / \sqrt{E}) + 0.4\%$ 
for electromagetic showers and
$\sigma_E / E = (50\% / \sqrt{E})$ for hadrons. 

The muon system, used for
the identification of muons and determination of their trajectories and
momenta, consists of five separate solid-iron toroidal magnets, together with
sets of proportional drift tube chambers. The muon system covers 
$|\eta|<3.3$. The material in the calorimeter and iron toroids combined varies
between 13 and 19 interaction lengths.
The achieved momentum resolution is
$\sigma_p / p = 0.2  + 0.003\, p$ (with $p$ measured in \gevc)
for the rapidity range $|\eta| < 3.3$.

\subsection{Top Quark Production at the Tevatron}

In $p\bar p$ collisions at $\sqrt{s} = 1.8$ TeV, the dominant top
quark production mechanism is $t\bar t$ pair production through $q\bar
q$ annihilation. Gluon-gluon fusion, which will be the main
production mechanism at LHC energies, contributes to about 10\% at
the Tevatron. The production of single top quarks through the
creation of a virtual $W$ boson is estimated to be about one order of
magnitude lower than the $t\bar t$ pair production at $\sqrt{s} = 1.8$ TeV.
 
During the Tevatron Run~I about $5\cdot10^{12}$ $p\bar p$ collisions
occured within the CDF and D\O\ detectors but only about 500 $t\bar t$
pairs have been produced. The top quark production
cross section is about ten orders of magnitudes lower than the total
inelatic cross section at the Tevatron. Comparing $\sigma_{t\bar t}$
to other physics processes like $W$ boson production 
shows that $\sigma_{t\bar t}$ is still three orders of magnitude lower
than the $W$ cross section. This means
the challenge in discovering and studying top quarks is to
separate them from backgrounds in hadron collisions.  

\subsection{Top Quark Decay Signature}
 
\begin{figure}[tbp]\centering
\begin{picture}(145,55)(0,0)
\centerline{
\epsfysize=6.0cm
\epsffile{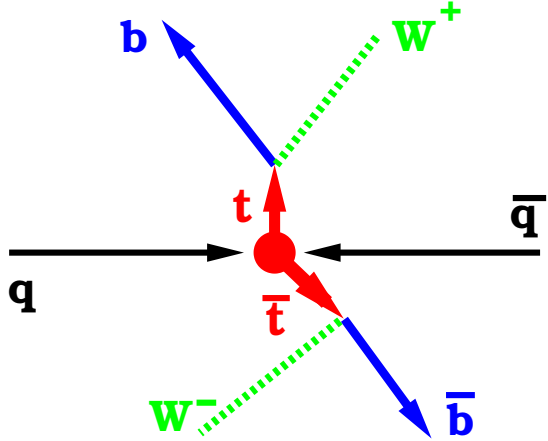}}
\end{picture}
\caption{Top decay signature within the Standard Model.}
\label{top_decay}
\end{figure}

Within the Standard Model, each of the pair produced top quarks decays
almost exclusively 
into a $W$ boson and a $b$ quark as shown in Fig.~\ref{top_decay}. The
$W$ boson decays into either a lepton-neutrino or 
quark-antiquark pair. The top decay signature depends primarily on the
decay of the $W$ boson. Events are classified by the number of
$W$'s that decay leptonically. In this context lepton refers to an electron
or a muon.

\begin{figure}[p]\centering
\begin{picture}(145,70)(0,0)
\put(-6,68){\large\bf (a)}
\put(74,68){\large\bf (b)}
\centerline{
\epsfysize=7.0cm
\epsffile{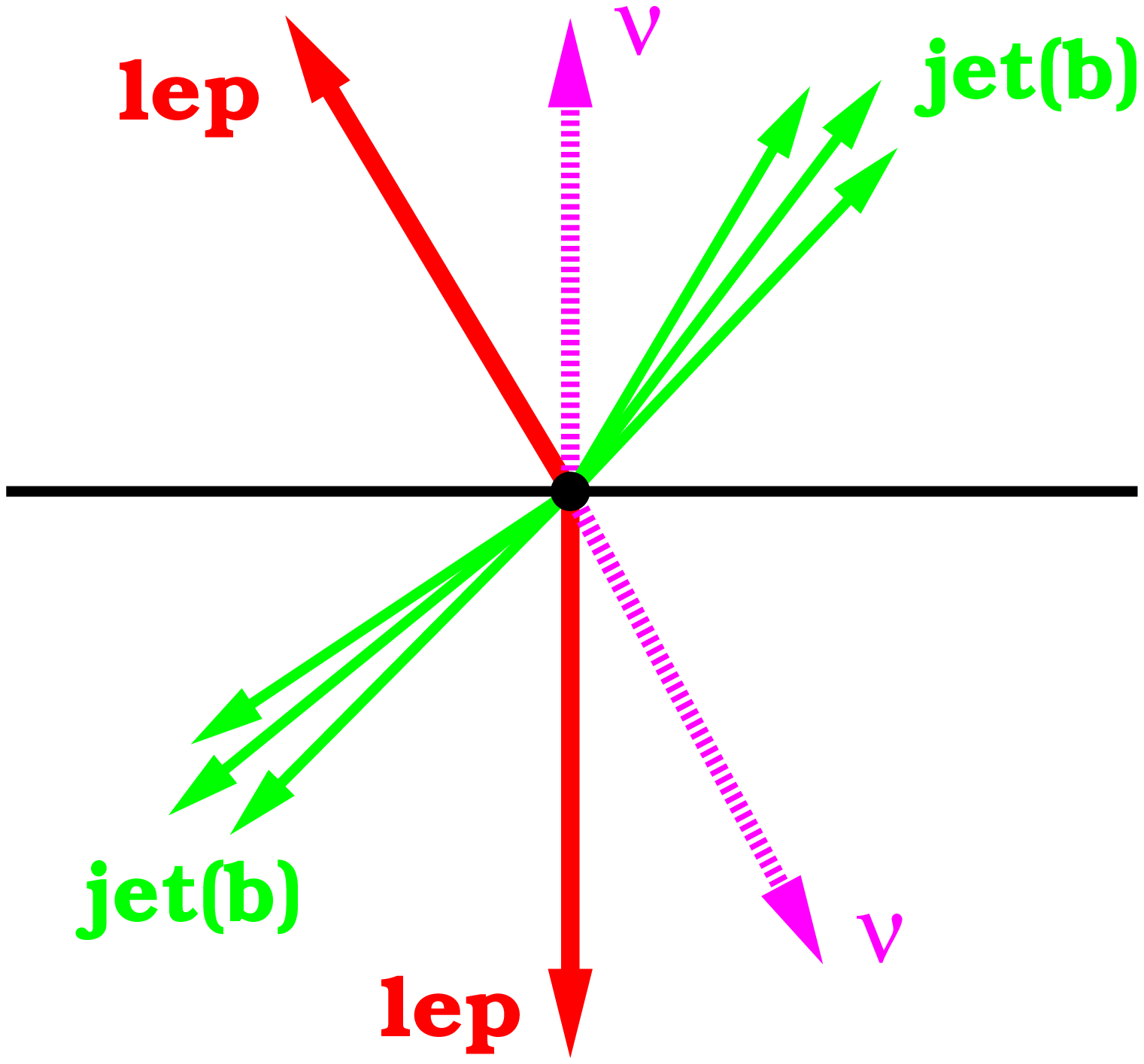}
\hspace*{0.5cm}
\epsfysize=7.0cm
\epsffile{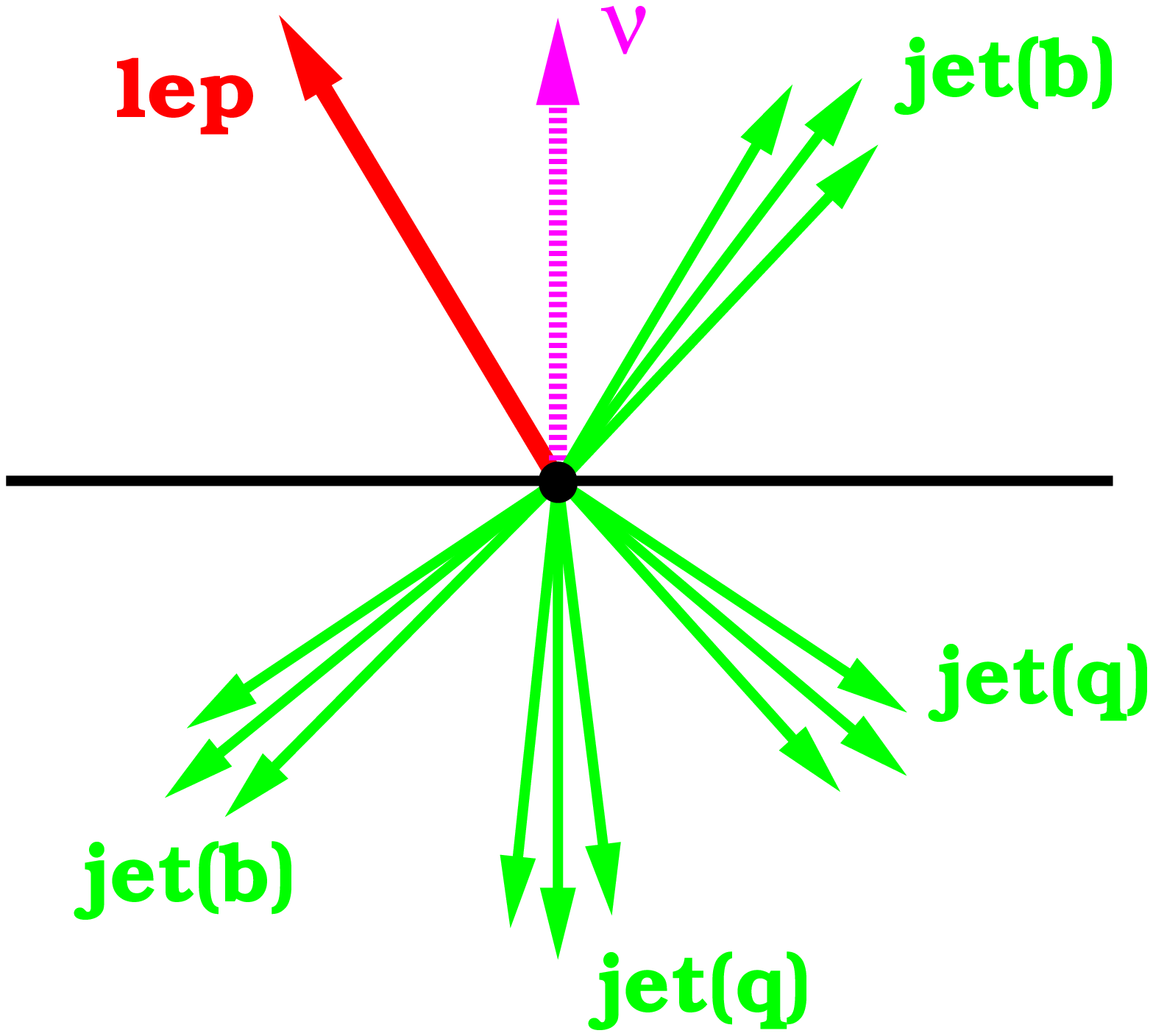}
}
\end{picture}
\caption{Top decay signature of (a) the dilepton channel and (b) the
lepton plus jets channel.}
\label{top_dec_sig1}
\end{figure}

If both $W$ bosons decay leptonically into $W \ra \ell\nu$, we call it the
`dilepton channel', where the final state consists of
$\ell^- \bar\nu \ell^+ \nu b\bar b$ as can be seen in
Figure~\ref{top_dec_sig1}a). This means there will be two leptons and
two jets originating from $b$ quarks in the event. Due to both $W$'s
decaying semileptonically, this top decay mode has a small branching
fraction of about 5\%. 
 
If one of the $W$ bosons decays leptonically into $W \ra \ell\nu$ and
the other into $W \ra q\bar q^{\prime}$, we call it the
'lepton plus jets channel', where the final state consists of
$\ell \nu q\bar q^{\prime} b\bar b$ as shown in
Figure~\ref{top_dec_sig1}b). 
In this case one lepton and four jets, where 
two jets originate from $b$ quarks, can be seen in the event. 
This decay mode happens in about 30\% of the time.
 
\begin{figure}[p]\centering
\begin{picture}(145,75)(0,0)
\centerline{
\epsfysize=7.5cm
\epsffile{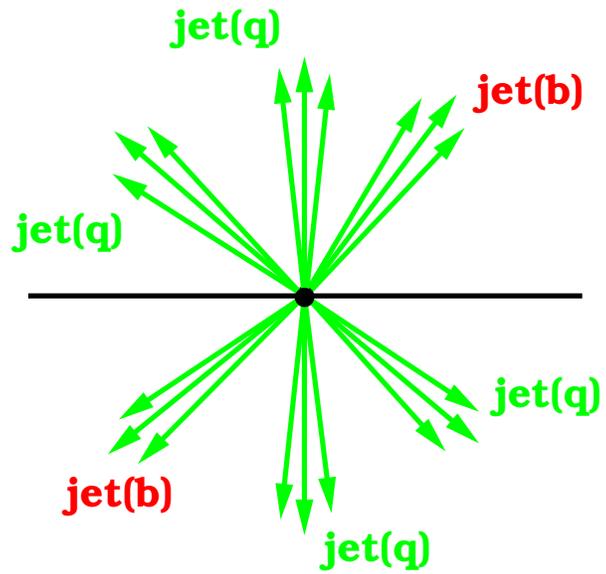}
}
\end{picture}
\caption{Top decay signature of the all hadronic channel.}
\label{top_dec_sig2}
\end{figure}

If both $W$ bosons decay into quark pairs as $W \ra q\bar q^{\prime}$,
we call it the `all hadronic channel'. The final state consists of
$q\bar q^{\prime} q\bar q^{\prime} b\bar b$ (see Fig.~\ref{top_dec_sig2}). 
In this case we can find six jets in the event, where
two jets originate from $b$ quarks. 
This top decay mode occurs at a large rate of about 44\%. 

In additon there
are about 21\% of $t\bar t$ decays to final states containing $\tau$
leptons. According to the $\tau$ decay these top decays are either
classified as dilepton or lepton plus jet events, if the tau decays
into $e$ or $\mu$ or as all hadronic events if the tau decays
hadronically. In the following we are going to use these decay channels
to discuss top decays at the Tevatron.
  
\subsection{The Top Dilepton Channel}
 
The signature of the dilepton channel (see Fig.~\ref{top_dec_sig1}a)
is two isolated high \Pt\ leptons ($e$ or $\mu$) and missing energy 
(\met) from the two neutrinos that escape the detector
unobserved. In addition, two or more jets can be found in the
event. The event selection also relies on kinematic requirements.
The dominant backgrounds are from $WW$ diboson
production plus jet 
activity where both $W$'s decay leptonically and from jet plus $Z$
production with $Z \ra \tau \tau$ followed by subsequent leptonic
$\tau$ decays. Events where the $e^+e^-$ or $\mu^+\mu^-$ invariant mass is
within  $75~\gevcc < m_{\ell\ell} < 105~\gevcc$ are considered as
$Z \ra \ell\ell$ candidates and removed in the event selection. Fake
leptons and Drell Yan production of lepton pairs are further sources of
background. These backgrounds are estimated from data as well as 
Monte Carlo predictions. 

The dilepton channel has a good signal to background ratio, but low
statistics. Due to the two neutrinos, this channel is not ideal for a
determination of the top quark mass. The dilepton event summary for
the CDF and D\O\ experiment is shown in Table~\ref{dilep_sum}.
Both experiments find a few events on small backgrounds in the $e\mu$,
$ee$ or $\mu\mu$, and $e$ or $\mu$ plus $\tau$ mode. 
The expected division of the dilepton signal events 
is consistent with the data observed by CDF and D\O.
For example, in the case of CDF 58\% $e\mu$, 27\% $\mu\mu$, and 15\% $ee$
events are expected.
In Tab.~\ref{dilep_sum} the expected
yield from $t\bar t$ production is also listed. 
It is based on determinations of the 
top cross section \cite{laenen} for a top quark mass of 175~\gevcc.

\begin{table}[tbp]
\begin{center}
\vspace*{-0.6cm}
  \begin{tabular}{|l||c|c|}
   \hline 
   Sample    &    D\O      &   CDF   \\
   \hline 
   \hline 
  $e \mu$   &     3         &   7      \\
   Background  &   0.4 $\pm$ 0.1        &  0.76 $\pm$ 0.21       \\
   Expected Yield  &  1.7 $\pm$ 0.3 & 2.4 $\pm$ 0.2\\
    \hline
  $e e$ or $\mu \mu$   &    2        &   2      \\
   Background  &   1.2 $\pm$ 0.4        &   1.23 $\pm$ 0.36       \\
   Expected Yield  &  1.4 $\pm$ 0.1  & 1.6 $\pm$ 0.2 \\
    \hline
  $e$ or $\mu + \tau$   &    2       &    4     \\
   Background  &        1.4 $\pm$ 0.5       &  1.96 $\pm$ 0.35       \\
   Expected Yield  &    1.4 $\pm$ 0.1         &   0.7 $\pm$ 0.1 \\
   \hline
\end{tabular}
\caption{Event summary for the dilepton channel. 
	The expected yield is based on determinations of the 
	top cross section \cite{laenen} for a top quark mass of 175~\gevcc.}
\label{dilep_sum} 
\end{center}
\vspace*{-0.5cm}
\end{table}

\subsection{The Top Lepton plus Jets Channel}
\label{toplepjet}
 
The signature of the lepton plus jets channel (see Fig.~\ref{top_dec_sig1}b)
is one isolated high \Pt\ lepton ($e$ or $\mu$), missing energy 
(\met) from the neutrino and four jets where two of them are from $b$ quarks.
The dominant backgrounds are from $W$ plus jet production including $W$
plus $b\bar b$ production. In this channel it is important to reduce
the backrounds ($S:B \approx 1:4$), where both experiments follow
different strategies. 

To reduce background, CDF tags the $b$ jets in the event through a
so-called `soft lepton tag' and a `SVX tag'. The first technique
identifies $b$ jets by searching for a lepton from $b \ra \ell X$ or 
$b \ra c \ra \ell X$ decays,
which have branching fractions of about 10\% each. 
Since these leptons typically have lower momenta
than the leptons from the primary $W$ decay, this technique is known as
`soft lepton tag' (SLT). It looks for electrons and muons 
by matching tracks from the central
drift chamber with
electromagnetic energy clusters in the calorimeter or track segments in the   
muon chambers. The \Pt\ threshold is at 2 \gevc. The
efficiency for SLT tagging a $t\bar t$ event is $(18\pm2)\%$, and the
typical fake rate per jet is about 2\%. Details of the SLT algorithm
can be found in Ref.~\cite{CDF_top_prd}

\begin{figure}[tbp]\centering
\begin{picture}(145,70)(0,0)
\centerline{
\epsfysize=7.0cm
\epsffile[10 10 600 740]{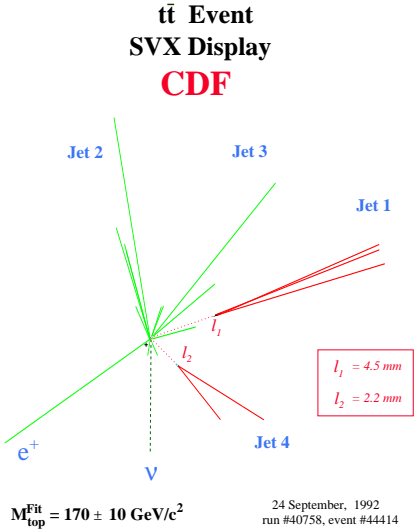}}
\end{picture}
\caption{Example of a SVX tagged event where both $b$ jets are SVX tagged.}
\label{svx_event}
\end{figure}

The second, more powerful $b$ tagging technique exploits the finite
lifetime of $b$ hadrons by searching for a secondary decay vertex with
CDF's silicon vertex detector. This technique is known as the `SVX
tag'. The algorithm begins by searching for displaced vertices
containing three or more tracks which satisfy a loose set of track
quality requirements. If no such vertices are found in an event,
two-track vertices that satisfy more stringent quality cuts are
accepted. A jet is defined to be tagged if it contains a secondary
vertex which is displaced from the primary vertex with a significance
of greater than three. The efficiency for SVX tagging a $t\bar t$
event is $(41\pm4)\%$, while the fake rate is only $\approx 0.5\%$. 
More information on the SVX tag can be found in
Ref.~\cite{CDF_top_prd,CDF_top_disc}
An example of a SVX tagged event display can be found in Fig.~\ref{svx_event}. 
Both $b$ jets are SVX tagged and well separated from
the primary interaction vertex by 2.2 mm and 4.5 mm, respectively.

The D\O\ experiment makes use of two different approaches to reduce
the background in the lepton plus jets channel. D\O\ uses kinematic
and topological cuts as well as $b$ tagging via soft muon tagging. The
first approach exploits the fact that the large top quark mass gives
rise to kinematically different events. Jets from $t\bar t$ decays tend to be
more energetic and more central than from typical QCD background
events. In addition $t\bar t$ events as a whole are more spherical 
while QCD jet production results in more planar event shapes. 
Top enriched data samples can therefore be selected with a set of
topological and kinematic cuts like the total hadronic activity in the
event, $H_T = \sum \Et^{\rm jet}$, which can be combined with the
aplanarity $\cal A$ of the $W$ plus jets system. 
A value of $\cal A =$ 0 indicates a planar event shape, while
$\cal A =$ 1/2 reflects a spherical event shape.
In addition, a third kinematic variable with discrimating power, the
total leptonic transverse energy, $\Et^L = \Et^{\rm lep} + \met$, is
also used. The second D\O\ approach uses $b$ tagging via muon tags  
through  $b \ra \mu X$ and $b \ra c \ra \mu X$ decays.
The typical fake rate for background events is at the $\approx 2\%$
level for tagging muons of $\Pt > 4 $ \gevc. For more details on both
techniques see Ref.~\cite{D0_top_disc}

The lepton plus jets event summary for
the CDF and D\O\ experiment is shown in Table~\ref{lepjet_sum}.
CDF finds 34 events with at least one SVX tag on a background of
$(8.0\pm1.4)$ events, while $(19.8\pm4.0)$ events are expected
from $t\bar t$ production, where a top quark mass of 175~\gevcc\ has
been assumed. The CDF events shape analysis is based on only 67 $p$b$^{-1}$.
 
\begin{table}[tbp]
\begin{center}
\vspace*{-0.6cm}
  \begin{tabular}{|l||c|c|}
   \hline 
   Sample    &    D\O     &   CDF   \\
   \hline 
   \underline{Event Shape:}   &            &         \\ 
   Observed   &   21       &   22      \\
   Background   &    9.2 $\pm$ 2.4       &  7.2 $\pm$ 2.1        \\
   Expected Yield  &   12.9 $\pm$ 2.1  &   - \\
   \hline
   \underline{Lepton tag (SLT):}   &            &         \\
   Observed   &    11      &  40       \\
   Background  &    2.5 $\pm$ 0.4    &  24.3 $\pm$  3.5      \\
   Expected Yield &   5.2 $\pm$ 1.0  &  9.6 $\pm$ 1.7  \\
   \hline
   \underline{Displaced Vertex (SVX tag):}   &            &         \\
   Observed   &    -      &    34     \\
   Background  &    -      &  8.0 $\pm$ 1.4       \\
   Expected Yield & - &   19.8 $\pm$ 4.0 \\
   \hline
\end{tabular}
\caption{Event summary for the lepton plus jets channel. 
	The expected yield is based on determinations of the 
	top cross section \cite{laenen} for a top quark mass of 175~\gevcc.}
\label{lepjet_sum} 
\end{center}
\vspace*{-0.5cm}
\end{table}

\subsection{The Top All Hadronic Channel}
 
The signature of the all hadronic channel (see Fig.~\ref{top_dec_sig2})
is nominally six jets where two of them are from $b$ quarks, no
leptons, and low \met. Since not all jets are always observed, a jet
multiplicity of at least five is required. 
In order to overcome the huge background from QCD multijet
production, $b$ tagging alone is not sufficient and kinematic cuts are
used in addition.
If the backgrounds can be controlled, the all hadronic channel would
be the ideal way to determine the top mass because no neutrinos are
present and all objects of the top decay are measured in the detector.

After a set of kinematic cuts is applied in the CDF all hadronic
analysis, at least five jets are required, where the leading jets have to
pass an aplanarity cut. In addition at least one jet has to be SVX
tagged. The efficiency of a SVX tag in a $t\bar t$ event is $(47\pm5)\%$ 
in the all
hadronic mode, slightly larger than in the lepton plus
jets channel due to the presence of additional charm tags from $W \ra
c\bar s$. CDF observes 192 events on a predicted background of
$137\pm11$ events as can be seen in Table~\ref{allhad_sum}. 

Recently, D\O\ also reported on a search for $t\bar t$ pairs in the
all hadronic channel. They require at least six jets within $|\eta|<2$
and use several additional kinematic quantities. Finally, a soft muon
tag has to be present in at least one of the jets. 15 $\mu$ tagged
events are observed for an expected background of $11\pm2$ events (see
Table~\ref{allhad_sum}).  

\begin{table}[tbp]
\begin{center}
\vspace*{-0.6cm}
  \begin{tabular}{|l||c|c|}
   \hline 
   Sample    &    D\O      &   CDF   \\
   \hline 
   Observed  &       15       &  192       \\
   Background  &     11 $\pm$ 2       &  137.1 $\pm$ 11.3       \\
   Expected Yield  &   4.5 $\pm$ 0.5  &  26.6 $\pm$ 9.1 \\
   \hline
\end{tabular}
\caption{Event summary for the all hadronic channel. 
	The expected yield is based on determinations of the 
	top cross section \cite{laenen} for a top quark mass of 175~\gevcc.}
\label{allhad_sum} 
\end{center}
\vspace*{-0.5cm}
\end{table}

In summary, we explored top quark production through three different top decay
modes, the dilepton, the lepton plus jets, and the all hadronic channel. 
We reviewed the current yield of top decays in these three decay
modes observed by the CDF and D\O\ experiment. As was true after the
discovery of 
the bottom quark, the top quark has been confirmed in different decay modes.
In the following, we use these top event candidates to measure
the fundamental 
quantities of the top quark: the top production cross section \sigtt\
and the top quark mass \mtop.

\subsection{The Top Production Cross Section}

The measurement of the top production cross section \sigtt\ is of
interest for several reasons. It checks QCD calculations of top
production, which have been performed by several 
groups \cite{laenen,nason,berger,catani},
and it provides an important benchmark for estimating top yields in
future high statistics experiments at the Tevatron and LHC. In
addition, a value of the top cross section significantly different
from the QCD prediction could indicate nonstandard top prodcution or
decay mechanisms. 

\begin{figure}[tbp]\centering
\begin{picture}(145,90)(0,0)
\centerline{
\epsfysize=9.0cm
\epsffile[100 130 600 660]{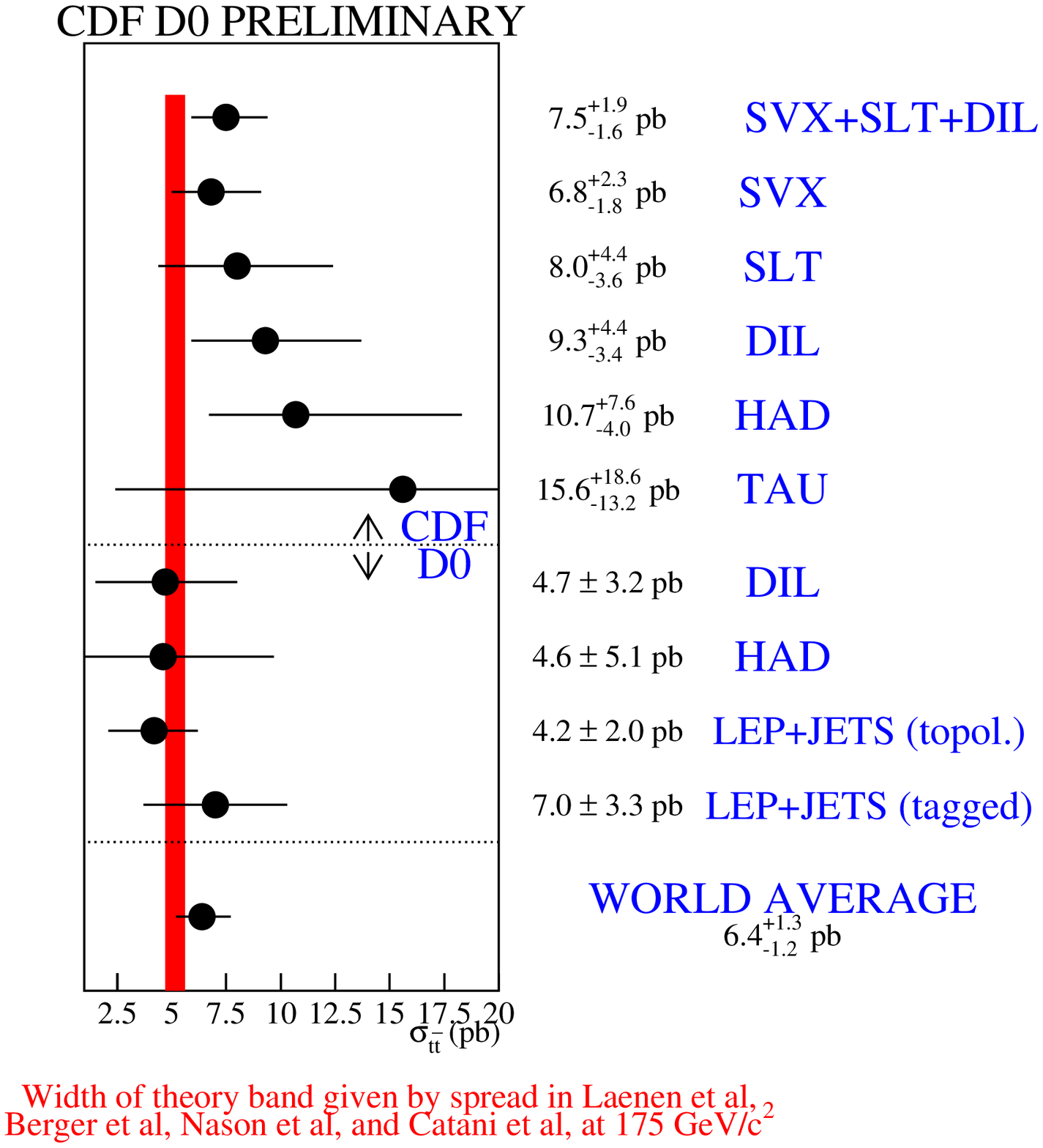}}
\end{picture}
\caption{Summary of the CDF and D\O\ top quark production cross
section measurements in different decay channels.}
\label{topxsec_sum}
\end{figure}

The measurement of the top production cross section \sigtt\ is
straight forward:
\begin{equation}
\sigtt = \frac{N_{\rm obs} - N_{\rm bkg}}{A\, {\cal L}}.
\label{eq:dilution}
\end{equation}
The number of predicted background events $N_{\rm bkg}$ is subtracted
from the number of observed top candidates $N_{\rm obs}$ and divided
by the acceptance $A$ of the sample selection and the integrated
luminosity $\cal L$ of the used data set. The measurement of \sigtt\
has been determined in each decay channel individually as detailed in
Fig.~\ref{topxsec_sum}. The results of the different \sigtt\
measurements from CDF and D\O\ can be compared to each other and to
theoretical predictions indicated by the dark band also shown in
Fig.~\ref{topxsec_sum}. The width of the theory band is given by the
spread in the theoretical predictions of Laenen et al, \cite{laenen}
Berger et al, \cite{berger} Nason et al, \cite{nason} and Catani et al
\cite{catani}. From all these measurements of \sigtt\ a world average
top production cross section 
\begin{equation}
\sigtt = (6.4 ^{+1.3} _{-1.2})\ p{\rm b}
\end{equation}
can be determined, which is slightly larger but in good agreement with
the theoretical predictions.

\subsection{Measurement of the Top Quark Mass}

The top quark mass \mtop\ is a fundamental parameter in the Standard Model. A
precise determination of \mtop\ is therefore the most
important measurement of the CDF and D\O\ experiment in Run~I. The
goal is to use
as many decay modes as possible and to measure \mtop\ as accurately as
possible. The use of several methods also allows cross checking of the
different techniques and studying of systematic uncertainties. An
incredible amount of work exploring these issues has been done by both
experiments. This resulted in significantly improved measurements of
\mtop\ since the discovery publications~\cite{CDF_top_disc,D0_top_disc}. 
The uncertainties on the top quark mass measurements by CDF and D\O\
have been improved by a factor of more than two. In such a short time,
this is a respectable amount of progress in understanding the newly discovered
top quark.  

\begin{figure}[tbp]\centering
\begin{picture}(145,70)(0,0)
\centerline{
\epsfysize=7.0cm
\epsffile[10 320 600 750]{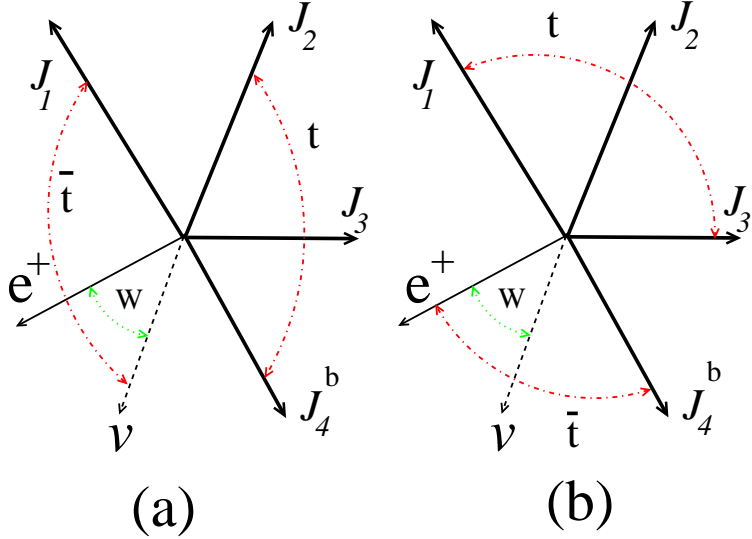}}
\end{picture}
\caption{Illustration of different assignments of the observed objects in
a $t \bar t$ decay to the originating partons.} 
\label{mapjet}
\end{figure}

The primary dataset for measuring the top quark mass is the lepton
plus jets sample. This is the decay mode with the most power. The
background in the all hadronic channel is too large and the statistics
of the dilepton
mode is too weak together with the additional complication of two
missing neutrinos. Thus, the preferred method of determining \mtop\ by
both experiments, is a constrained fit to the lepton plus 4-jet events
arising from the process $t\bar t \ra W b W \bar b \ra
\ell \nu J_1 J_2 J_3^b J_4^{\bar b}$. The task is to assign the observed
particles, jets, and \met\ in these events one-to-one to the decay
products of the $t$ and $\bar t$. However, the problem remains that we
don't know a priori how to map the observed jets to the partons from
the $t\bar t$ decay. This is illustrated in Figure~\ref{mapjet}, where
for example in Fig.~\ref{mapjet}a) jet $J_1$ is combined with the
lepton and neutrino to originate from one top quark, while jets
$J_2$ through $J_4$ are assigned to the $\bar t$ quark. 
Fig.~\ref{mapjet}b) shows another possible assignment, which might as
well be the result of a $t\bar t$ decay. In total there are 12
possible jet-parton assignements which are reduced to six combinations
if one $b$ jet is tagged. If there are two $b$ tags in the event,
there are still two possible combinations from assigning 
the two $b$ jets to the $t$ or $\bar t$ quark.
In additon, this combinatorics gets doubled, because there is a twofold
ambiguity for $p_z^{\nu}$ since only the transverse component of the
missing energy \met\ is reconstructed by the experiments. 

To select the best combination CDF and D\O\ use a likelihood method
that exploits the many constraints in the system. Each event is fitted
individually to the hypothesis that three of the jets come from one
top quark through its decay into $W b$, and that the lepton, \met, and
the remaining jet come from 
the other top quark. In addition, $b$ tagged jets are assigned
as $b$ quarks in the fit, the invariant dijet mass has to equal the
$W$ boson mass, and $m_t = m_{\bar t}$ is required. CDF chooses the
solution with the best fit 
$\chi^2$, while D\O\ takes a weighted average of the three best
solutions, where each solution has to satisfy a minimum $\chi^2$ cut.
The result is a distribution of the best fit top mass from each
candidate event. The final value of the top mass is then extracted by
fitting this distribution to a set of Monte Carlo templates from $t\bar t$ 
production and background. There are several reasons why a MC
simulation of the kinematic fitting is necessary. At first, there is
the need to relate the jet energies to parton energies, where the
experimental smearing, predominantly in the jet energies, has to be
taken into account. In addition, the four most energetic jets in the
event may not be directly related to the $t\bar t$ decay products, if
for example a hard gluon radiation produced an energetic jet. Finally,
the fit solution with the lowest $\chi^2$ may not have the correct jet
assignment. 

\subsubsection{CDF Top Quark Mass Measurement}

CDF has a good signal to background ratio in the 
lepton plus jets channel, where one of the jets is
SVX tagged. This sample of lepton plus 4-jet events with at least one
SVX tag provided the original top mass measurement at CDF \cite{CDF_top_disc}.
Recent optimization studies indicate a reduced error on \mtop, if the
tagged events are subdivided into different tagging classes and a set
of no-tagged events is added. The following four data samples are used
for this optimized determination of the top quark mass: (a) events
with one SVX tag (15 events), (b) events with two SVX tags (5 events),
(c) events with a SLT tag but no SVX tag (14 events), and (d) events
with no tag but all four leading jets having $E_{\rm t} > 15$ GeV (48
events). 

\begin{figure}[tbp]\centering
\begin{picture}(145,70)(0,0)
\put(-6,53){\large\bf (a)}
\put(88,53){\large\bf (b)}
\centerline{
\epsfysize=7.0cm
\epsffile[20 150 550 720]{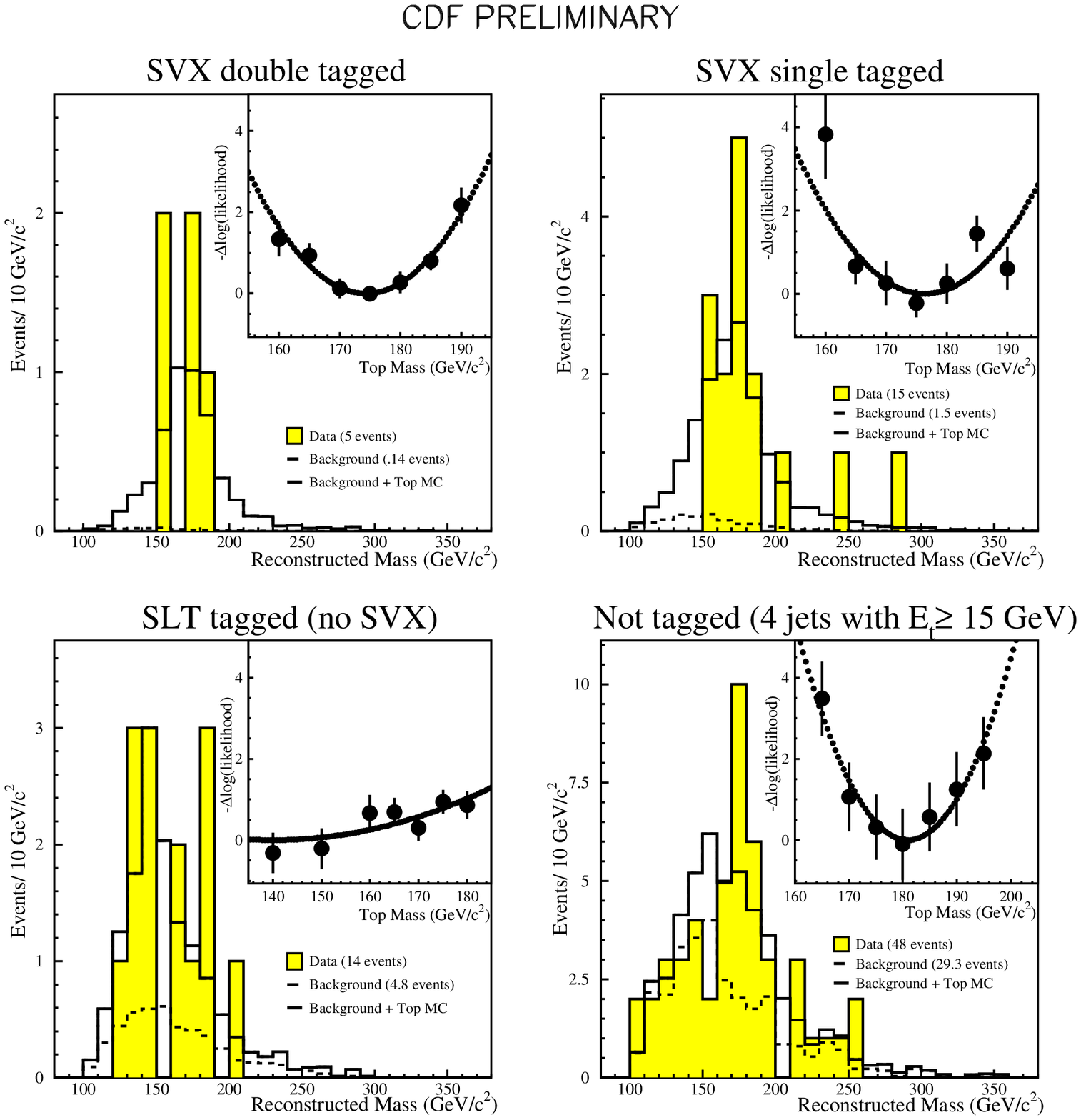}
\hspace*{0.5cm}
\epsfysize=7.0cm
\epsffile[30 160 520 690]{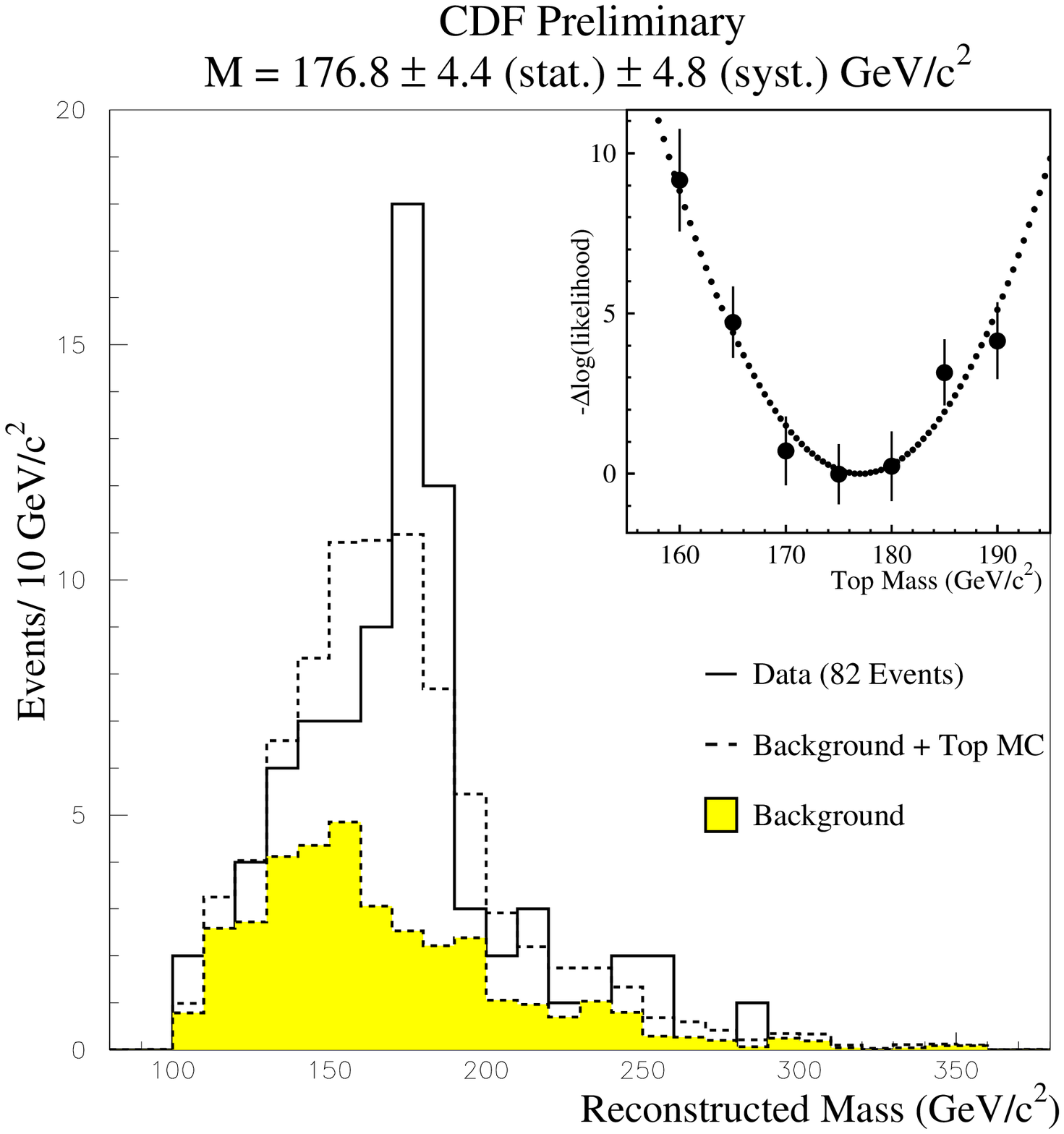}
}
\end{picture}
\caption{CDF top mass distributions: (a) For the four subsamples
discussed in the text. 
The shaded area represents the
data, while the solid line is the predicted sum of background plus
expected signal. The dashed line just represents the background
contribution. The inserts show the obtained likelihood function versus
top mass. 
In (b) the combined top mass
distribution is shown.
The solid line represents the
data, while the dashed line is the predicted sum of background plus
expected signal. The shaded area represents the background
contribution. The insert shows the obtained likelihood function versus
top mass.}
\label{top_mass_cdf}
\end{figure}

These four subsamples are orthogonal to each other and a top
mass can be extracted from each sample individually. These top mass
distributions for the four samples are shown separately in
Figure~\ref{top_mass_cdf}a). For each of the subsamples, an expected
background fraction was estimated, and predicted top signal
distributions were obtained from Monte Carlo events at several values
of \mtop. A likelihood fit was performed for each subsample, which is
shown in the inserts of the distributions in
Fig.~\ref{top_mass_cdf}a). Since the likelihood values from the four
subsamples are essentially independent, a combined result is obtained
by taking the product of the four likelihood values.  
The sums of the
four distributions, as well as the likelihood product versus top mass,
are shown in Fig.~\ref{top_mass_cdf}b), where the solid line represents the
data, while the dashed line is the predicted sum of background plus
expected signal. The shaded area just represents the background
contribution. 

From the distribution of the likelihood product versus \mtop\ a top mass value
of  
$\mtop~=~(176.8\pm4.4\pm4.8)$ \gevcc\ has been extracted.  
The systematic errors are detailed in Table~\ref{top_mass_syst_cdf}.
The largest contribution ($\pm 3.6$ \gevcc) comes from soft gluon radiation
plus 
jet energy scale uncertainties in estimating the true jet energy from the
observed energy. The systematic error due to hard gluon effects
reflects the uncertainty in how often one of the four leading jets is
associated with a hard gluon rather than directly with a parton in
a top decay. All other systematic errors contribute less than 1.5 \gevcc\
(see Tab.~\ref{top_mass_syst_cdf}).

\begin{table}[tbp]
\begin{center}
\vspace*{-0.6cm}
  \begin{tabular}{|c|c|}
   \hline
   CDF Lepton plus Jets Mass Systematics & $\Delta \mtop$   \\
   \hline \hline
    Soft Gluon + Jet $E_{\rm t}$ Scale & 3.6 \gevcc\  \\
    Different MC Generators   &   1.4 \gevcc\  \\
    Hard Gluon Effects   &  2.2 \gevcc\ \\
    Kin. \& Likelihood Fitting Method   &  1.5 \gevcc\   \\
    $b$-tagging Bias     &  0.4 \gevcc\  \\
    Background Spectrum  &  0.7 \gevcc\   \\
    Monte Carlo Statistics & 0.8 \gevcc\   \\
   \hline \hline
    Total                &  4.8 \gevcc\   \\
   \hline
  \end{tabular}
\caption{Systematic uncertainties of the CDF top quark mass measurement.}
\label{top_mass_syst_cdf}
\end{center}
\vspace*{-0.5cm}
\end{table}

\subsubsection{D\O\ Top Quark Mass Measurement}

\begin{figure}[tb]\centering
\begin{picture}(145,70)(0,0)
\put(-6,58){\large\bf (a)}
\put(84,58){\large\bf (b)}
\centerline{
\epsfysize=7.0cm
\epsffile[60 105 555 685]{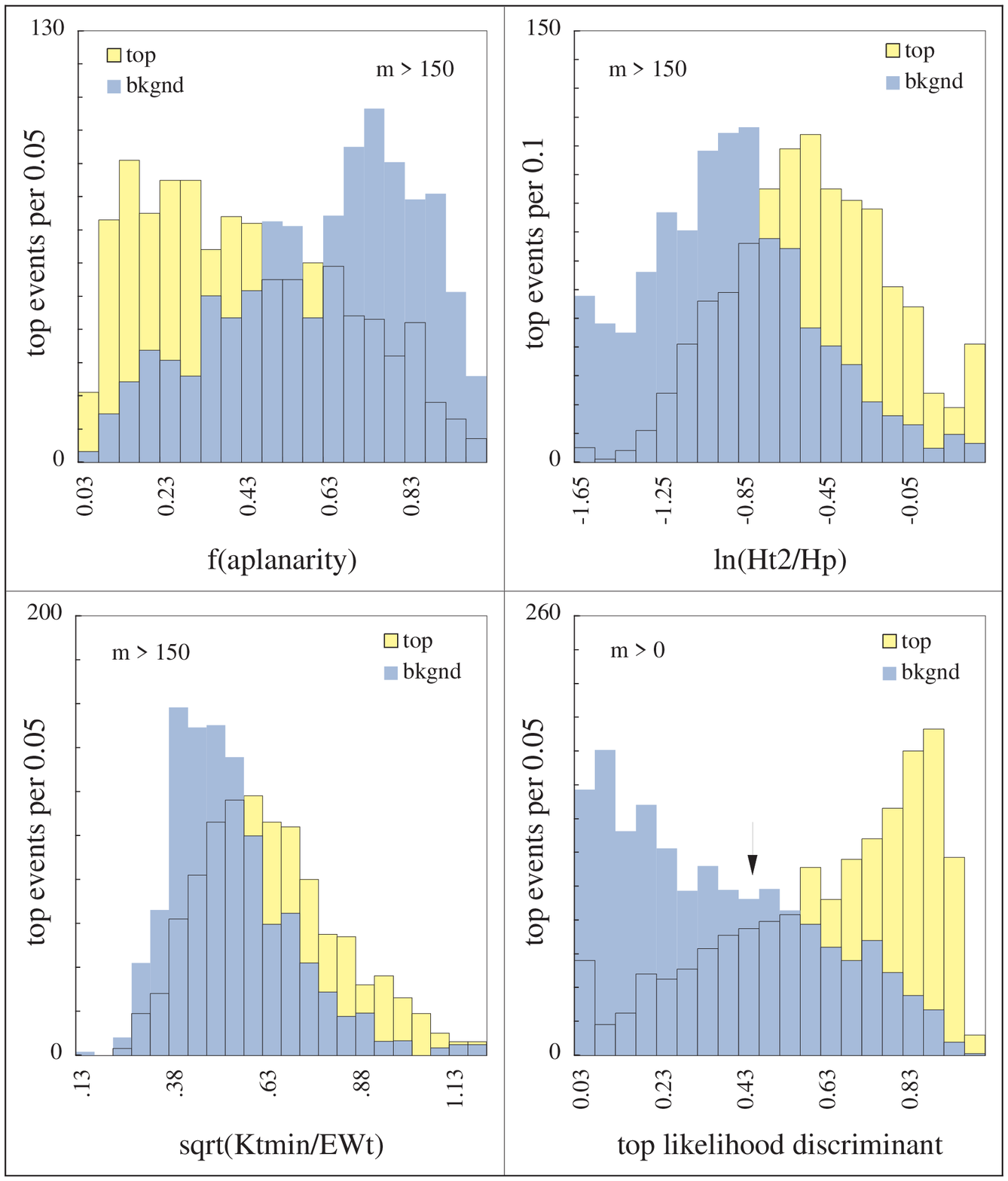}
\hspace*{0.5cm}
\epsfysize=7.0cm
\epsffile[5 5 565 565]{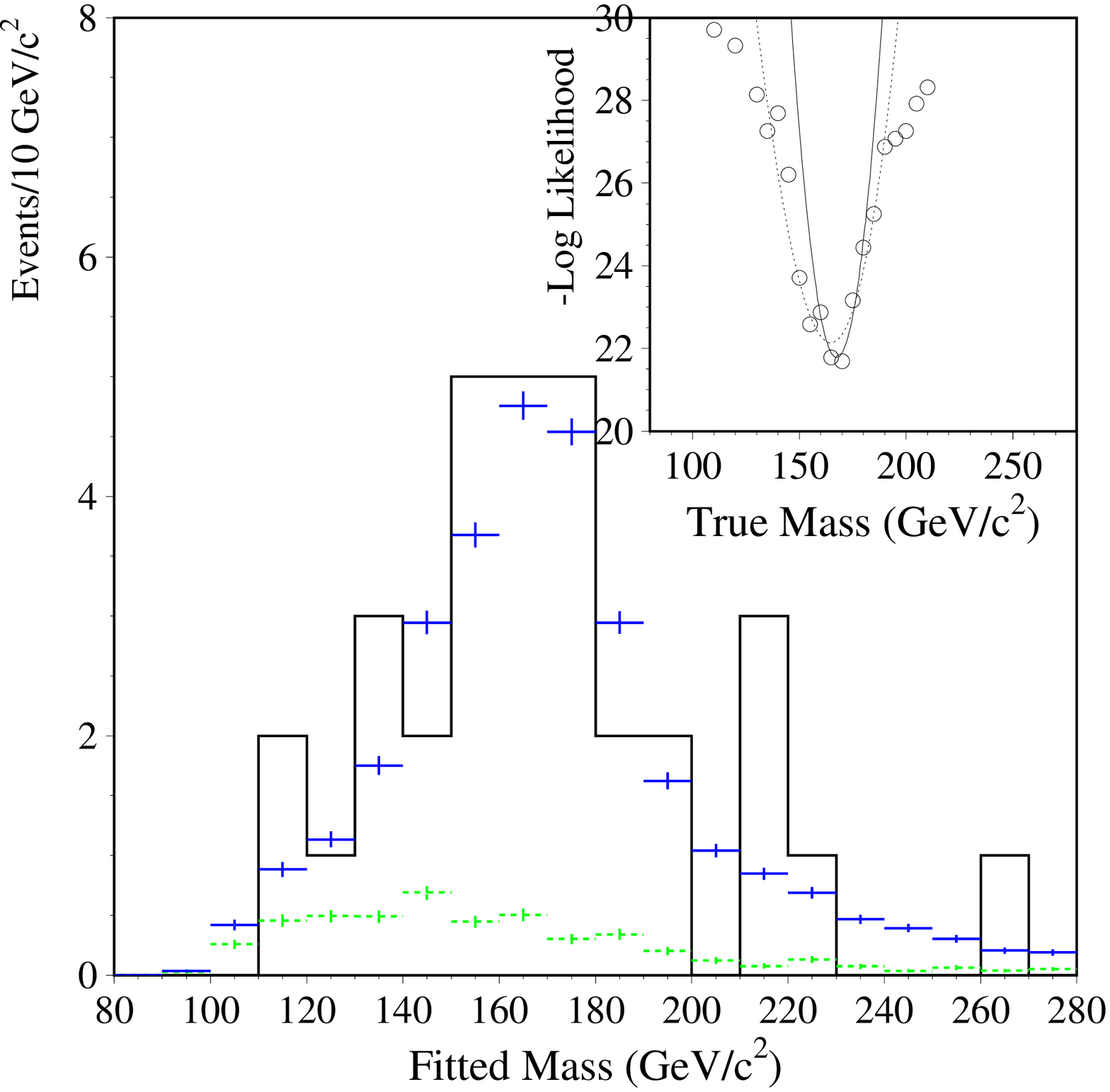}
}
\end{picture}
\caption{D\O\ top mass measurement: (a) Distribution of MC top (light) and
background (dark) lepton plus 4-jet events for different kinematic
parameters like aplanarity, $H^{\prime}_{T_2}$,
$K^{\prime}_{T_{min}}$, and the combined top likelihood discriminat
$\cal D$ (from top left to bottom right). (b) Observed (solid) top
quark mass distribution, where the solid crosses represent the prediction
for background plus signal. The dashed crosses are just the
background expectation.
The curves in the insert are used to determine the top
quark mass and its statistical error. }
\label{top_mass_d0}
\end{figure}

As already discussed, the power to extract a top quark mass from the lepton
plus 4-jet 
events depends on the suppression of the background. The D\O\
experiment combines four kinematic variables into one 
top likelihood discriminant $\cal D$, which provides a distinct
separation power between top signal and background without biasing the
analysis. For example, 
a straight cut on the total hadronic energy $H_T$
would push both background and signal
distributions towards higher values of \mtop\ and make the background
look like signal.
The top likelihood discriminant $\cal D$ combines \met,
the aplanarity $\cal A$ of the $W$ plus jets system, 
the fraction of the $E_{\rm t}$
of the $W$ plus jets system which is carried by the $W$ ($H_{T2}^{\prime}$),
and the $E_{\rm t}$ weighted RMS $\eta$ of the $W$ and jets
($K^{\prime}_{T_{min}}$). Some of the distributions of these
variables, as well as the combined top likelihood discriminant $\cal
D$ are shown in Fig.~\ref{top_mass_d0}a). As can be seen, there is 
separation between a top signal predicted from Monte Carlo (light shaded
area) and 
background expectations (dark shaded area). After applying a cut of
${\cal D} > 0.43$, 32 events have been selected for the fit of
\mtop. 

The distribution of the fitted top mass values is shown in
Fig.~\ref{top_mass_d0}b) as the solid histogram. The 
solid crosses represent the prediction for background plus expected
signal, which is 
in good agreement with the data. The
dashed crosses in Fig.~\ref{top_mass_d0}b) show the background
expectation only. A measurement of the top quark mass of 
$\mtop~=~(168\pm8\pm8)$ \gevcc\ 
has been extracted from this distribution.
The insert in Fig.~\ref{top_mass_d0}b) shows the obtained likelihood
as a function of true \mtop\ as well as the curves used to determine the top
quark mass and its statistical error. 
The systematic errors on the D\O\ top mass measurement are detailed in
Table~\ref{top_mass_syst_d0}. 
The largest systematic uncertainty results from the jet energy
correction.

\begin{table}[tbp]
\begin{center}
\vspace*{-0.6cm}
  \begin{tabular}{|c|c|}
   \hline
   D\O\ Lepton plus Jets Mass Systematics & $\Delta \mtop$   \\
   \hline \hline
   Jet Energy Correction  &  7.3 \gevcc\ \\
   Monte Carlo Model  &  3.3 \gevcc\ \\
   Fitting Method  &  2.0 \gevcc\ \\
   \hline \hline
    Total                &  8 \gevcc\   \\
   \hline
  \end{tabular}
\caption{Systematic uncertainties of the D\O\ top quark mass measurement.}
\label{top_mass_syst_d0}
\end{center}
\vspace*{-0.5cm}
\end{table}

\subsubsection{World Average Top Quark Mass}

An attempt can be made to combine the CDF and D\O\ top quark mass
measurements to a world average top mass. For this purpose the
reported mass measurements from the lepton plus jets channel have been
used. CDF measured $\mtop~=~(176.8\pm4.4\pm4.8)$~\gevcc, while the D\O\
measurement results in $\mtop~=~(169\pm8\pm8)$~\gevcc. In the determination of
the 
world average top mass the conservative assumption has been made that all the
systematic errors, except for the energy scale, the $b$ tagging bias, and
the Monte Carlo statistics, are fully correlated between both
experiments. This results in a world average top quark mass of
\begin{equation}
\mtop = (175.0 \pm 3.9 \pm 4.5)\ \gevcc.
\end{equation}
It is worthwhile to note that knowing the top quark mass with a
combined statistical and systematic error of $\pm 6$ \gevcc\ 
(a 3.4\% relative error) is a big
accomplishment of both Tevatron experiments in Run~I. 

\subsubsection{Summary of Top Quark Mass Measurements}

\begin{figure}[tbp]\centering
\begin{picture}(145,90)(0,0)
\centerline{
\epsfysize=9.0cm
\epsffile[130 180 580 650]{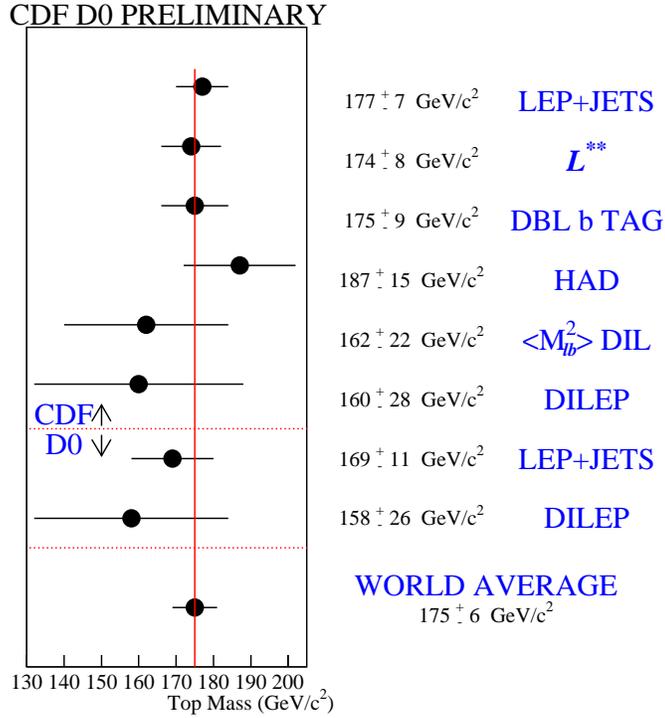}}
\end{picture}
\caption{Summary of the CDF and D\O\ top quark mass
measurements in different decay channels.}
\label{top_mass_sum}
\end{figure}

A summary of all top
quark mass measurements performed by CDF and D\O\ is given in
Fig.~\ref{top_mass_sum}. Top mass measurements 
from other top decay modes like the dilepton
channel or the all hadronic channel are also shown.
They are in good
agreement with the measurement from the lepton plus 4-jet samples,
which have been discussed in the previous sections in more
detail. This decay mode represents the most 
precise determination
of the top mass for each experiment.

\begin{figure}[tbp]\centering
\begin{picture}(145,80)(0,0)
\centerline{
\epsfysize=8.0cm
\epsffile[70 180 520 680]{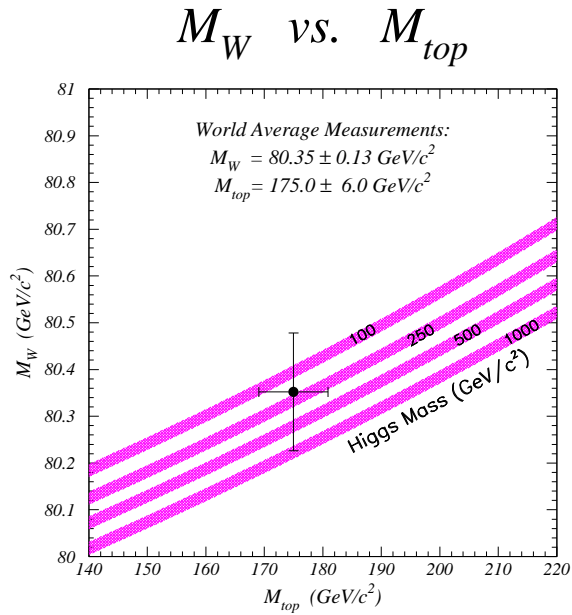}}
\end{picture}
\caption{Relation between the top quark mass and the $W$ boson mass. The
dependence on the Higgs mass is shown through the different bands for several
assumptions of $m_{\rm Higgs}$.}
\label{mw_mtop}
\end{figure}

The knowledge of the top quark mass plays an important role in
calculations of radiative corrections that relate electroweak
parameters. For example, higher order radiative corrections relate the
$W$ boson mass $m_W$ to the top quark mass \mtop. This relationship
also depends on the mass of the Higgs boson $m_{\rm Higgs}$, which can also
participate in these higher order loops. The relationship between
$M_W$ and \mtop\ is
displayed in Fig.~\ref{mw_mtop}. 
The well known dependence on the Higgs mass is shown through the different
bands for several assumptions of $m_{\rm Higgs}$.
Although $m_W$ and \mtop\ are
precisely known, the sensitivity on the Higgs boson mass is still too
poor. A precise measurement of the top quark mass together with $m_W$
is therefore a high 
priority of both collider experiments in Run~II.

\newpage
\section{{\boldmath $B$} Physics at a Hadron Collider}
\label{b_sec}

The principal interest in studying $B$ hadrons in the context of the Standard
Model arises from the fact that $B$ hadron decays provide valuable information
on the weak mixing matrix, the Cabibbo-Kobayashi-Maskawa (CKM)
matrix~\cite{ckm}. In fact, $B$~decays determine five of the nine CKM matrix
elements: $V_{cb}$, $V_{ub}$, $V_{td}$, $V_{ts}$, and $V_{tb}$. The future
interest in $B$ physics certainly lies in the study of $CP$ violation in the
system of neutral $B$ mesons, which will be discussed further in
Sec.~\ref{outlook_cp}.   

Traditionally, $B$ physics has been the domain of $e^+ e^-$ machines, but
already the UA\,1 collaboration has shown that $B$ physics
is feasible at a hadron collider \cite{bfeasi}.
However, the combination of a better mass resolution and vertex
detection enables the CDF experiment
to perform a broader $B$ physics program. The D\O\ experiment has also
published several $B$ physics results \cite{bphysd0},
but due to the lack of a precision momentum measurement of charged particles
within a magnetic field and the absence of a precision micro vertex
detector, D\O\ is 
not ideally suited to do $B$ physics. Since we concentrate in this
presentation on recent results from $B$~hadron lifetimes and time dependent
$B^0 \bar B^0$ oscillations, we shall report only on measurements from the
CDF experiment.    

One advantage of $B$ physics at a hadron collider compared to an $e^+ e^-$
machine at the $\Upsilon(4S)$ is that all $B$ hadron species are
produced. Another advantage in doing $B$ physics at a hadron colliders can
be seen by comparing the $B$ production cross section, which is about 1~$n$b
at the $\Upsilon(4S)$, while at the $Z^0$ pole $\sigma(B\bar B)~\approx~6~n$b. 
However, at the Tevatron the $b$ quark production cross
section is quite large with $\sigma_{b} \sim 50~\mu$b within the central region of
rapidity less than 1 ($|\cos\theta| < 1$). This is a huge cross section which
resulted in about $5\cdot 10^9\ b\bar b$ pairs being produced in Run~I within
the Tevatron detectors. But the total
inelastic cross section is still about three orders of magnitude
larger. This puts 
certain requirements on the trigger system to find $B$ decay products. In
addition the total integrated $b$ quark production cross section is a rapidly
falling cross section, which drops by about two orders of magnitude for a
$b$ quark $\Pt^b$ greater than about 8 \gevc\ compared to $\Pt^b$ greater
than about 
20 \gevc. This means, in terms of trigger thresholds for $b$ decay products,
one likes to go as low as possible in \Pt\ in order to increase the amount of
recorded $B$ triggers. Of course the experiment's DAQ bandwidth is the limiting
factor.


All $B$ physics triggers at CDF are based on leptons.
Dilepton and
single lepton triggers both exist. Exploiting the steeply falling $b$ quark
production cross section, 
CDF was able  
to maintain low \Pt\ thresholds throughout Run~I without increasing the
experiments deadtime during data taking. 

CDF's dilepton triggers consist of a dimuon trigger with $\Pt > 2$
\gevc\ for both muon legs, and an $e\mu$ trigger with $\Pt^{\mu} > 3$ \gevc\
and $E_{\rm t}^e > 5$ GeV. 
The dimuon trigger is the source of CDF's $J/\psi$ sample. Both
dilepton trigger samples are also starting points for $B$ mixing analyses.
The thresholds for the single lepton triggers are
higher with $\Pt > 7.5$ \gevc\ for muons and $E_{\rm t} > 8$ GeV for
electrons. Analyses
involving semileptonic $B$ decays are based on these single lepton datasets.
The given \Pt\ thresholds are
representative for Run~Ib but similar for Run~Ia.

\begin{figure}[tbp]\centering
\begin{picture}(145,70)(0,0)
\put(15,60){\large\bf (a)}
\put(130,60){\large\bf (b)}
\centerline{
\epsfysize=7.1cm
\epsffile[30 150 520 660]{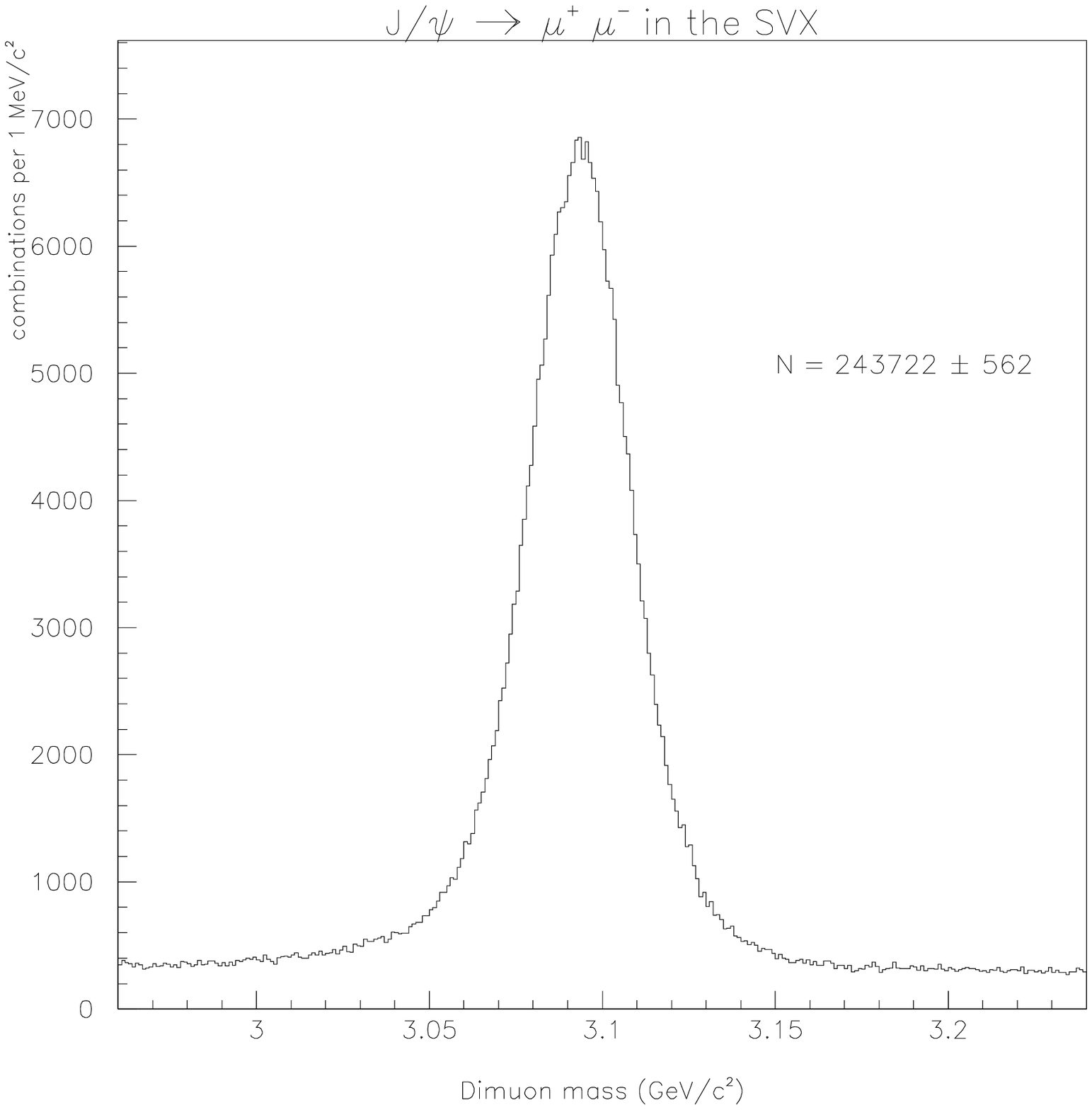}
\hspace*{0.5cm}
\epsfysize=7.1cm
\epsffile[30 150 520 660]{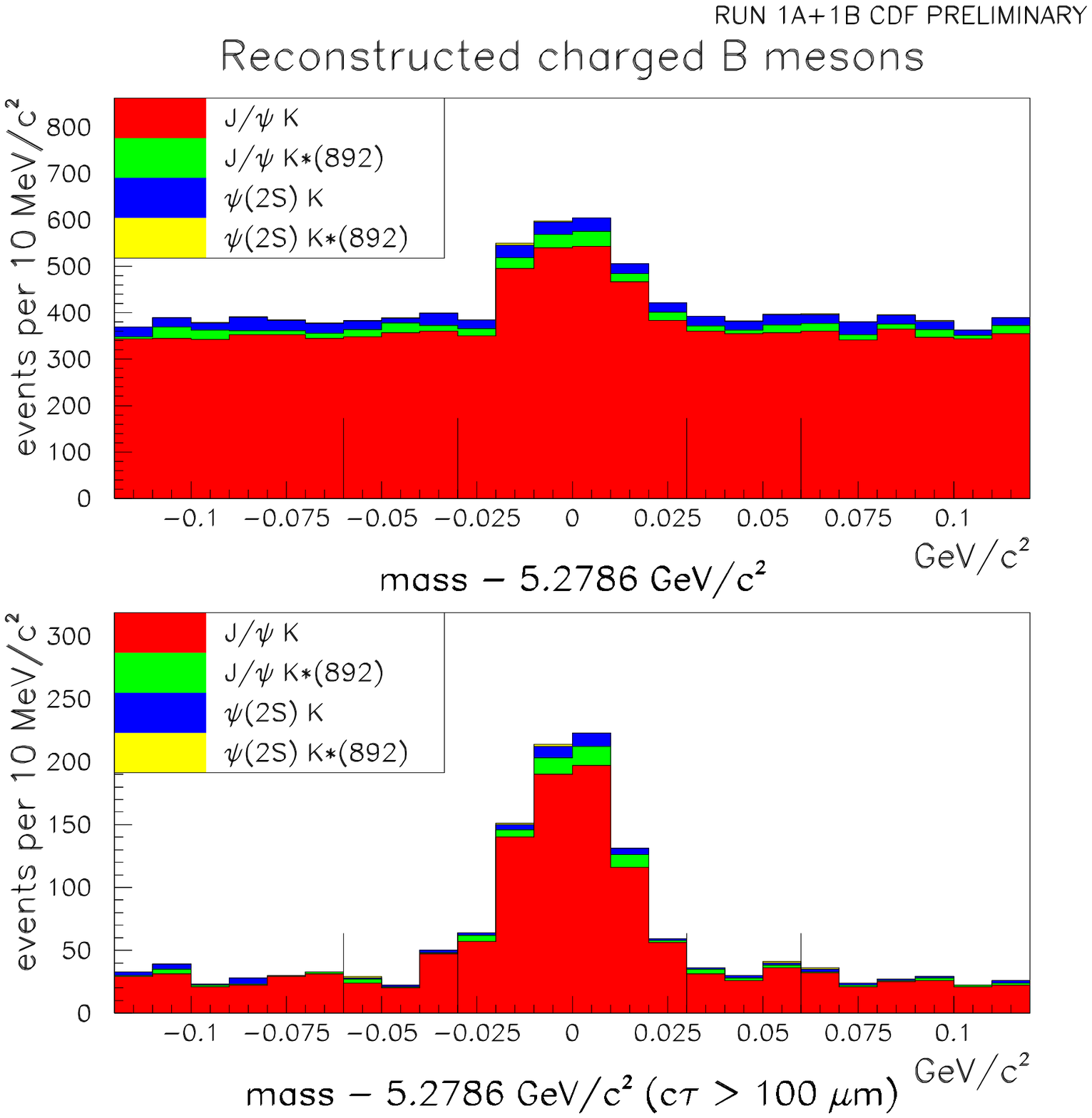}
}
\end{picture}
\caption{Invariant mass distribution of (a) dimuon trigger in the
$J/\psi$ mass region and (b) $B^+ \ra J/\psi K^+$ candidates without
(top) and with a $c\tau > 100\ \mu$m requirement (bottom).}
\label{bdemo}
\end{figure}

The basis of CDF's $B$ physics program are the good tracking and
vertexing capabilities of the CTC and the SVX. This is demonstrated in
Figure~\ref{bdemo}. The invariant dimuon mass from CDF's dimuon
trigger sample is shown in Fig.~\ref{bdemo}a). 
From all Run~I data a prominent signal 
of about 240,000 $J/\psi$ candidates
with both muons in the SVX
can be found on low
background. The mass
resolution of the observed signal is about 16 \mevcc. One handle
to reduce backgrounds 
with the help of CDF's silicon vertex detector is demonstrated in 
Fig.~\ref{bdemo}b). Here, $J/\psi$ candidates are paired with another
track in the event, which is assumed to be a kaon, in order to fully
reconstruct $B^+ \ra J/\psi K^+$ decays. 
The top plot shows the invariant $J/\psi K^+$ mass
distribution, where a $B^+$ signal is observed on a large
background. This background is drastically reduced after a displaced 
$B$~vertex with e.g.~$c\tau > 100~\mu$m as measured in the SVX is required
(see Fig.~\ref{bdemo}b). 

\subsection{{\boldmath $B$} Hadron Lifetimes}

The lifetimes of $B$ hadrons are fundamental properties of these
particles and can be used to test theoretical models of heavy flavour
decays. Predictions for $B$ hadron lifetimes and their ratios have
been made by several groups~\cite{bigi,neubert}.   
Bigi~et~al.~\cite{bigi} predicts the charged $B$ lifetime to be longer
than the $B^0$ lifetime by about 5\%
\begin{equation}
\frac{\tau(B^+)}{\tau(B^0)} \simeq 1.0 + 0.05\cdot
	 \frac{f_B^2}{(200~{\rm MeV})^2},
\end{equation}
and expects the lifetime of the $\Lambda_b^0$ baryon to be shorter than
$\tau(B^0)$ with a ratio of $\tau(\Lambda_b^0)/\tau(B^0)$ not smaller
than about 0.9. On the other hand using the heavy quark expansion,
Neubert~\cite{neubert} obtains: 
\begin{eqnarray}
\frac{\tau(B^+)}{\tau(B^0)} & = & 1.0 + {\cal O}(1/m_b^3), \\
\frac{\tau(\Lambda_b^0)}{\tau(B^0)} & = & 0.98 + {\cal O}(1/m_b^3),
\end{eqnarray}
where the estimate for $\tau(\Lambda_b^0)/\tau(B^0)$ includes
corrections that arise at order $1/m_b^2$. Although these ratios 
might appear to be close to unity, Neubert argues that the $1/m_b^3$
corrections might be large due to phase space
enhancement from effects involving the $u,\,d$ spectator quark. He
concludes that theoretical uncertainties allow a lifetime ratio in the
range between 0.8 and 1.2. 

This subject is controversal, and best
solved by precisely measuring all the $B$~hadron lifetimes. As we shall
see, the uncertainties on the individual $B$~lifetimes have reached a level of
a few percent. In the following, we review the latest $B$~hadron
lifetime results from CDF and also compare them to other experiments.

\subsubsection{{\boldmath $B^+\ {\rm and}~B^0$} Lifetimes with
    Fully Reconstructed {\boldmath $B$} Mesons}  

\begin{figure}[tbp]\centering
\begin{picture}(145,70)(0,0)
\put(55,60){\large\bf (a)}
\put(130,60){\large\bf (b)}
\centerline{
\epsfysize=7.0cm
\epsffile[30 150 520 650]{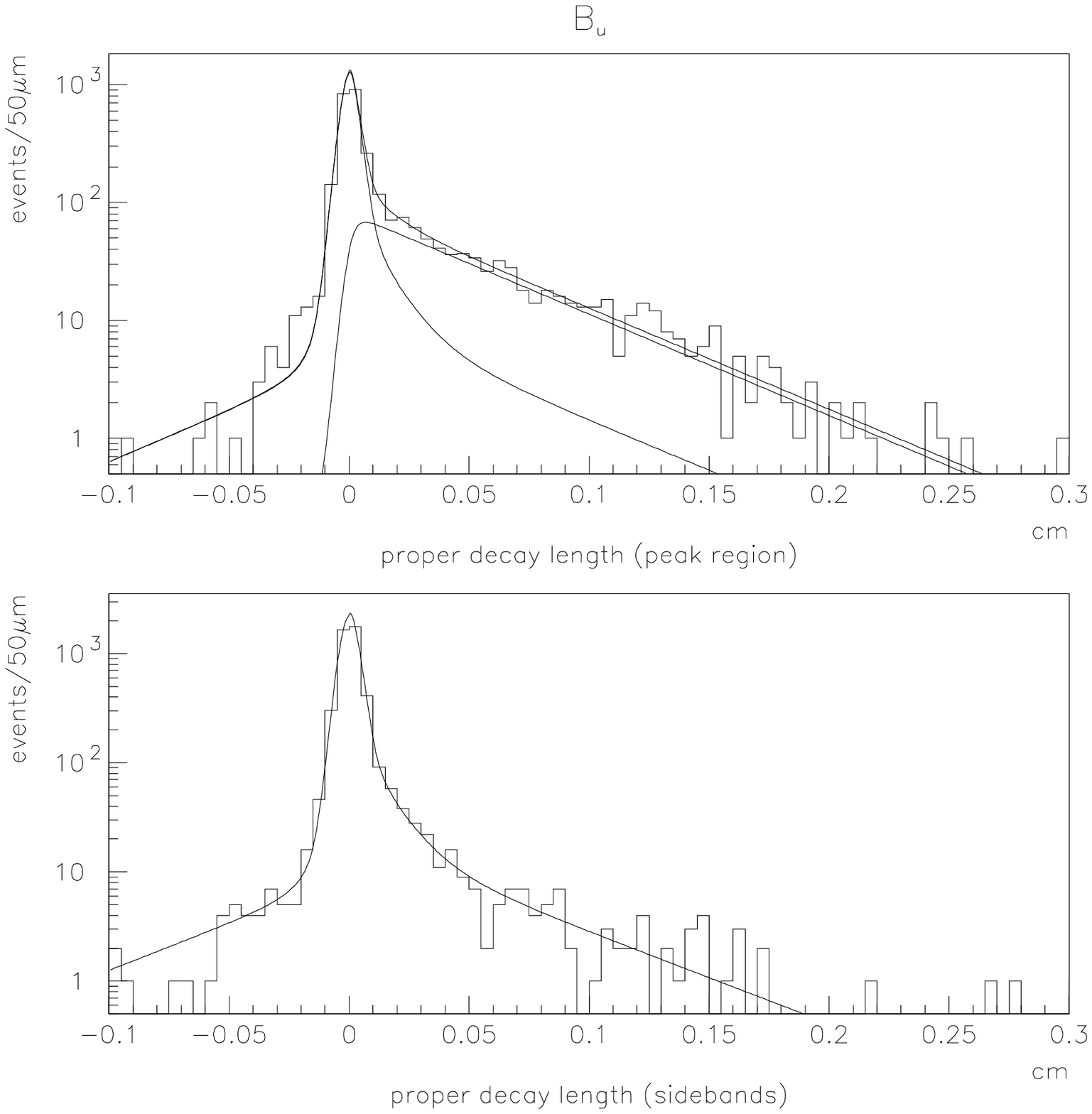}
\hspace*{0.5cm}
\epsfysize=7.0cm
\epsffile[30 150 520 650]{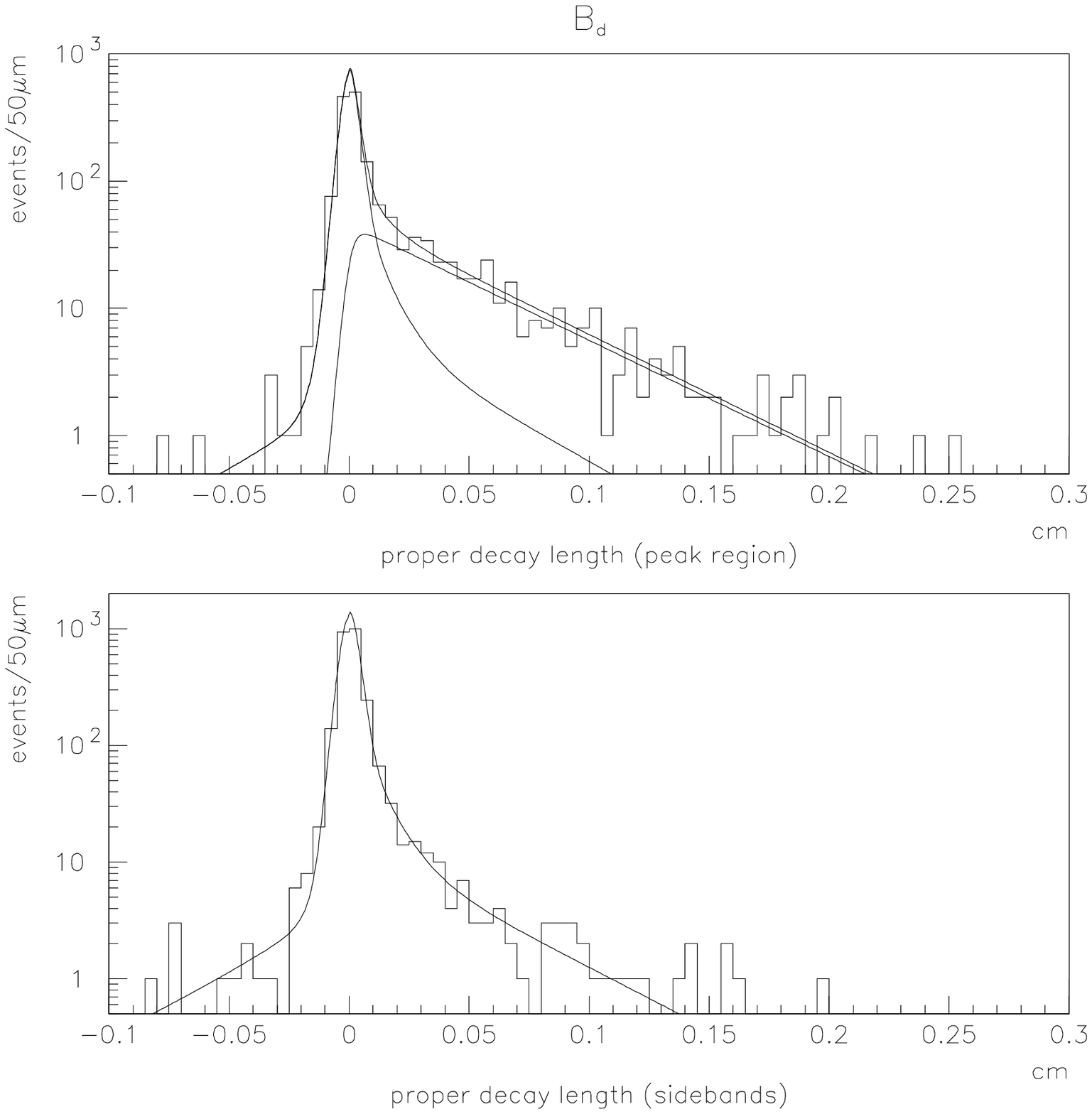}
}
\end{picture}
\caption{Proper time distributions of (a) charged and (b) neutral $B$
mesons, which were fully reconstructed through $B \ra \Psi${\bf K}. 
The bottom plots represent the background $c\tau$ distributions as
obtained from the $B$ sidebands.}
\label{blife:exclusive}
\end{figure}

The analysis principle for the 
$B$ lifetime measurement using fully reconstructed $B^+$ and $B^0$
mesons is as follows.  All Run~I 
dimuons forming a $J/\psi$
candidate, as shown in Fig.~\ref{bdemo}a), are used.
All possible decay modes 
$B \ra \Psi${\bf K} have been investigated, where $\Psi$ is mainly a $J/\psi$
but can also be a $\psi(2S) \ra \mu\mu$. 
{\bf K} represents the different kaon states
$K^+$ and $K^*(892)^+$ for the charged $B$ lifetime measurement, as well as
$K^0_S$ and $K^*(892)^0$ for the $\tau(B^0)$ determination. 
The vertex and mass constrained $J/\psi$
candidates are vertexed with the {\bf K} candidates yielding the
two-dimensional decay length \lxy. Together with the known $B$
transverse momentum \Pt, the proper time distributions, shown in
Fig.~\ref{blife:exclusive} for (a) charged and (b) neutral $B$
candidates, are obtained.
The bottom $c\tau$ distributions represent the background as
obtained by fitting the $B$ sideband regions to a gaussian with
exponential tails. Using this background shape, an unbinned likelihood fit of 
the signal,
which is assumed to be an exponential convoluted with a Gaussian, is
performed.   
The following lifetimes are obtained:
\begin{eqnarray}
\tau(B^+) = (1.68 \pm 0.07 \pm 0.02)\ p{\rm s} \\
\tau(B^0) = (1.58 \pm 0.09 \pm 0.02)\ p{\rm s} \\
\tau(B^+)/\tau(B^0) = 1.06 \pm 0.07 \pm 0.01.
\end{eqnarray}

The exclusive $B$ lifetime measurement is still statistics dominated.
One way to increase the number of $B$ candidates is to not fully
reconstruct the 
$B$ meson. This is done in the semi-exclusive analysis 
described in the next section.

\subsubsection{{\boldmath $B^+\ {\rm and}~B^0$} Lifetimes with
    Partially Reconstructed {\boldmath $B$} Mesons}  
\label{partblife}
 
The $B$ lifetime analysis using partially reconstructed $B$ mesons
exploits the semi\-leptonic decays $B \ra D^{(*)}\ell X$ and starts with
the single lepton trigger data. In a cone around the trigger
electron or muon, $D^{(*)}$ meson candidates are reconstructed through their
decay modes: \\
\hspace*{0.5cm} 1.~$D^0 \ra K^- \pi^+$, where the $D^0$ is not from a
$D^{*+}$, \\
\hspace*{0.5cm} 2.~$D^{*+} \ra D^0 \pi^+,\ D^0 \ra K^- \pi^+$, \\
\hspace*{0.5cm} 3.~$D^{*+} \ra D^0 \pi^+,\ D^0 \ra K^- \pi^+\pi^+\pi^-$, \\
\hspace*{0.5cm} 4.~$D^{*+} \ra D^0 \pi^+,\ D^0 \ra K^- \pi^+ \pi^0$, 
where the $\pi^0$ is not reconstructed. \\
The $D^{(*)}$ candidates are intersected with the
lepton to find the $B$ decay vertex. Since the $B$ meson is not fully
reconstructed, its $c\tau_B$ cannot be directly determined. A correction
has to be applied to scale from the 
$D^{(*)} \ell$ momentum to $\Pt(B)$. This $\beta\gamma$ correction is
obtained from a Monte Carlo simulation.   

\begin{figure}[tbp]\centering
\begin{picture}(145,70)(0,0)
\centerline{
\epsfysize=7.0cm
\epsffile[68 183 560 712]{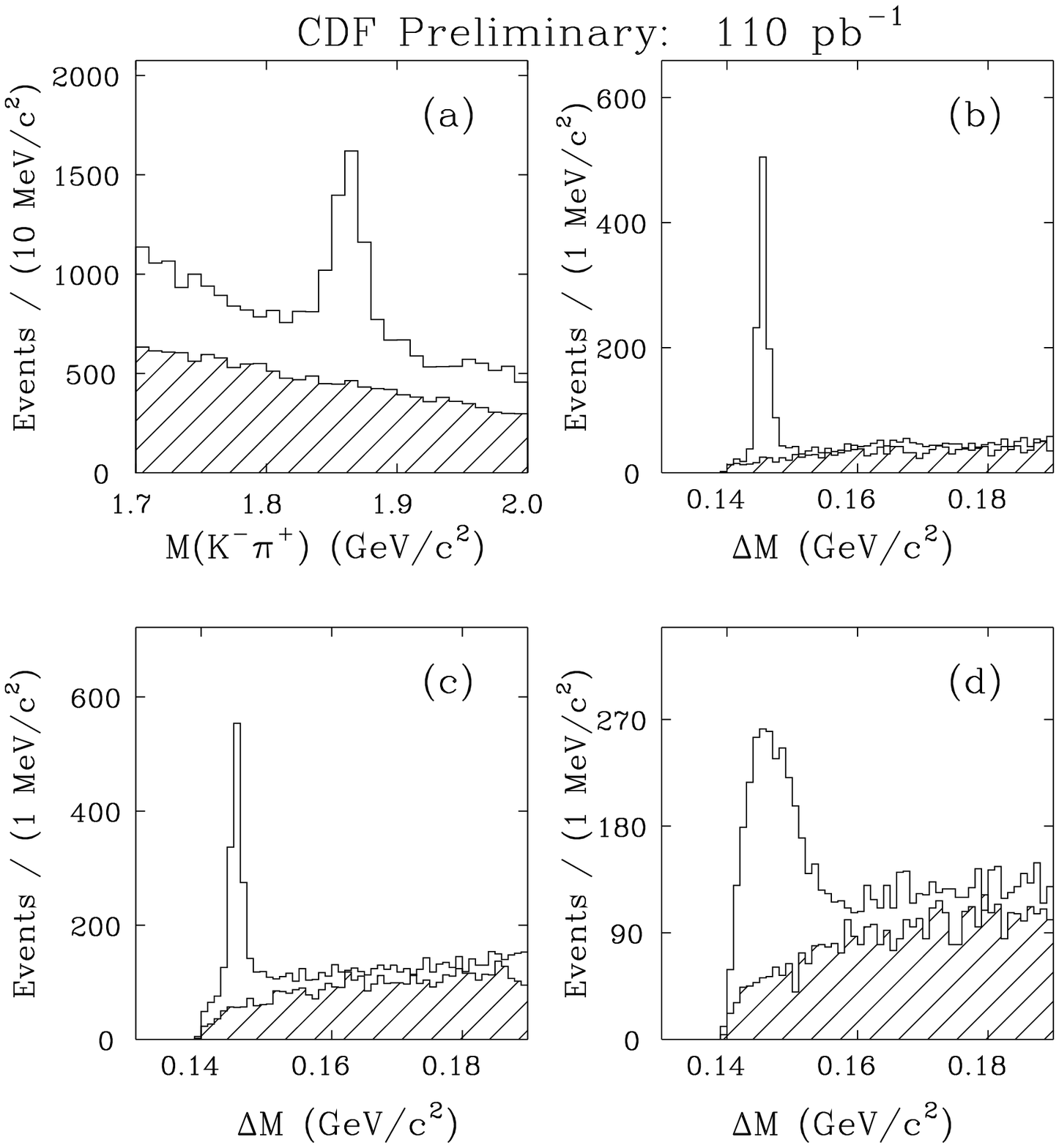}
\hspace*{0.5cm}
\epsfysize=7.0cm
\epsffile[30 188 543 690]{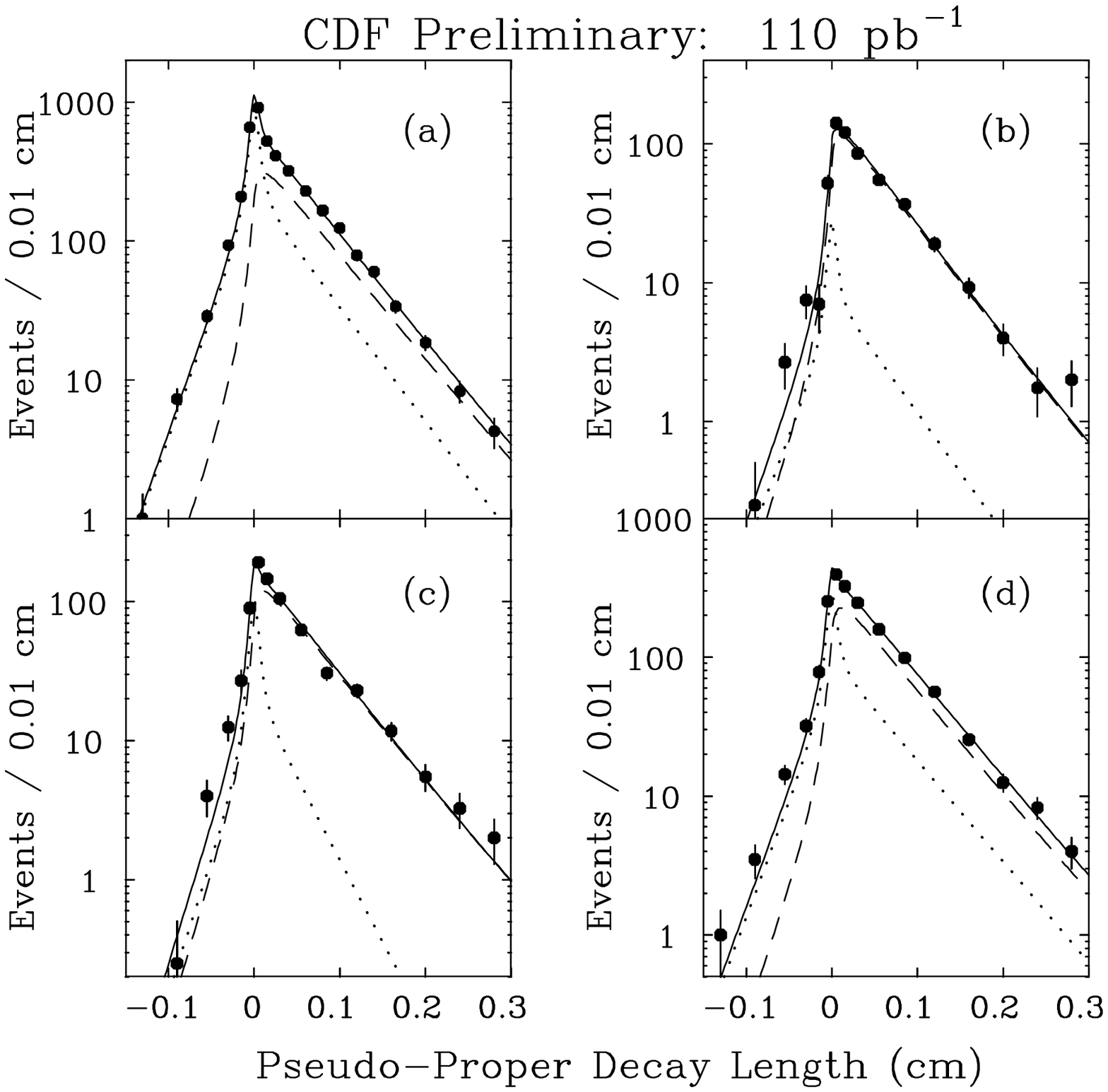}
}
\end{picture}
\caption{Invariant mass distribution of $D^{(*)}$ candidates (left
hand side) and proper time distributions (right hand side) from the
$B$ lifetime analysis using partially reconstructed $B \ra D^{(*)}\ell X$ 
decays with
(a) $D^0 \ra K^- \pi^+$, where the $D^0$ is not from a $D^{*+}$,
(b) $D^{*+} \ra D^0 \pi^+,\ D^0 \ra K^- \pi^+$,
(c) $D^{*+} \ra D^0 \pi^+,\ D^0 \ra K^- \pi^+\pi^+\pi^-$, and
(d) $D^{*+} \ra D^0 \pi^+,\ D^0 \ra K^- \pi^+ \pi^0$.}
\label{blife:semiex}
\end{figure}

The final $D^{(*)}$ candidates can be seen on the left hand side of
 Fig.~\ref{blife:semiex} 
for (a)~$D^0 \ra K^- \pi^+$, where the $D^0$ is not from a $D^{*+}$,
(b) $D^{*+} \ra D^0 \pi^+,\ D^0 \ra K^- \pi^+$,
(c) $D^{*+} \ra D^0 \pi^+,\ D^0 \ra K^- \pi^+\pi^+\pi^-$, and
(d) $D^{*+} \ra D^0 \pi^+,\ D^0 \ra K^- \pi^+ \pi^0$. Although the
resolution of the $D^{*+}$ mass peak is worse in mode (d) compared to
the other channels, it is still good enough to be used
in this analysis. Note, the charm signals in Fig.~\ref{blife:semiex} are
quite clean and rather competitive with $D^{(*)}$ signals found at $e^+e^-$
machines.  This
demonstrates the feasibility of $B$ physics in a hadron collider environment
without using $J/\psi$'s.

The obtained lifetime distributions from $\ell^+ \bar D^0$ and $\ell^+ D^{*-}$ 
are used to determine the individual $B^+$ and $B^0$ lifetimes. 
A $\ell^+ \bar D^0$ combination usually originates from a charged $B$ meson
while 
$\ell^+ D^{*-}$ comes from a $B^0$. This simple picture is complicated by the
existence of $D^{**}$ states which are the source of $\bar D^0$ ($D^{*-}$)
mesons that originate from a decay $B^0 \ra D^{**-} \ell^+,\ D^{**-}
\ra \bar D^0 X$
($B^+ \ra \bar D^{**0} \ell^+,\ \bar D^{**0} \ra D^{*-} X$). This cross talk from
$D^{**}$ resonances has been decomposed using Monte Carlo. A combined
lifetime fit, as shown on the
righthand side of Fig.~\ref{blife:semiex}, yields the following $B$ lifetimes: 
\begin{eqnarray}
\tau(B^+) = (1.64 \pm 0.06 \pm 0.05)\ p{\rm s} \\
\tau(B^0) = (1.48 \pm 0.04 \pm 0.05)\ p{\rm s} \\
\tau(B^+)/\tau(B^0) = 1.11 \pm 0.06 \pm 0.03.
\end{eqnarray}

\subsubsection{{\boldmath $B^+\ {\rm and}~B^0$} Lifetime Comparison}

\begin{figure}[tbp]\centering
\begin{picture}(145,90)(0,0)
\centerline{
\epsfysize=9.0cm
\epsffile[15 250 550 765]{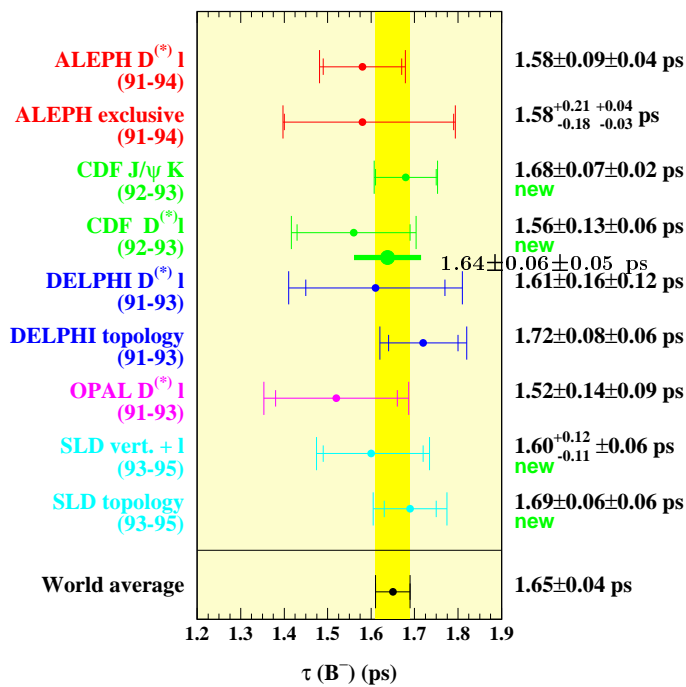}
}
\end{picture}
\caption{Comparison of the CDF $B^+$ lifetime measurements with other
experiments.} 
\label{blife:compbp}
\end{figure}

\begin{figure}[tbp]\centering
\begin{picture}(145,90)(0,0)
\centerline{
\epsfysize=9.0cm
\epsffile[20 250 530 765]{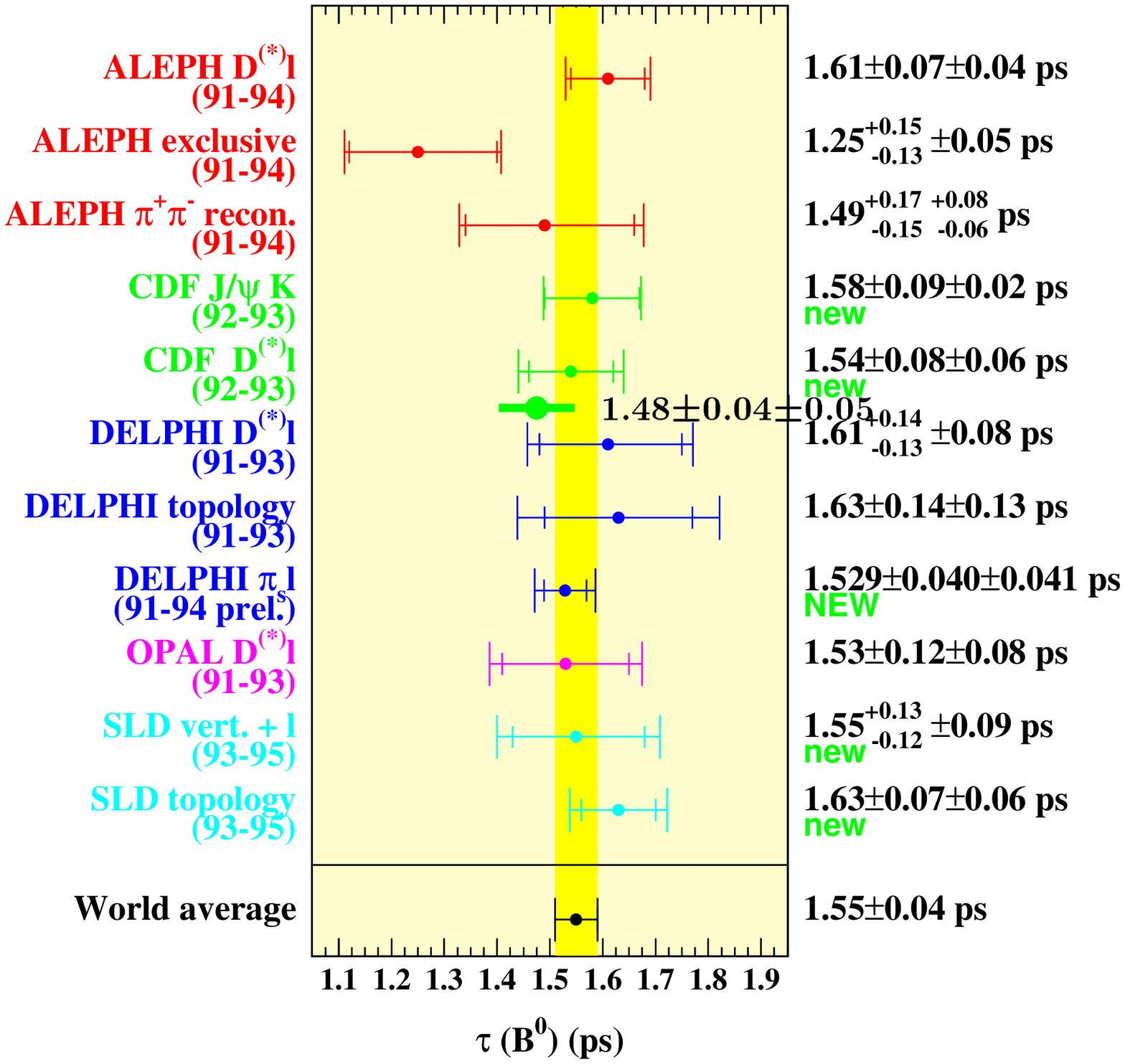}
}
\end{picture}
\caption{Comparison of the CDF $B^0$ lifetime measurements with other
experiments.} 
\label{blife:compb0}
\end{figure}

\begin{figure}[tbp]\centering
\begin{picture}(145,80)(0,0)
\centerline{
\epsfysize=8.0cm
\epsffile[23 255 505 765]{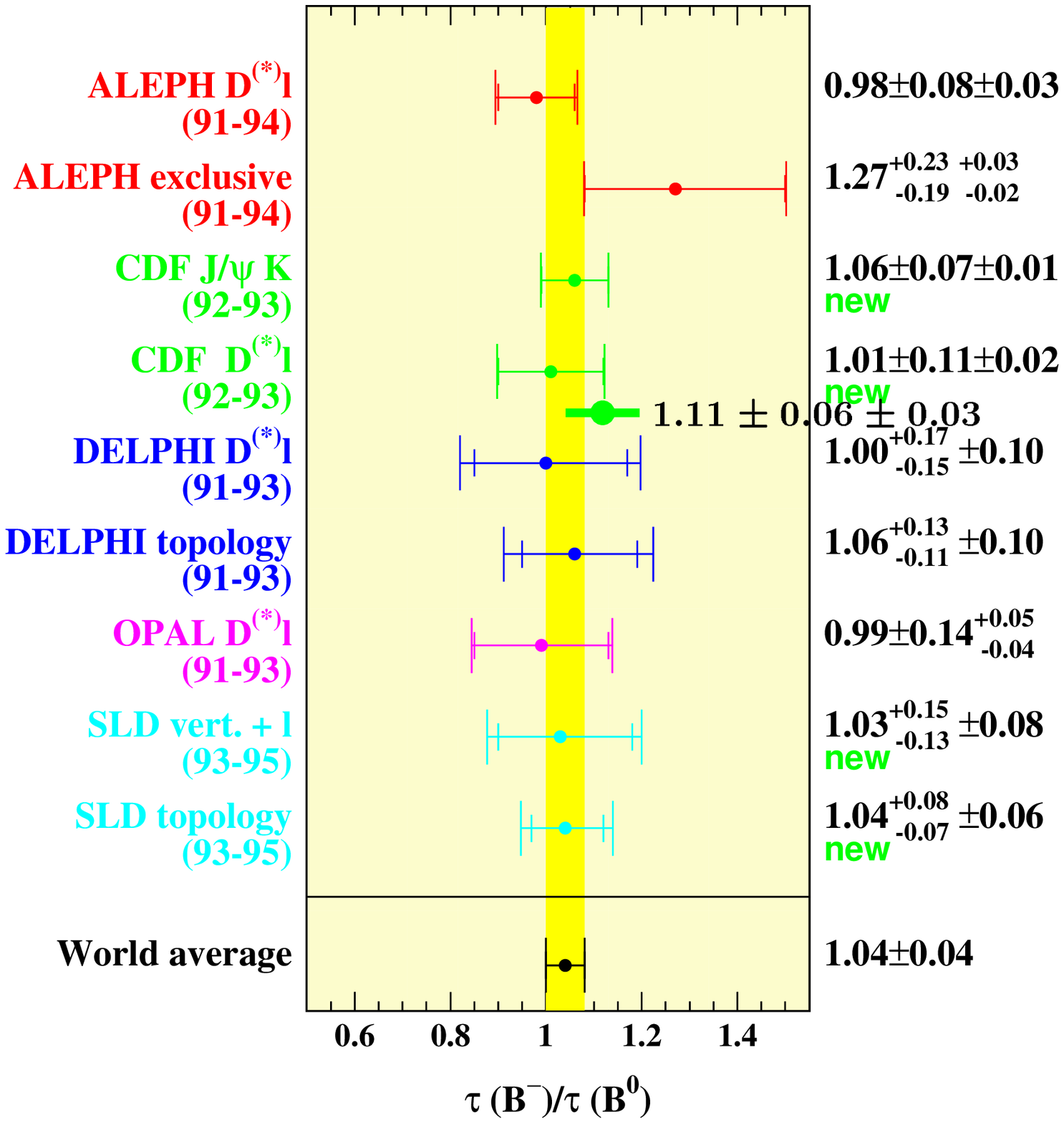}
}
\end{picture}
\caption{Comparison of the CDF $\tau(B^+)/\tau(B^0)$ lifetime ratio
measurements with other experiments.} 
\label{blife:compratio}
\end{figure}

A comparison of the CDF $B^+$ and $B^0$ lifetime measurements with other
experiments, as presented at 
the 28th International Conference on
High Energy Physics, Warsaw, Poland
(ICHEP'96)~\cite{richman},
can be found in Figures~\ref{blife:compbp} and \ref{blife:compb0},
respectively. A comparison of the CDF lifetime ratio measurements is
presented in Figure~\ref{blife:compratio}.
The $B$ lifetime analysis using
partially reconstructed $B \ra D^{(*)}\ell X$ decays
is a new result comprising
the full
Run~I statistics and updates the 1992-93 measurement.
These new measurements have not yet 
been available at the time of the 
ICHEP'96 conference and have been added
to Figures~\ref{blife:compbp} through \ref{blife:compratio}. This
comparison shows that the CDF $B$ lifetime measurements are
competitive with the results from the $Z$ pole at LEP and SLC.

The combination of the presented $B$ lifetime results using fully
and partially reconstructed $B$ decays yields the following combined CDF
$B$ lifetime average:
\begin{eqnarray}
\tau(B^+) = (1.66 \pm 0.05)\ p{\rm s} \\
\tau(B^0) = (1.52 \pm 0.06)\ p{\rm s} \\
\tau(B^+)/\tau(B^0) = 1.09 \pm 0.05.
\end{eqnarray}
The error is the sum of statistical and systematic errors, where the
correlated systematic errors are taken into account. 
It appears that the lifetime ratio measurement is different from unity
by almost two standard deviations. 

\subsubsection{{\boldmath $B$} Lifetimes: {\boldmath \Bs} Meson}   

\begin{figure}[tbp]\centering
\begin{picture}(145,70)(0,0)
\put(18,58){\large\bf (a)}
\put(70,58){\large\bf (b)}
\centerline{
\epsfysize=7.0cm
\epsffile[60 220 504 720]{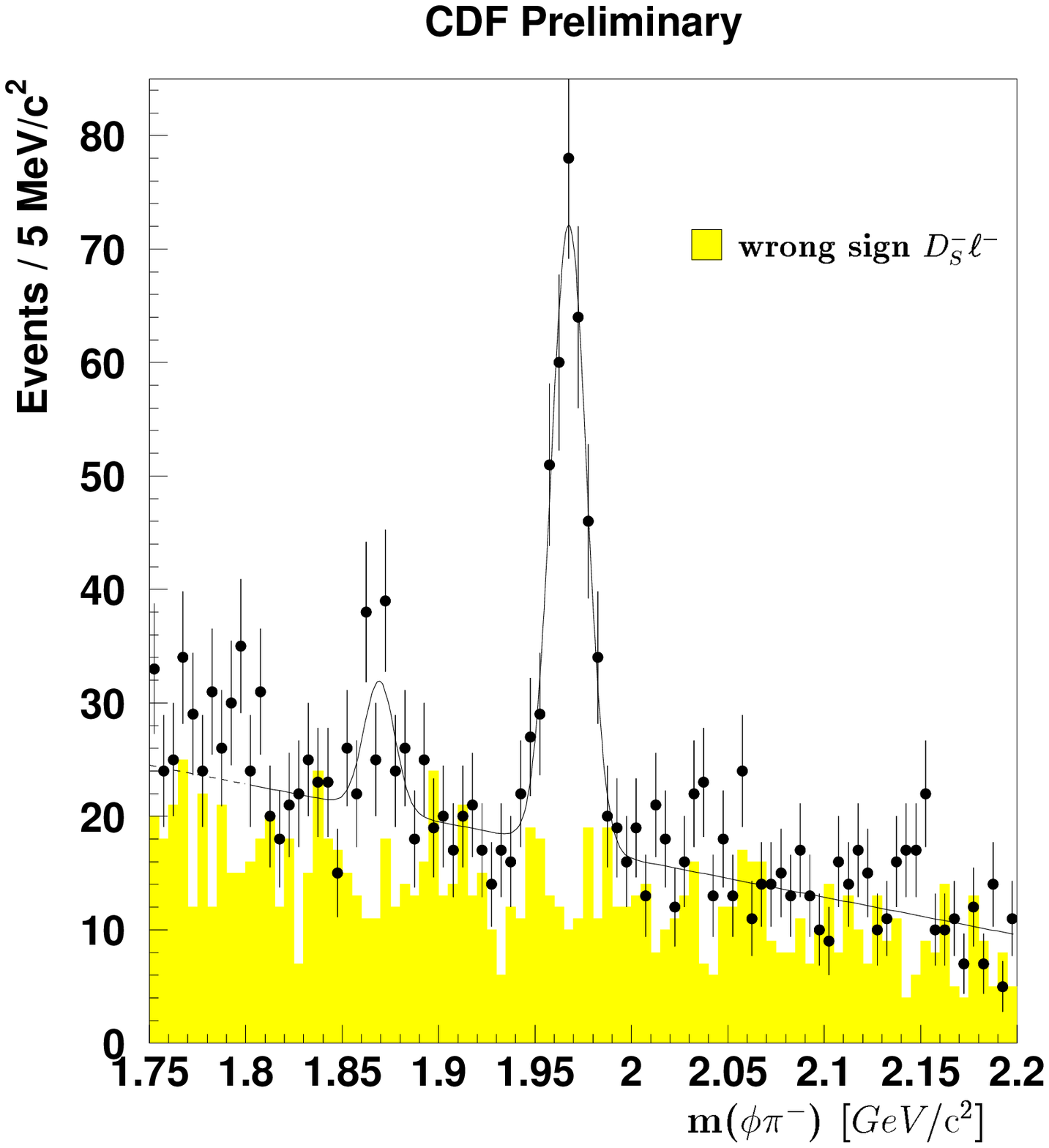}
\hspace*{0.5cm}
\epsfysize=7.0cm
\epsffile[30 10 520 510]{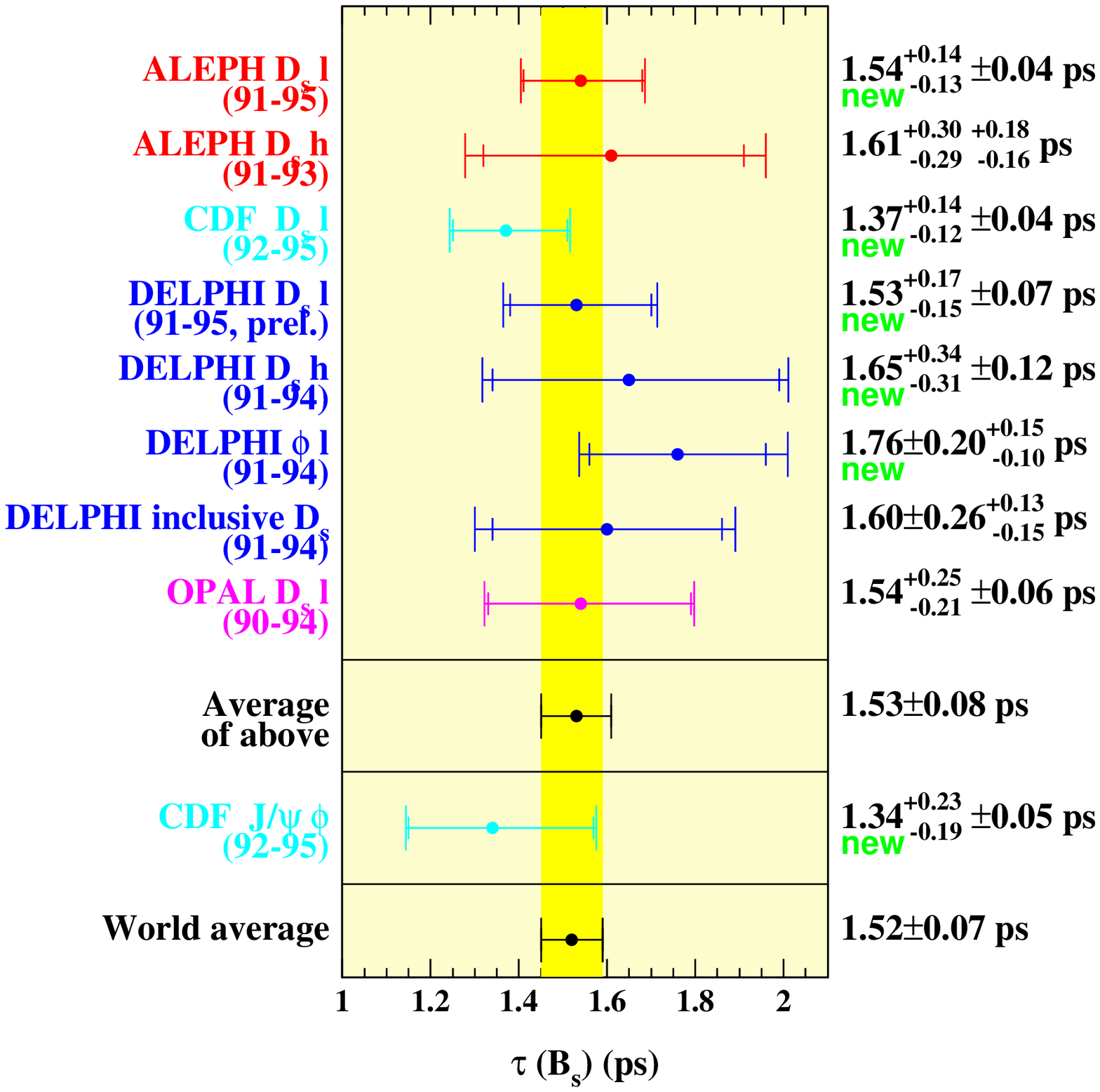}
}
\end{picture}
\caption{\Bs\ lifetime measurement: (a) Invariant $\phi\pi^-$ mass
distribution for right sign \Dsl\ combinations. The shaded histogram
shows the wrong sign distribution. (b)
Comparison of the CDF \Bs\ lifetime measurement with other
experiments.} 
\label{blife:bs}
\end{figure}

The lifetime of the \Bs\ meson is measured at CDF using the semileptonic decay 
$\Bs \ra \Dsl \nu X$, where the \Ds\ is reconstructed through its decay mode 
$\Ds \ra \phi \pi^{-}$, $\phi \rightarrow K^+ K^-$. The analysis
starts again with the single lepton ($e,\,\mu$) trigger data searching
for $\Ds \ra \phi \pi^-$ candidates in a cone around the lepton.   
The \Ds\ candidates are intersected with the lepton to find the \Bs\
decay vertex from where the analysis follows the description of the $B$
lifetime measurement using partially reconstructed $B$ mesons (see
Sec.~\ref{partblife}). 
A signal of $(254\pm21)$ \Dsl\ candidates is found as shown
in Fig.~\ref{blife:bs}a), where the $\phi\pi^-$ invariant mass
distribution for right sign \Dsl\ combinations is plotted. 
The shaded histogram shows the wrong sign $\Ds\ell^-$ distribution.  
Using these events, the \Bs\ meson lifetime is determined to be
\begin{equation}
\tau(\Bs) = (1.37^{\ + 0.14}_{\ - 0.12} \pm 0.04)\ p{\rm s}. 
\end{equation}
A comparison of the CDF \Bs\ lifetime measurement with other
experiments at LEP and SLC can be found in Fig.~\ref{blife:bs}b).

\subsubsection{{\boldmath $B$} Lifetimes: {\boldmath $\Lambda_b^0$} Baryon}   

\begin{figure}[tbp]\centering
\begin{picture}(145,70)(0,0)
\put(15,58){\large\bf (a)}
\put(75,63){\large\bf (b)}
\centerline{
\epsfysize=7.0cm
\epsffile[50 155 522 622]{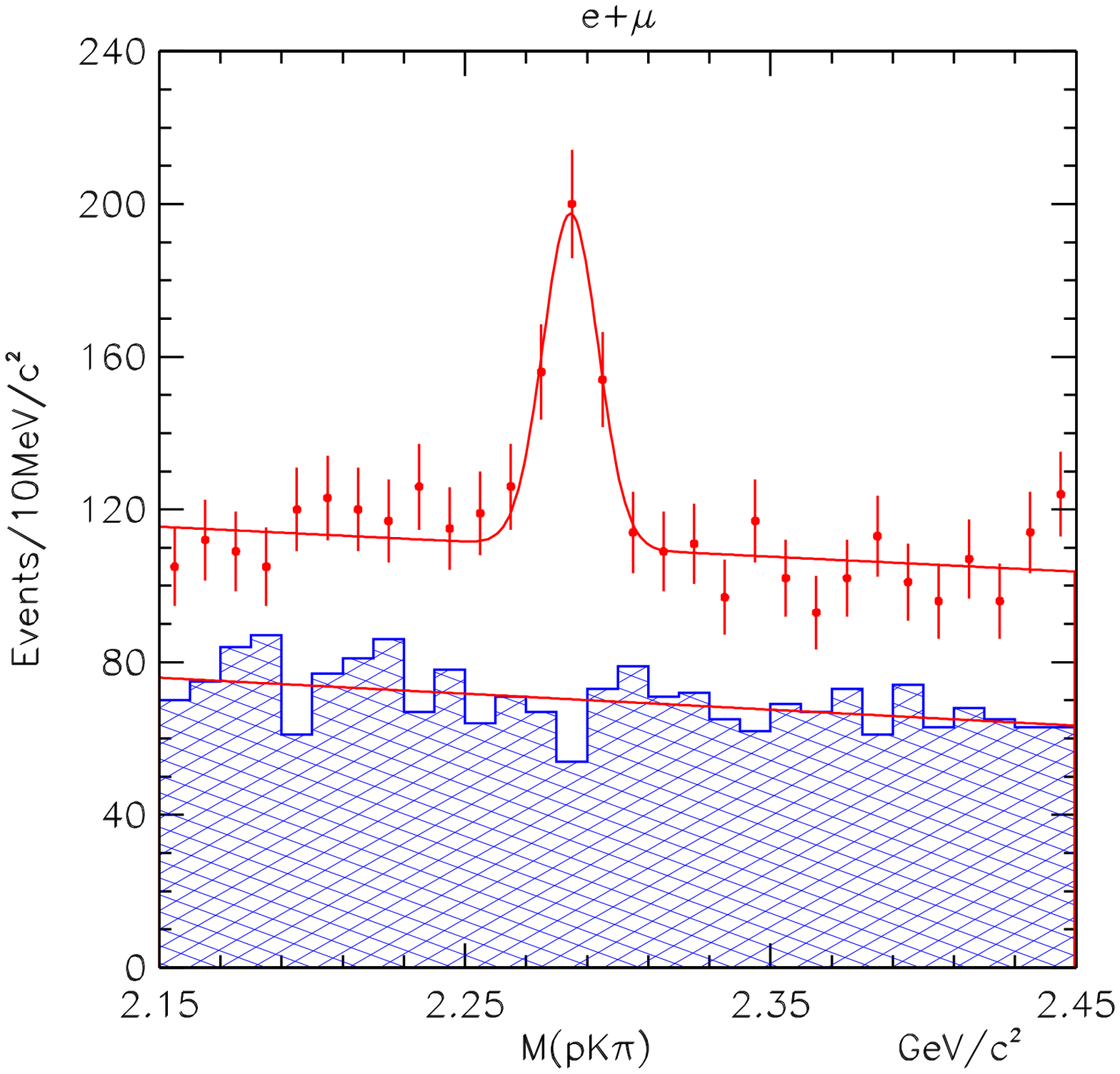}
\hspace*{0.5cm}
\epsfysize=6.5cm
\epsffile[70 5 440 340]{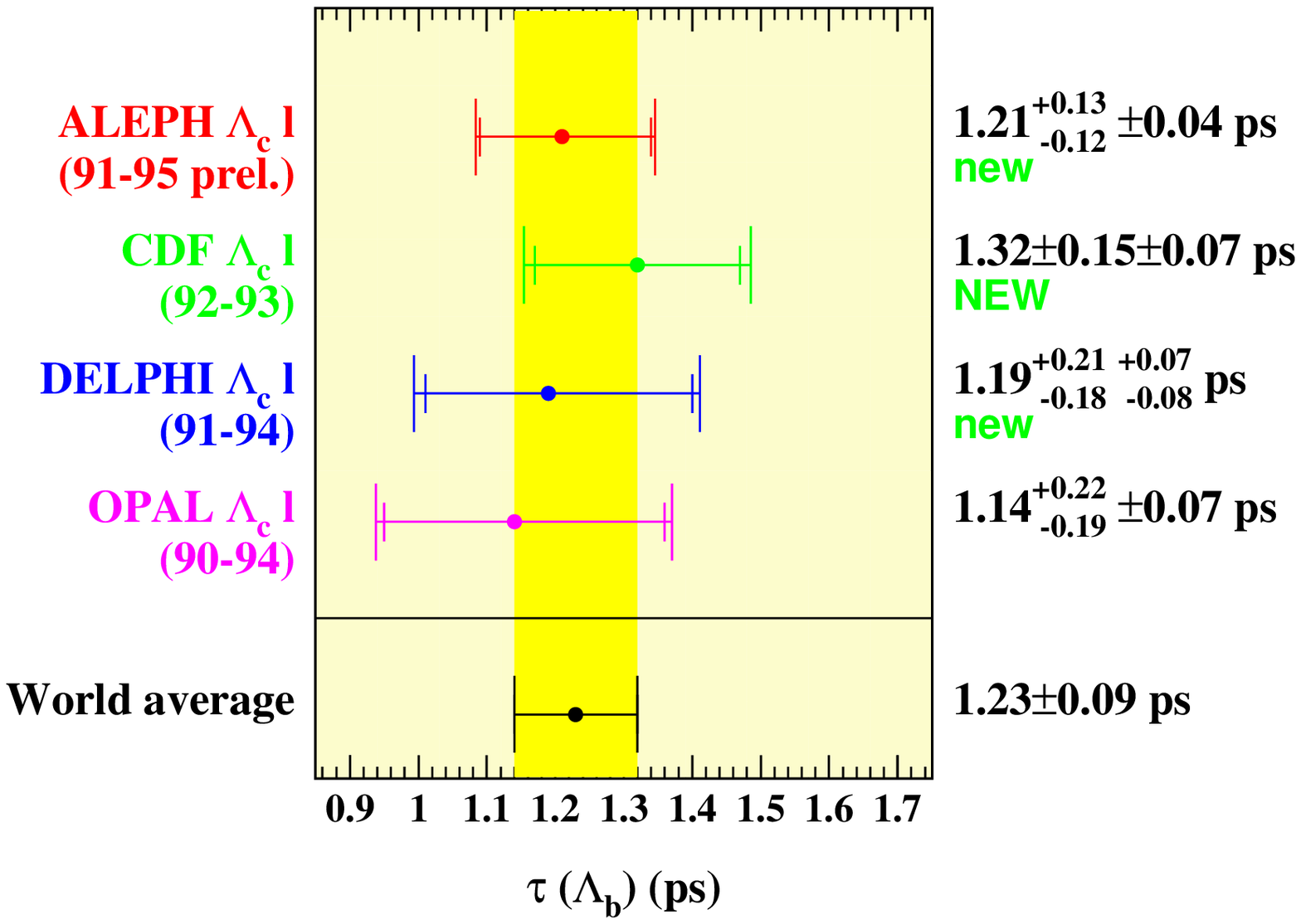}
}
\end{picture}
\caption{$\Lambda_b^0$ lifetime measurement: (a) Invariant $pK\pi$ mass
distribution for right sign $\Lambda_c^+\ell^-$ combinations. 
The shaded histogram
shows the wrong sign distribution. (b)
Comparison of the CDF $\Lambda_b^0$ lifetime measurement with LEP results.}
\label{blife:lamb}
\end{figure}

The analysis principle for the $\Lambda_b^0$ lifetime measurement is
very similar to the \Bs\ lifetime analysis at CDF. The $\Lambda_b^0$
baryon is reconstructed through the semileptonic decay 
$\Lambda_b^0 \ra \Lambda_c^+ \ell^- \nu X$, with the subsequent decay
$\Lambda_c^+ \ra p K^-\pi^+$.
The analysis again
uses the single lepton trigger data searching
for $\Lambda_c^+ \ra p K^-\pi^+$
candidates in a cone around the lepton.   
The $\Lambda_c^+$ candidates are intersected with the lepton to find the
$\Lambda_b^0$
decay vertex.
A signal of $(197\pm25)$ $\Lambda_c^+$ candidates is obtained 
as shown in Fig.~\ref{blife:lamb}a), where the $p K\pi$ invariant mass
distribution for right sign $\Lambda_c^+ \ell^-$ combinations is plotted. 
The shaded histogram shows the wrong sign $\Lambda_c^- \ell^-$ distribution.  
Using these events, the $\Lambda_b^0$ lifetime is determined to be
\begin{equation}
\tau(\Lambda_b^0) = (1.32 \pm 0.15 \pm 0.07)\ p{\rm s} 
\end{equation}
A comparison of the CDF $\Lambda_b^0$ lifetime measurement with
LEP results can be seen in Fig.~\ref{blife:lamb}b). The CDF
$\Lambda_b^0$ lifetime is competitive with the LEP measurements in
precision, but tend to be longer compared to the LEP
results.

\subsubsection{{\boldmath $B$} Lifetimes: Summary}   

A summary of the $B$ lifetime measurements at CDF is given in
Table~\ref{blife:sum}. As we have seen, CDF's $B$ lifetime results are
very competitive with the LEP and SLC measurements where a precision
of a few percent is reached.  
Although CDF's measurement of the $B^+/B^0$ lifetime ratio appears to be
different from unity by almost two standard deviations, 
the precision is still not yet sufficient to distinguish between
theoretical approaches.
The $\Lambda_b^0$ lifetime lies closer to the $B^0$ lifetime 
at CDF with a ratio of $\tau(\Lambda_b^0)/\tau(B^0) = 0.87\pm0.11$,
while the LEP experiments report a ratio of
$0.78\pm0.04$~\cite{richman}. 
Theory favours the value for this ratio to be closer to 0.9-1.0 in
good agreement with the CDF measurement.

\begin{table}[b]
\begin{center}
\vspace*{-0.6cm}
\begin{tabular}{|rcl|}
\hline
\multicolumn{3}{|c|}{CDF $B$ Lifetime Summary} \\
\hline
\hline
 $\tau(B^+)$ & = & $(1.66 \pm 0.05)\ p$s \\ 
 $\tau(B^0)$ & = & $(1.52 \pm 0.06)\ p$s \\ 
 $\tau(B^+)/\tau(B^0)$ & = & $1.09 \pm 0.05$ \\ 
\hline
 $\tau(\Bs)$ & = & $(1.37 \pm 0.14)\ p$s \\ 
 $\tau(\Lambda_b^0)$ & = & $(1.32 \pm 0.17)\ p$s \\ 
\hline
\end{tabular}
\caption{Summary of CDF $B$ hadron lifetime results.}
\label{blife:sum} 
\end{center}
\vspace*{-0.5cm}
\end{table}

\subsection{{\boldmath $B\bar B$} Oscillations}

In the Standard Model $B\bar B$ mixing occurs through the electroweak box
diagram, where the dominant contribution is through the top quark loop.
The size of the oscillation is expressed in terms of
the mixing parameter $x = \Delta m / \Gamma$, where $\Delta m$ is the
difference in mass between the two $B$ meson eigenstates and $\Gamma$ refers to
the average lifetime of both $B$ states $\tau_B = \hbar/\Gamma$. 
For a beam initially pure in $B^0$ mesons (at $t=0$), the numbers of $B^0$ and
$\bar B^0$ mesons at proper time $t$, 
$N(t)_{B^0 \rightarrow B^0}$ and $N(t)_{B^0 \rightarrow \bar{B}^0}$,
respectively, are given by:
\begin{eqnarray}
N(t)_{B^0 \rightarrow B^0} 
        = \frac{1}{2\tau_B} e^{-t/\tau_B}(1+\cos\Delta m_d\, t) \\
N(t)_{B^0 \rightarrow \bar{B}^0} 
        = \frac{1}{2\tau_B} e^{-t/\tau_B}(1-\cos\Delta m_d\, t).
\end{eqnarray}

Measurements of the frequencies of $B^0$ and \Bs\ oscillations can
potentially constrain the magnitudes of the CKM matrix elements
$V_{td}$ and $V_{ts}$, where in the ratio of $\Delta m_d / \Delta m_s$ 
several theoretical uncertainties cancel out
\begin {equation}
  \frac{\Delta m_d}{\Delta m_s} = \frac{m_{B^0}}{m_{\Bs }}\,
        \frac{\eta_{B^0}^{\rm QCD}}{\eta_{\Bs }^{\rm QCD}}\, 
	\frac{f^2_{B^0} B_{B^0}} 
        {f^2_{\Bs} B_{\Bs}}\, \frac{|V_{td}|^2}{|V_{ts}|^2}. 
\end{equation}
Here, $f_B$ is the weak $B$ decay constant, $B_B$ the
bag parameter of the $B$ meson, and $\eta^{\rm QCD}$ are QCD corrections which
are in the order of one.

In general, a time dependent mixing measurement requires the knowledge
of the flavour of the $B$ meson at production and at decay, as well as
the proper decay time of the $B$ meson.
Experimentally, the flavour of the $B$ meson is determined at the
time of its decay from the observed decay products like the charge of
the lepton from a semileptonic $B$ decay.  The flavour
at production time can be determined in various ways, employing
either the second $b$-flavoured hadron in the event, or the charge
correlation with particles produced in association with the
$B$ meson. We report about two recent time dependent $B\bar B$ mixing
results from 
CDF, which exploit these two ways of tagging the $B$ flavour at the origin.

\subsubsection{{\boldmath $B\bar B$} Mixing in {\boldmath $e\mu$} Events}

For this analysis the $e\mu$ trigger data are used, where both leptons
are assumed to come from the semileptonic decay of both $b$ hadrons in
the event: $b_1 \ra eX$ and $b_2 \ra \mu X$. This means, the flavour of
the $B$ meson at decay is tagged by its semileptonic decay, while the
semileptonic decay of the other $b$ hadron in the event tags the $B$
flavour at production. The requirement $m_{e\mu} > 5$ \gevcc\ ensures
that both leptons originate from two $b$ hadrons and not from a
sequential decay of one $b$ hadron: $b \ra c\, \ell_1  X$, with $c \ra
\ell_2 X$.  

The principle of this analysis is to search for an inclusive secondary
vertex associated with one of the leptons. The decay length of this
vertex and the momenta of the tracks associated with the lepton
provide an estimate of the $c\tau$ of the $B$ meson. The boost
resolution for this technique is about 21\% for electrons and
about 24\% for muons. 
In order to search for an inclusive
secondary vertex, a modified version of the SVX $b$ tagging alogrithm,
which was successfully used in the search for $b$ tags in top quark
events (see Sec.~\ref{toplepjet}), has been used. This algorithm has
been tuned for high efficiency near $c\tau = 0$, with the efficiency
reaching a plateau of about 40\% for $c\tau > 500~\mu$m according to
a MC study.

\begin{figure}[tbp]\centering
\begin{picture}(145,70)(0,0)
\put(14,55){\large\bf (a)}
\put(90,44){\large\bf (b)}
\centerline{
\epsfysize=7.0cm
\epsffile[5 22 511 550]{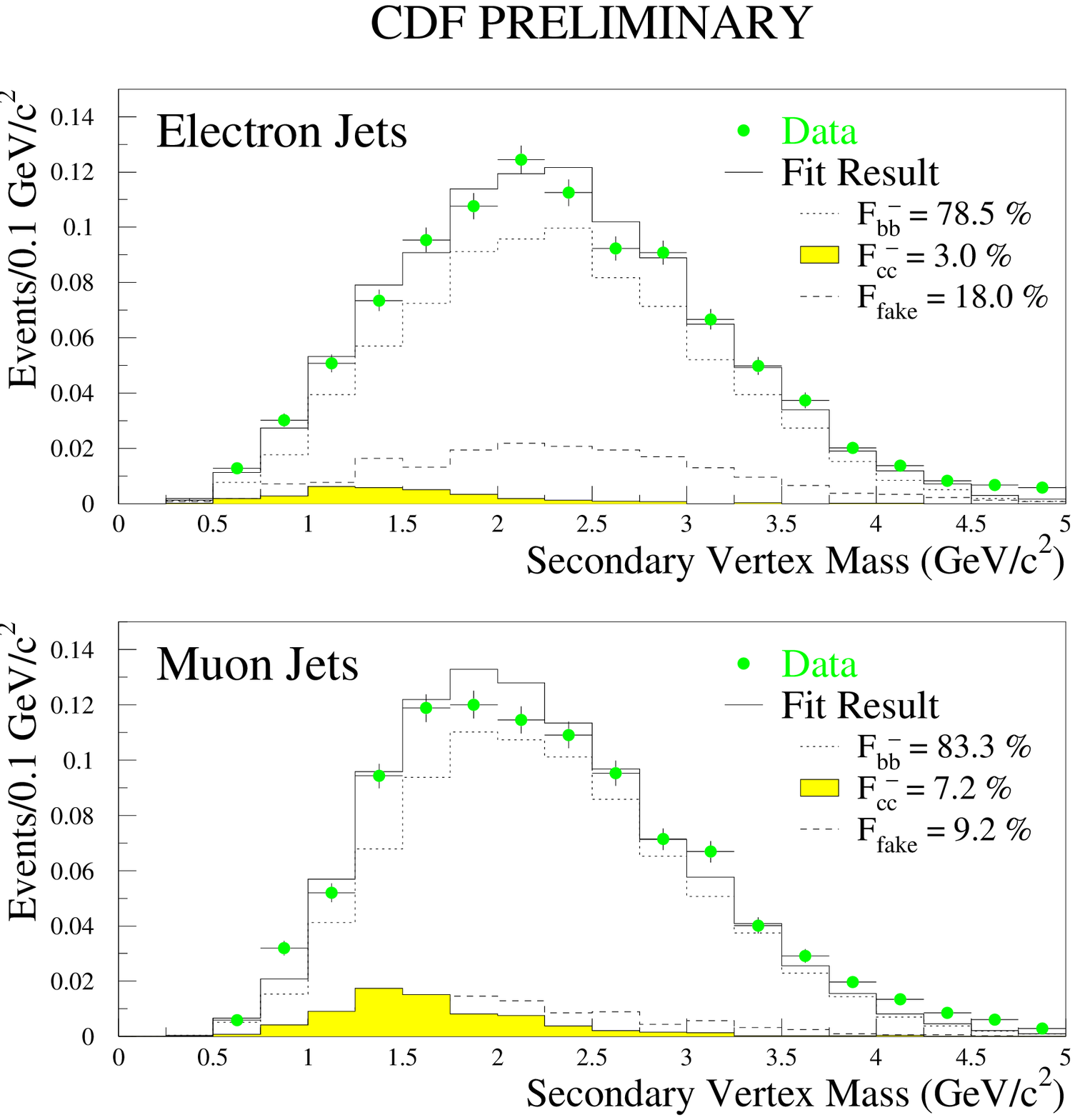}
\hspace*{0.5cm}
\epsfysize=7.0cm
\epsffile[5 22 511 550]{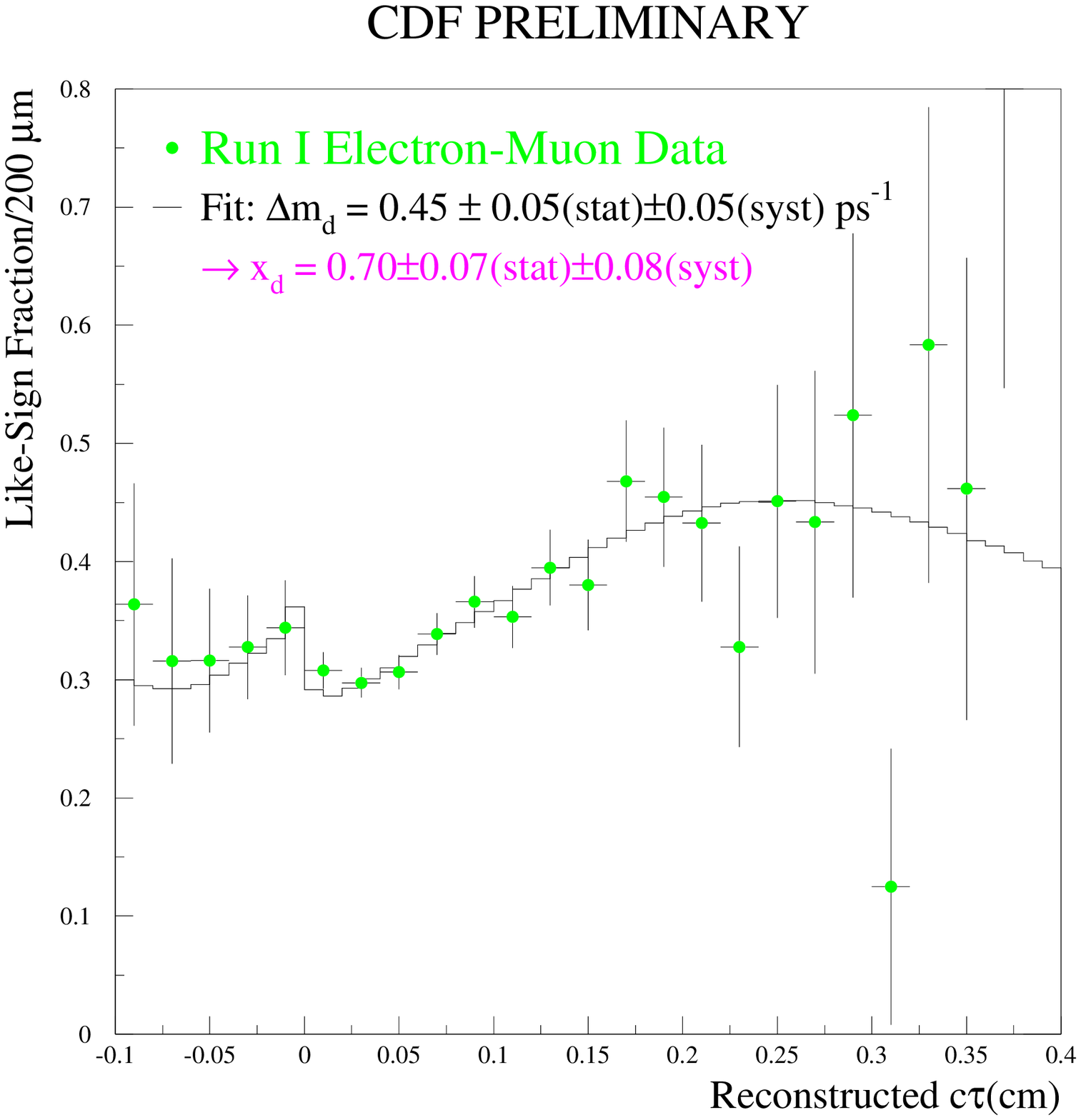}
}
\end{picture}
\caption{(a) Example of the determination of the sample composition in
the $e\mu$ mixing analysis using the invariant mass of the secondary
vertex. The fitted fractions from $b\bar b$, $c\bar c$, and fake events
are shown. 
The
terms `$e$ Jet' and `$\mu$ Jet' indicate that the secondary vertex is
associated with the electron or muon. In (b) the fit to the like-sign
fraction versus $c\tau$ is 
displayed.} 
\label{bmixemu}
\end{figure}

The important task of this analysis is to determine the sample
composition, the fraction of events which come from $b\bar b$
decays with respect to events from $c\bar c$ or background events with
at least one fake lepton.  
We find to a good approximation that the fake electron events are a
subset of the fake muon events due to the higher \Pt\ cut. Other
backgrounds arise from sequential $b \ra c \ra \ell$ decays. The
sample composition has been estimated from several kinematic
quantities, like the \ptrel\ distribution or the invariant mass of
the tagged secondary vertex. Here, \ptrel\ is defined as the
transverse momentum of the lepton with respect to the highest \Pt\
track in a cone around the lepton. An example of the determination of
the sample composition is shown in Fig.~\ref{bmixemu}a). The fitted
fractions from $b\bar b$, $c\bar c$, and fake events are displayed. The final
sample composition is given in Table~\ref{emu_sample}, which shows
that more than 80\% of the events originate from $b\bar b$ decays. 
   
\begin{table}[tbp]
\begin{center}
\vspace*{-0.6cm}
\begin{tabular}{|l|c|c|} \hline
Component & $e$ Jet & $\mu$ Jet \\ 
\hline \hline 
{Fake e with Real $\mu$}   &  $\leq 1\%$    & $\leq 1\%$  \\ 
{Fake $\mu$ Fraction} & $(15\pm 4)\%$ & $(7\pm 3)\%$ \\ \hline 
{$c\bar c$ events} &  $(2\pm 2)\%$ &  $(4\pm 3)\%$ \\ \hline 
{$b\bar b$ events} & $(83\pm 5)\%$ & $(89\pm 4)\%$ \\ \hline
{Sequential $e$ }  &  $(8.8\pm 1.3)\%$ &  $(7.9\pm 1.2)\%$ \\
{Sequential $\mu$ }      & $(13.6\pm 2.0)\%$ & $(16.5\pm 2.5)\%$ \\ \hline
\end{tabular}
\caption{Final sample composition of the $e\mu$ mixing analysis. The
terms `$e$ Jet' and `$\mu$ Jet' indicate that the secondary vertex is
associated with the electron or muon. The sequential fractions are
part of the $b\bar b$ component.}
\label{emu_sample} 
\end{center}
\vspace*{-0.5cm}
\end{table}

From a fit to the like-sign lepton fraction as a function of $c\tau$ the
mixing frequency $\Delta m_d$ is extracted as shown in
Fig.~\ref{bmixemu}b). The fit includes components for direct 
and sequential $b$ decays, $c\bar c$, and fake events.  
In about 16\% of the events with a secondary vertex around one lepton,
a secondary vertex is also found around the other lepton. These events
enter the like-sign fraction distribution twice and we allow for a
statistical correlation between the two entries. The final fit result~is
\begin{equation}
\Delta m_d = (0.45 \pm 0.05 \pm 0.05)\ p{\rm s}^{-1}, 
\end{equation}
where the dominant systematic error arises from the uncertainty in 
the sample composition.

\subsubsection{{\boldmath $B\bar B$} Mixing in $\ell D^{(*)}$ Events}
\label{sec:sst}

\begin{figure}[tbp]\centering
\begin{picture}(145,70)(0,0)
\put(10,60){\bf (a)}
\put(44,60){\bf (b)}
\put(10,25){\bf (c)}
\put(44,25){\bf (d)}
\put(87,61){\bf (e)}
\centerline{
\epsfysize=7.0cm
\epsffile[40 10 510 505]{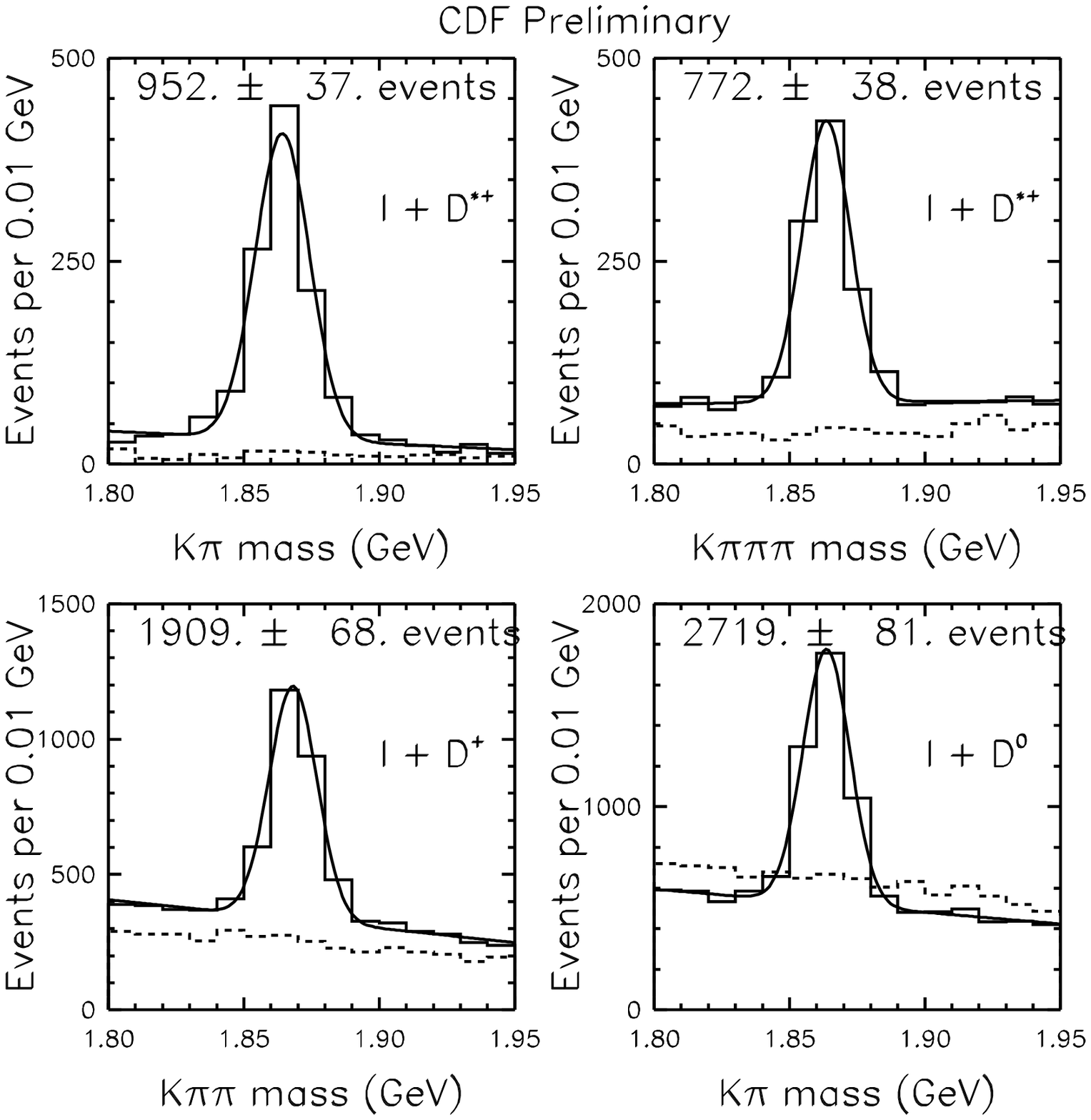}
\hspace*{0.5cm}
\epsfysize=7.0cm
\epsffile[10 10 510 505]{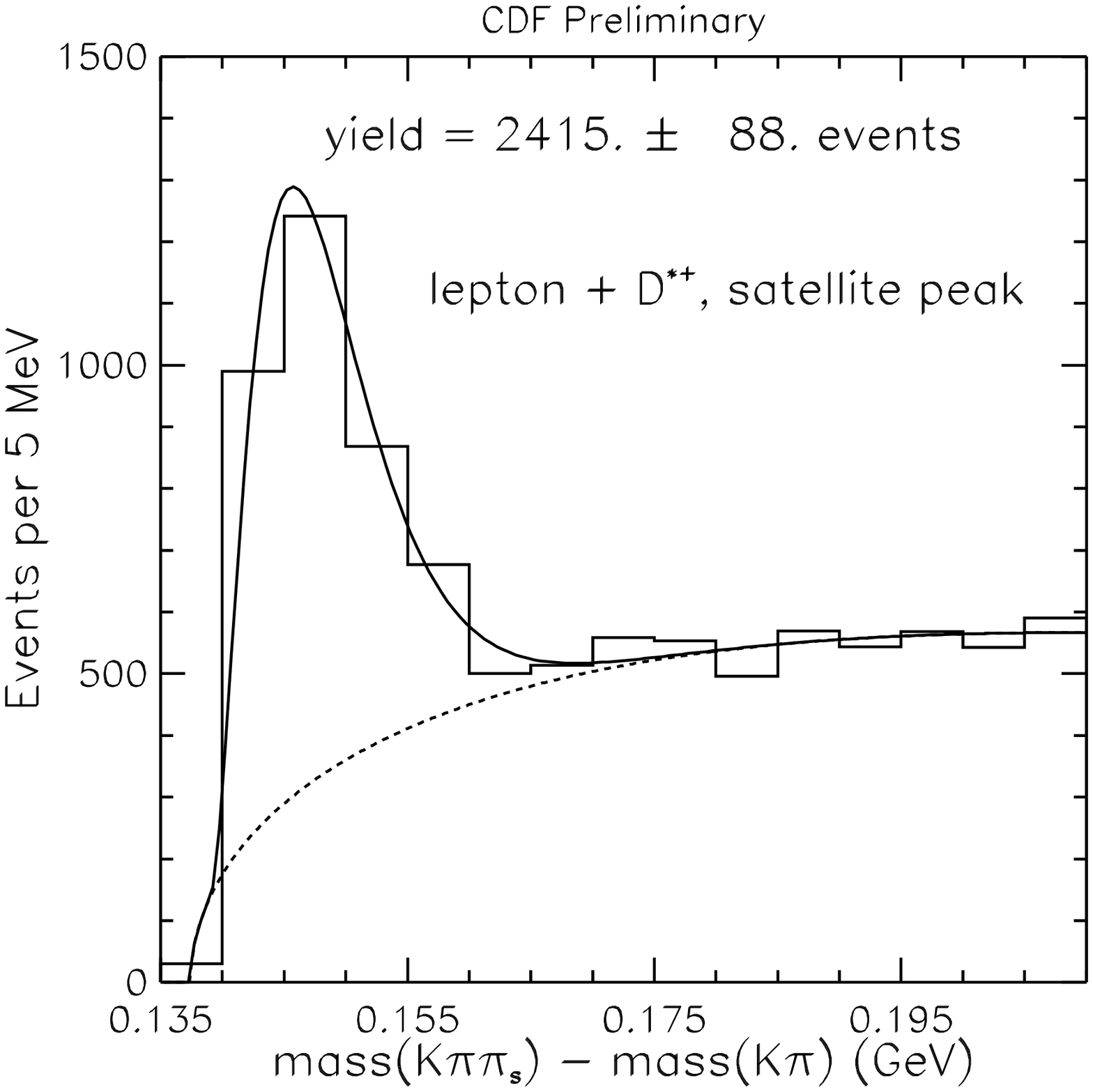}
}
\end{picture}
\caption{Invariant mass distribution of $D^{(*)}$ candidaites used in the
$B$ mixing analysis using partially reconstructed $B \ra D^{(*)}\ell X$ 
decays: 
(a) $\ell D^{*+}$ with $D^{*+} \ra D^0 \pi^+$, $D^0 \ra K^- \pi^+$,
(b) $\ell D^{*+}$ with $D^{*+} \ra D^0 \pi^+$, $D^0 \ra K^- \pi^+\pi^+\pi^-$, 
(c) $\ell D^+$ with $D^+ \ra K^- \pi^+ \pi^+$, 
(d) $\ell D^0$ with $\ell D^0 \ra K^- \pi^+$, where the $D^0$ is not from
a $D^{*+}$, and  
(e) $\ell D^{*+}$ with $D^{*+} \ra D^0 \pi^+$, $D^0 \ra K^- \pi^+ \pi^0$,
where the $\pi^0$ is not reconstructed.}
\label{bmixdlep}
\end{figure}

For this analysis $B$ mesons are reconstructed through 
their semi\-leptonic decays $B \ra D^{(*)}\ell X$
(see also Sec.~\ref{partblife}). The analysis starts with
the single lepton trigger data and reconstructs $D^{(*)}$ meson
candidates in a cone around the trigger electron or muon in the
following channels: 
$$
\begin{array}{lll}
\bar B^0 \ra D^{*+}\ell^-\nu, & 
	D^{*+} \ra D^0 \pi^+, &
	D^0 \ra K^- \pi^+, \\
 & & D^0 \ra K^- \pi^+\pi^+\pi^-, \\
 & & D^0 \ra K^- \pi^+ \pi^0,\ (\pi^0\ {\rm not\  reconstructed}) \\
\bar B^0 \ra D^+\ell^-\nu, & 
	D^+ \ra K^- \pi^+ \pi^+, &  \\
B^- \ra D^0\ell^-\nu, & 
	D^0 \ra K^- \pi^+ & ({\rm veto}\ D^{*+}\ {\rm candidates}).  \\
\end{array}
$$
Tracks with impact parameters
significantly displaced from the primary vertex are selected in order
to decrease combinatorial backgrounds. The
signals are identified as peaks in the invariant mass spectra of the
charm decay products as shown in Fig.~\ref{bmixdlep} for the different
$D^{(*)}$ decay modes.

The $D^{(*)}$ candidates are intersected with the
lepton to find the $B$ decay vertex. Since the $B$ meson is not fully
reconstructed, its $c\tau_B$ cannot be directly obtained. The boost of
the $B$ meson is determined from the observed decay
products and a $\beta\gamma$ correction is applied as
obtained from a Monte Carlo simulation.   

In order to tag the $B$ flavour at production, we use a `same side
tagging' (SST) algorithm, which exploits the correlation between the
$B$ flavour and the charge of tracks from either the fragmentation
process or $B^{**}$ resonances~\cite{Rosner}.
In this analysis no attempt is made to
differentiate the sources of correlated pions.
To study the correlation between the flavour of the $B$ meson and
the charged particles produced in association with it, we consider
all tracks that are within an $\eta$-$\phi$ cone of radius $0.7$ centered
around the direction of the $B$ candidate.  Since the $B$ meson is only
partially reconstructed, we approximate this direction with the
momentum sum of the lepton and charm hadron.

The tracks considered as tags should be consistent with the hypothesis
that they originate from the fragmentation chain or the decay of
$B^{**}$ mesons, i.e. that they originate from the primary vertex of the
event.  All tracks with transverse momentum
$\Pt>0.4$~GeV/$c$ are therefore required to
satisfy $d_0/\sigma_{d_0}<3$, where 
$d_0$ is the distance of closest approach of the track trajectory to
the estimated $B$~production position, and $\sigma_{d_0}$ is the estimated
error on this quantity.

String fragmentation models indicate that the velocity of the fragmentation
particles, that we seek for our tag, is close to the velocity of the $B$ meson.
Similarly, pions from $B^{**}$ decays should also have a velocity that is
close to the velocity of the $B$ meson.  In particular, the relative-transverse
momentum ($\ptrel$) of the particle with respect to
the combined momentum of the $B$ momentum plus particle momentum,
should be small. 
Of the candidate tracks, we select as the tag the track that has the minimum
component of momentum $\ptrel$ orthogonal to the momentum sum of that track,
the lepton, and the $D$~meson.
The efficiency for finding such a tag is about $72\%$ for this algorithm.

Since we know the flavour of the $B$ meson at decay from the
$D^{(*)}\ell$ signature, we compare the number of right-sign ($N_{RS}$)
correlations to the number of wrong-sign ($N_{WS}$) tags as a function
of $c\tau$. For the $B^0$ meson we expect the asymmetry $A(t)$ to be:
\begin{equation}
A(t) =  \frac{N_{RS}(t) - N_{WS}(t)}{N_{RS}(t) + N_{WS}(t)} =
	D\cdot\cos(\Delta m_d\,t),
\end{equation}
where $D$ is the dilution of the same side tagging algorithm. $D$ is also
often expressed in terms of the mistag fraction $w$ as $D = 1 - 2\, w$.
In our analysis we fit for both $\Delta m_d$ and $D$.

\begin{figure}[tbp]\centering
\begin{picture}(145,70)(0,0)
\put(63,60){\large\bf (a)}
\put(96,54){\large\bf (b)}
\centerline{
\epsfysize=7.0cm
\epsffile[10 10 510 505]{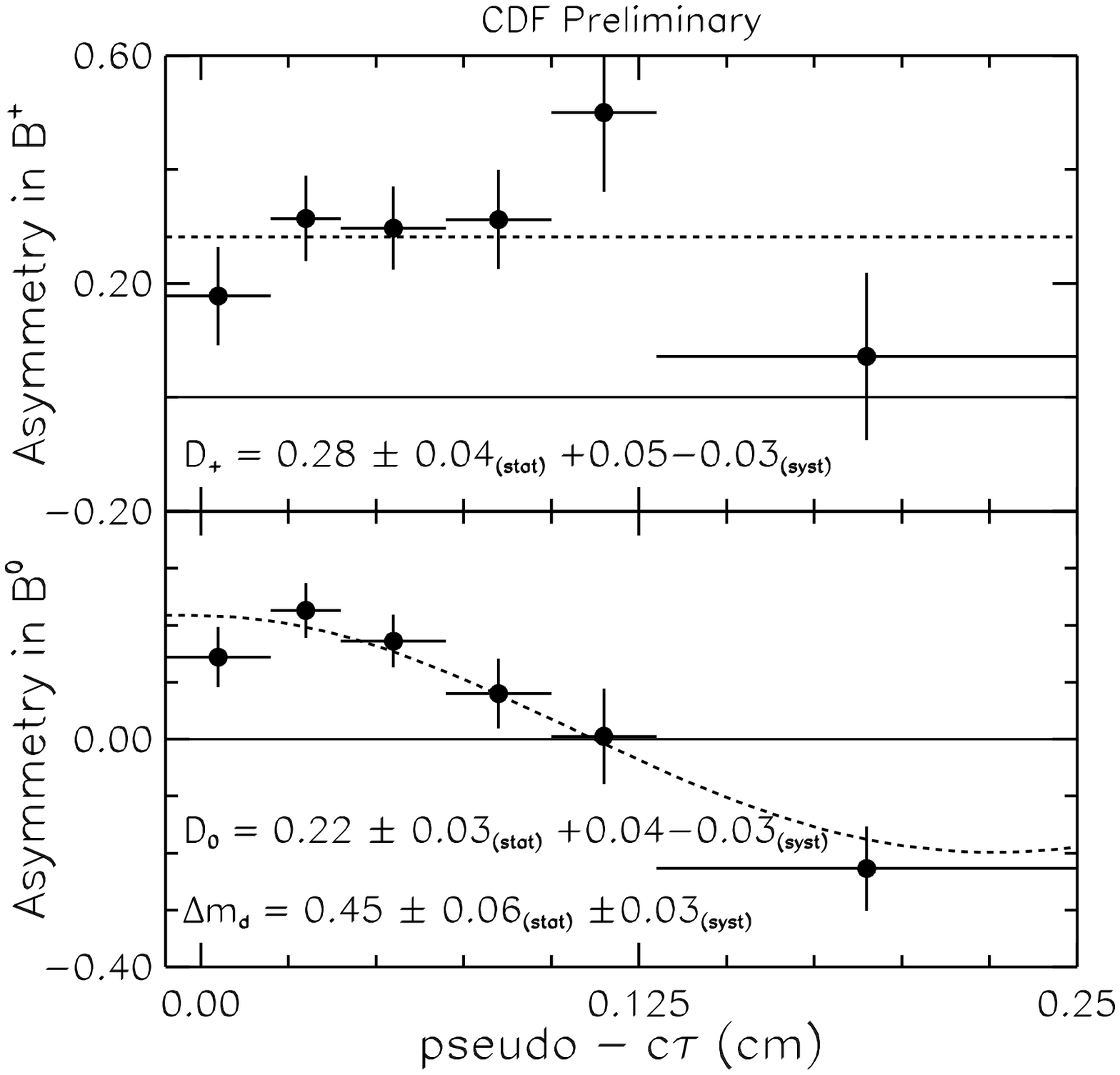}
\hspace*{0.5cm}
\epsfysize=7.5cm
\epsffile[40 30 453 530]{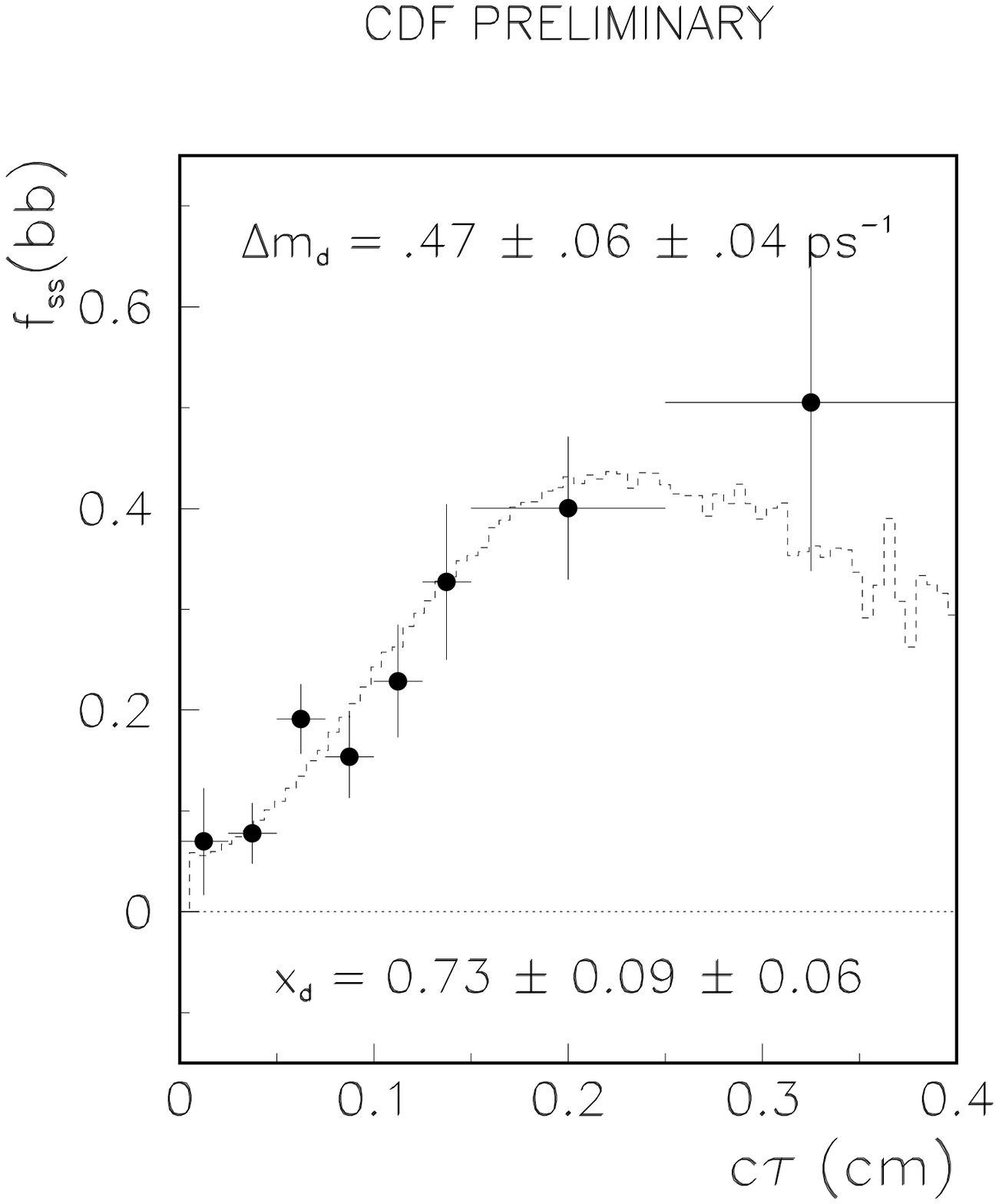}
}
\end{picture}
\caption{(a) Time dependent asymmetry for the $D^{(*)}\ell$ mixing
analysis, where the top plot shows the measured asymmetry for $B^+$
while the bottom plot represents the asymmetry for $B^0$.
(b) Fitted like-sign fraction versus $c\tau$ for a $B$ mixing analysis using
a jet charge and soft lepton tag.} 
\label{bmix:petar_owen}
\end{figure}

To obtain the asymmetry for $B^0$ and $B^+$ mesons, we correct for the
fact that each $D^{(*)}\ell$ signal has contributions from both
neutral and charged $B$ mesons via $D^{**}$ decays. We correct for
this cross talk by performing a fit bin by bin in $c\tau$. The inputs
to the fit are the raw asymmetries as measured in each sample for a
given $c\tau$ bin, and parameters describing the $D^{**}$ composition
in semileptonic $B$ decays. We fit the corrected $B^0$ asymmetry as a
function of $c\tau$ to a cosine convoluted with the $c\tau$ resolution
function, and extract the mixing frequency $\Delta m_d$ as well as the dilution
$D$ of the same side tagging algorithm as shown in
Fig.~\ref{bmix:petar_owen}a). We also determine the asymmetry for the
charged $B$ meson, which is flat in $c\tau$ as expected. We measure       
\begin{equation}
\Delta m_d = (0.45 \pm 0.06 \pm 0.03)\ p{\rm s}^{-1}, 
\end{equation}
and find $D = 0.22\pm0.04^{+0.04}_{-0.03}$. We also determine the
effective tagging 
efficiency to be $\ed = (3.4\pm1.0^{+1.2}_{-0.9})\%$. The dominant
systematic error arises from the uncertainty in the fraction of $D^{**}$ in
semileptonic $B$ decays. 

In summary, we have reported two measurements of $\Delta m_d$ at CDF
using $D^{(*)}\ell$ events with a same side tag as well as $e\mu$
dilepton events.
There are more time dependent $B\bar B$ oscillation measurements
in preparation 
at CDF, like a mixing analysis using a 
soft lepton tag (see Sec.~\ref{toplepjet}) and a jet charge tag as shown
in Fig.~\ref{bmix:petar_owen}b). The result from this analysis is
\begin{equation}
\Delta m_d = (0.47 \pm 0.06 \pm 0.04)\ p{\rm s}^{-1}. 
\end{equation}
Since there is an event overlap of about 10\% between this analysis
and the $e\mu$ analysis, a combined CDF average has not yet been
determined, but CDF's mixing results start to become competitive with the LEP
measurements~\cite{gibbons}.

\section{A Brief Look to the Future}
\label{outlook_sec}

In Run~I the luminosity of the Tevatron was limited by the
antiproton current.  
The Fermilab accelerator complex is undergoing an upgrade to produce an
order of magnitude higher luminosities in the Tevatron. The
largest change will be to replace the Main Ring with the new Main
Injector, which will be housed in a new tunnel. The Main Injector will
provide higher proton intensity onto the antiproton production target,
and larger aperture for the antiproton transfer into the Tevatron. 
After the completion and comissioning of the Main Injector, 
the Tevatron is scheduled to deliver luminosity again in summer of 1999.
The centre-of-mass energy will then be at 2.0 TeV.
Luminosities of $2.0\cdot
10^{32}$ cm$^{-2}$s$^{-1}$ will be reached yielding an integrated
luminosity of 2~$f$b$^{-1}$ delivered to the collider experiments
within two years.
The physics projections for Run~II presented here assume 2~$f$b$^{-1}$
of integrated luminosity.  
In addition to the Tevatron upgrade the CDF and D\O\ experiments will
undergo major detector upgrades, which 
are described in detail elsewhere~\cite{cdfup,d0up}. 

\subsection{Top Physics in Run~II}

In Run~II both collider detectors will
be equipped with new and improved silicon vertex detectors which will
for example enhance the efficiency to detect top decays. 
In addition, the Tevatron's centre-of-mass energy will be at 2.0 TeV, 
which will increase the top quark production cross section by 40\%. 
The estimated
yield per experiment will be about 160 dilepton events, 990 lepton
plus 4-jet events with one or more $b$ tags and 
for the top mass measurement
about 240 lepton plus
4-jet events where both $b$ jets are tagged. This will result in a statistical
uncertainty on the top quark mass measurement of 1-2~\gevcc. The top
production cross section will be known with a precision of better than
10\%. In addition, single top production can be studied, top
polarization measurements can be performed, and rare top decays will
be searched for. A more detailed overview of
top quark physics in Run~II can be found in Ref.~\cite{fermi2000}

\subsection{{\boldmath $B$} Physics in Run~II}
\label{outlook_cp}

In this section we concentrate on reviewing 
the future of $B$~physics at CDF by 
summarizing the prospects of measuring $CP$ violation
in Run~II. The $B$~physics goal is to 
measure the $CP$ asymmetry in \bjpsi\ and \bpipi\ determining \stb\ 
and \sta, respectively.
CDF plans to also
look for $CP$ violation in $\bs \ra \ds K$ and $B \ra D K$ probing 
\stg.

CDF has the advantage of being an existing experiment that took plenty of data
in Run~I. We can use these data to study the ingredients for
a future $CP$ violation measurement eg. in \bjpsi. One input 
is the knowledge of the expected number of $J/\psi K^0_S$ events, which can be
extrapolated from the $J/\psi K^0_S$ yield in CDF's 
current data. The second ingredient is the knowledge of the $B$ flavour at
production. For this purpose CDF studies several $B$ flavour tagging methods
at a hadron collider environment. The figure of merit to compare flavour
tagging algorithms is the effective tagging efficiency \ed, where
$\varepsilon$ is the efficiency of how often a flavour tag is
applicable and $D$ 
is the dilution $D = (N_{RS} -N_{WS}) / (N_{RS} + N_{WS}) = 1 - 2\, w$. 
Here, $N_{RS}$ and
$N_{WS}$ are the numbers of right and wrong sign tags, while $w$ is the mistag
fraction (see also Sec.~\ref{sec:sst}). 

\begin{figure}[tbp]\centering
\begin{picture}(145,70)(0,0)
\centerline{
\epsfysize=7.0cm
\epsffile[10 10 510 530]{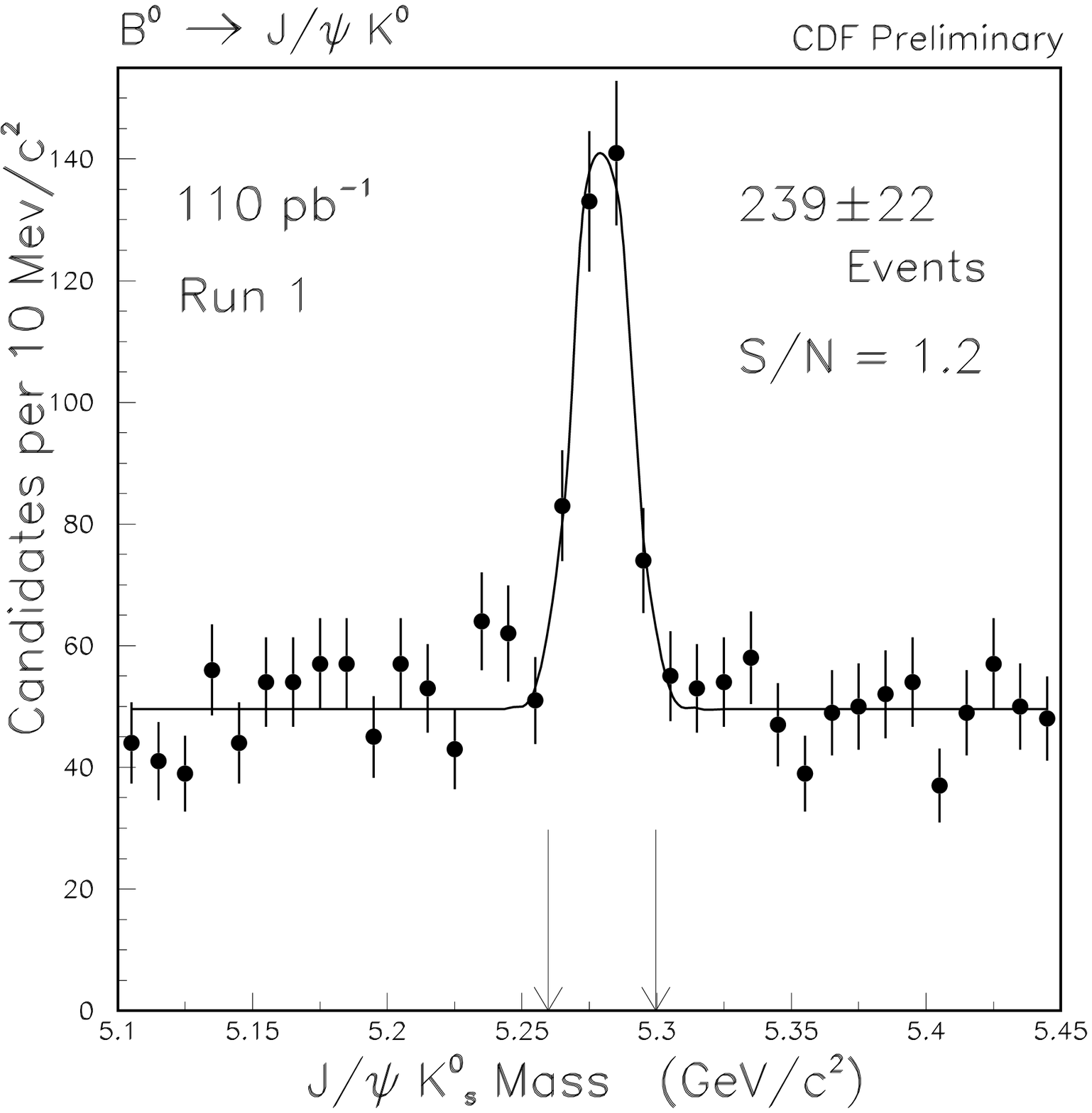}
}
\end{picture}
\caption{Run~I signal of \bjpsi\ at CDF.} 
\label{runik0s}
\end{figure}

Figure~\ref{runik0s} shows the invariant
$J/\psi K^0_S$ mass distribution from CDF's current data corresponding to
about 110 $p$b$^{-1}$. About 240 signal events with a signal-to-noise ratio 
better than 1:1 are observed. This is currently the world's largest
sample of \bjpsi\ 
and serves as proof that \bjpsi\ decays can be fully reconstructed 
in a hadron collider environment in a well understood way.

\subsubsection{{\boldmath $B$} Flavour Tagging Studies in Run~I}

Several $B$ flavour tagging methods are studied with CDF data. One is
referred to as `same side tagging' (SST) which is described in more
detail in Sec.~\ref{sec:sst}. 
This method exploits charge correlations between $B$ mesons and charged
particles 
produced in the fragmentation of the $b$ quarks.
Such correlations are expected to arise from particles
produced in the fragmentation chain and from decays of the $L=1$
$B$ mesons (the $B^{**}$ mesons)~\cite{Rosner}.
Another way of tagging the flavour of a $B$ meson at production is to exploit
the flavour of the other $B$ 
meson in the event.
This can be done through a $B$ semileptonic decay (lepton tagging) or
by counting 
the charge of the other $b$ jet (jet charge tagging). 
CDF has preliminary results on the effective tagging efficiency of these
three methods.

Using a high statistics sample of partially reconstructed $B$ mesons from
$\ell D^0$ ($B^+$ signature) and $\ell D^+/\ell D^{*+}$ ($B^0$ signature) 
combinations, charge correlations between the $B$ candidate and tracks in
its vicinity are studied as detailed in Sec.~\ref{sec:sst}.
In the case of charged $B$ mesons
we measure $\varepsilon D^2 = (5.7 \pm 1.5 ^{+2.0}_{-1.2})\%$ 
($D= (28 \pm 4 ^{+5}_{-3})\%$) for the same side tagging algorithm. 
For the neutral $B$ meson we obtain $\varepsilon D^2 = (3.4 \pm 1.0
^{+1.2}_{-0.9})\%$ with
$D= (22 \pm 3 ^{+4}_{-3})\%$. The proof that this tagging method actually
works is shown in Fig.~\ref{bmix:petar_owen}a), where the
asymmetry of right minus wrong sign tags normalized to the sum of both is
plotted versus the $c\tau$ of the $B$ meson. The top plot is for $B^+$
candidates 
and shows a flat behavior as expected. The bottom plots shows the same
distribution for $B^0$ mesons, and exhibits an oscillatory
behaviour as expected from $B \bar B$ mixing. This means applying same side
tagging to a sample of partially reconstructed $B$ mesons results in a
measurement of time dependent $B^0$ mixing. 
We also studied same side tagging with a sample of fully reconstructed $B \ra
J/\psi K^{(*)}$ decays.  
The results of 
$\ed = (4.0 \pm 1.9)\%$ ($D= (33 \pm 8)\%$) and 
$\ed = (1.5 \pm 3.0)\%$ ($D=~(19\pm19)\%$) 
for $B^+$ and $B^0$ mesons, respectively, are somewhat limited by statistics.

CDF also studied opposite side lepton tagging and obtains an effective
tagging efficiency of $\ed = (0.6\pm0.1)\%$ for 
soft muon tagging and $\ed = (0.3 \pm 0.1)\%$ for soft electron tagging.
The tagging algorithms are similar to the ones used in the top quark
search as described in
Sec.~\ref{toplepjet}. 
The preliminary result for the effective  
tagging efficieny of jet charge tagging is $(1.0 \pm 0.3)\%$, which makes
use of information from the silicon vertex detector. 
The proof that jet charge and soft lepton tagging actually
works is shown in Fig.~\ref{bmix:petar_owen}b),
where the time dependent $B$ mixing measurement using these two tags
is displayed.
Combining the existing
measurements on $B$ flavour tagging from Run~I results in an effective tagging
efficiency of $\approx 3.4\%$. 

\subsubsection{{\boldmath $CP$} Asymmetry in 
		{\boldmath $\bjpsi:\ \stb$}}

For the measurement of \stb\ in \bjpsi\ CDF expects about 15,000 $J/\psi K^0_S$
events. This event number will be obtained with a lower muon trigger
threshold of $\Pt > 1.5\ \gevc$ compared to about 2.0 \gevc\ in Run~I,
improved muon coverage, and by also triggering 
on $J/\psi \ra e^+e^-$. The effective tagging efficiencies are expected to
improve with the upgraded detector. We expect $\ed \approx 2\%$
for lepton tagging due to a
better coverage for the lepton identification. For same side tagging
we expect $\ed \approx 2\%$ from a cleaner selection of fragmentation
tracks with SVX~II. 
Finally, for jet charge tagging we expect $\ed \approx 3\%$ from an improved
purity of the algorithm with 3-dimensional vertexing and the extented
coverage of 
SVX~II. Considering the overlap of all three tags by combining them,
we expect a total \ed\ of about 5.5\% resulting in an uncertainty on \stb\ of 
$\Delta \stb = 0.09$. 

\subsubsection{{\boldmath $CP$} Asymmetry in 
		{\boldmath $\bpipi:\ \sta$}}

The key to measure the $CP$ asymmetry in \bpipi\ is to trigger on this decay
mode in hadronic collisions. CDF plans to do this with a three level
trigger system. 
On Level~1 two oppositely charged tracks with $\Pt > 2\ \gevc$ found with a
fast track processor yield an accept rate of
about 16~kHz. This will be reduced to about 20~Hz on Level~2 using impact
parameter information ($d > 100\ \mu$m). On Level~3 the full event information
is available further reducing the trigger rate to about 1~Hz. With this
trigger we expect about 10,000 \bpipi\ events in 2 $f$b$^{-1}$. Assuming the
same effective tagging efficiency of $5.5\%$ we expect an uncertainty
on \sta\ of  
$\Delta \sta = 0.10$. 
Backgrounds from $B \ra K \pi$ and $B \ra K K$ decays can be extracted from
the untagged signal by making use of the invariant mass distribution
as well as CDF's 
d$E$/d$x$ capability in the central tracking chamber.

\subsubsection{{\boldmath $CP$} Asymmetry in 
		{\boldmath $\bs \ra \ds K:\ \stg$}}

The $CP$ asymmetry in \stg\ completes the test of the unitarity triangle. The
angle $\gamma$ can be probed via the decay $\bs^0 \ra \ds^- K^+$ and $\ds^+
K^-$, where both the mixed and the unmixed amplitudes can decay to the same
final $\ds^+ K^-$ state. The interference between both amplitudes results in
the weak phase $\gamma$, but also in a relative QCD phase which is expected
to be small but a priori unknown. This measurement requires an all
hadronic trigger with similar requirements as for \bpipi. We expect an
overall efficiency times acceptance of $\approx 3\cdot10^{-4}$. Because of
rapid $\Bs$ oscillations a time dependent analysis is required but only a
small sample of tagged events is expected at CDF in Run~II. 

The $CP$ asymmetry in \stg\ can also be explored via $B \ra D K$. In
this case the decays are self tagging and a time integrated analysis
can be performed, but the theoretical uncertainties are large.
Thus, performing a
measurement of \stg\ through both modes will be a challenge at CDF
in Run~II. 

\section{Conclusion}
\label{conclude_sec}

In this article we have reviewed recent heavy flavour physics results from
the Tevatron $p\bar p$~collider at Fermilab.
We summarized the status of top quark physics at
CDF and D\O. 
The top production cross section has been measured to be 
$$
\sigtt = (6.4 ^{+1.3} _{-1.2})\ p{\rm b},
$$
and the top quark mass is know with a precision of 
$$
\mtop = (175.0 \pm 3.9 \pm 4.5)\ \gevcc.
$$ 
We also discussed 
recent $B$ physics results from the CDF collaboration.
We summarized CDF's
$B$ hadron lifetime measurements, which are very competitive with the
LEP and SLC results, and discussed latest time dependent $B\bar B$ mixing
results from CDF. 

We also reviewed future prospects of top and $B$ physics at the Tevatron.
In Run~II, starting in 1999, both experiments will measure the 
top quark mass with a statistical uncertainty of 1-2~\gevcc. 
The top
production cross section will be known with a precision of better than
10\%. The future prospects of $B$ physics at CDF will concentrate on
the discovery of $CP$ violation in the $B$ system.
CDF expects to measure
\stb\ in \bjpsi\ with a precision of $\Delta \stb = 0.09$. CDF will also 
search for $CP$ violation in \bpipi\ and expects to measure \sta\ with
a precision of  
$\Delta \sta = 0.10$. A measurement of \stg\ will be challenging in Run~II. 

\subsubsection*{Acknowledgements}
\noindent
It is a pleasure to thank all friends and colleagues from the CDF
and D\O\ collaboration for their excellent work and their help in
preparing this talk. Special thanks go to L.~Galtieri, J.~Lys, M.~Shapiro, and
B.~Winer. A constant source of inspiration and support is my wife Ann,
who I would like to thank for her continuous understanding about the life of
a physicist.   

I would like to thank the
CDF and D\O\ technical support staff at all CDF and D\O\ institutions for their
hard work and dedication.
I also thank the Fermilab Accelerator Division for their 
hard and successful work in commissioning the machine for this physics
run.
The CDF and D\O\ experiments are supported by
the U.S.~Department of Energy;
the U.S.~National Science Foundation;
the Istituto Nazionale di Fisica Nucleare, Italy;
the Ministry of Science, Culture, and Education of Japan; 
the Natural Sciences and Engineering Research Council of Canada;
the National Science Council of the Republic of China;
the Commissariat \`a L'Energie Atomique in France, 
the Ministry for Atomic Energy and the Ministry of Science and
Technology Policy in Russia, 
CNPq in Brazil,
the Departments of Atomic Energy and Science and Education in India,
Colciencias in Colombia,
CONA-CyT in Mexico,
the Ministry of Education, Research Foundation and KOSEF in Korea,
and the A. P. Sloan Foundation.

I gratefully acknowledge 
the organizers of this inspiring meeting which was a genuine pleasure
to attend. I am sorry for the late submission of this manuscript but
as Albert Einstein already pointed out, time is relative:
{\em ``Put your hand on a
hot stove for a minute, and it seems like an hour. Sit with a pretty girl for 
an hour, and it seems like a minute. THAT's relativity.''}

\end{document}